\documentclass[prd,aps,superscriptaddress,nofootinbib,amsmath,amssymb,showpacs]{revtex4}
\usepackage{graphicx}
\usepackage{dcolumn}
\usepackage{bm}
\usepackage{multirow}
\usepackage{adjustbox}
\usepackage{textcomp}
\usepackage{mathtext}
\usepackage{hyperref}
\graphicspath{{./eps/}}

\begin{document}

\title{{\Large Second class currents and T violation in quasielastic neutrino and antineutrino scattering from 
nucleons}}
\author{A. \surname{Fatima}}
\affiliation{Department of Physics, Aligarh Muslim University, Aligarh-202002, India}
\author{M. Sajjad \surname{Athar}}
\email{sajathar@gmail.com}
\affiliation{Department of Physics, Aligarh Muslim University, Aligarh-202002, India}
\author{S. K. \surname{Singh}}
\affiliation{Department of Physics, Aligarh Muslim University, Aligarh-202002, India}
 
\begin{abstract}
 The effect of the second class currents with and without time reversal invariance has been studied in the 
 quasielastic production of nucleons and hyperons induced by neutrinos and antineutrinos from the nucleons. The 
 numerical results are presented for the total scattering cross section~($\sigma$) as well as for the longitudinal, 
 perpendicular and transverse components of the polarization of the final baryons ($p$, $n$, $\Lambda$, $\Sigma^-$, 
 $\Sigma^0$) and muon produced in the quasielastic (anti)neutrino-nucleon scattering induced by the weak charged 
 current. In the case of the production of $\Lambda$ hyperon, which is the most suitable candidate for making the 
 polarization measurements, we have also calculated the $Q^2$ dependence of the polarization observables and the 
 differential scattering cross section ($d\sigma/dQ^2$). The measurement of the polarization observables and their 
 $Q^2$ dependence provides an independent way to determine the nucleon-hyperon transition form factors at high $Q^2$ 
 which can provide tests of the symmetries of the weak hadronic currents like G-invariance, T invariance and SU(3) 
 symmetry. 
 \end{abstract}
\maketitle

\section{Introduction}
The study of the second class currents~(SCC)~\cite{Weinberg:1958ut} in the weak interaction phenomenology has been 
pursued for a long time in the low energy processes of $\beta$ decays and muon capture from the nucleons and 
nuclei~\cite{Holstein, Wilkinson:2000gx,Wilkinson:2000xq, Hardy:2004id, Minamisono:2011zz, Triambak:2017jpw} but there 
are very few experiments in the high energy region of elastic and quasielastic (anti)neutrino scattering where such 
studies have been made~\cite{Ahrens:1988rr, Baker:1981su, Belikov:1983kg}. In the strangeness sector, there are some 
studies done on the semileptonic decays of polarized hyperons where the polarization and/or the angular correlations of 
the final leptons are analyzed to see the effect of the SCC~\cite{Cabibbo:2003cu, Hsueh:1988ar, Oehme:1971rd, 
AlaviHarati:2001xk, Garcia:1971wc, Eimerl:1974ht, Pritchett:1974bb, Oehme:1972fa}. Theoretically, however, there are 
many attempts in the past where the strength of the SCC couplings have been calculated in various models proposed for 
the nucleon structure~\cite{Kellett:1974kh, Chen:1976dq, Holstein:1976mw, Dominguez:1979xp, Donoghue:1981uk, 
Barik:1985in, LieSvendsen:1987ai, Weber:1988ca, Carson:1987gb, Hwang:1988fp, Jena:1991cx, Shiomi:1996np}. Most of these 
studies have been done assuming the validity of the time reversal invariance~(TRI) which requires the strength of the 
SCC couplings to be real as in the case of the first class currents~\cite{commins, Marshak, Pais:1971er, 
LlewellynSmith:1971uhs, Henley:1969uz}. However, if the strength of the SCC is assumed to be described by a quantity 
which is complex or purely imaginary then it implies the violation of TRI leading to the time reversal violating~(TRV) 
observables~\cite{Marshak,Pais:1971er, LlewellynSmith:1971uhs, Henley:1969uz, Cannata:1970br, DeRujula:1970ek, 
Adler:1963, Berman:1964zza, Fujii2, Fujii1, Cabibbo:1964zza, Glashow:1965zz, Okamura:1971pn, Ketley, Egardt}. In past, 
there have been some theoretical calculations to study the effect of the SCC with or without assuming TRI in the 
(anti)neutrino scattering from the nucleons~\cite{Cannata:1970br, DeRujula:1970ek, Adler:1963, Berman:1964zza, Fujii2, 
Fujii1, Cabibbo:1964zza, Glashow:1965zz, Okamura:1971pn, Ketley, Egardt, Block:1965zol, Block:NAL}, and suggestions to 
experimentally study these TRV observables in the $\Delta S = 0$ sector~\cite{Javannic:NAL, Block:1965zol, Block:NAL} in
the bubble chamber experiments at CERN and FNAL. However, no attempts were made due to the difficulties associated with 
the measurements of these observables in performing such experiments. Whereas, there are few experiments done to study 
SCC in (anti)neutrino scattering in the $\Delta S=0$ sector assuming TRI~\cite{Ahrens:1988rr, Baker:1981su, 
Belikov:1983kg}, while there is one experiment in the $\Delta S = 1$ sector done at CERN~\cite{Erriquez:1978pg} where 
the SCC have been studied in the (anti)neutrino scattering with as well as without assuming TRI.

In modern times, very intense neutrino and antineutrino beams are being used to do (anti)neutrino experiments at 
various laboratories around the world with advanced detector technology~\cite{Cao:2017hno, Brailsford:2018dzn, 
Adams:2018gbi, Tufanli:2017mwt, Longhin:2017tfq, Jediny:2017yiq, Patrick:2018gvi, Abe:2016ero, Antonello:2015lea}. 
These experiments are making measurements of various observables i.e. the differential cross sections and the angular 
as well as the energy distributions of the leptons and baryons produced in the final state. The results are interpreted 
using the standard model (SM) of the electroweak interactions assuming the absence of the second class currents and the 
validity of TRI. It seems to be the appropriate time to examine the feasibility of making measurements of those 
observables which enable us to study the effect of the SCC with and without the assumption of TRI. The present work is 
done with the aim of identifying the suitable observables in the quasielastic scattering of neutrinos and antineutrinos 
from the nucleons where the effect of the SCC with or without TRI make important contributions and can be studied 
experimentally in the near future.

Generally, the effect of the SCC is small in the differential cross sections and the angular as well as the energy 
distributions of the final leptons and baryons produced in the $\nu_e (\bar{\nu}_e)$ and $\nu_\mu (\bar{\nu}_\mu)$ 
scattering off the nucleons. This is because the leading order contributions of the second class currents to the 
differential cross sections, which arise due to the interference of the first and second class currents, are 
linear in their coupling strengths, are either proportional to $m_l^2/M^2$ ($m_l$ being the lepton mass and M is the 
nucleon mass) and/or $\Delta^2/M^2$ ($\Delta$ being the mass difference between the initial and final baryons) and are, 
therefore, almost negligible in the reactions induced by $\nu_e (\bar{\nu}_e)$ and $\nu_\mu (\bar{\nu}_\mu)$ in the 
$\Delta S = 0$ sector. The contributions which are independent of $m_l^2/M^2$ and $\Delta^2/M^2$ depend quadratically 
on the strength of the SCC and are small unless the SCC couplings are enormously large~\cite{Akbar:2015yda, Day:2012gb}.
Moreover, their dependence on the momentum transfer is $O(Q^2/M^2)$ or higher and their contribution is quantitatively 
significant only at very high $Q^2$, where the differential cross sections become very small making it difficult to 
isolate the contribution of the SCC. The situation is different in the case of the polarization observables of the 
leptons and baryons produced in the final state where the leading order contributions linear in the strength of the SCC 
couplings also contain terms which are independent of $m_l^2/M^2$ and $\Delta^2/M^2$. Moreover, in the $\Delta S = 1$ 
sector where hyperons are produced, the contributions of $\Delta^2/M^2$ terms are not negligible and could be important 
in favorable kinematics. In view of this, various calculations on the polarization observables of muons and baryons 
produced in the final state of the (anti)neutrino and electron induced weak processes have been done by many 
authors~\cite{Marshak,Pais:1971er, LlewellynSmith:1971uhs, Henley:1969uz, Cannata:1970br, DeRujula:1970ek, Adler:1963, 
Berman:1964zza, Fujii2, Fujii1, Cabibbo:1964zza, Glashow:1965zz, Okamura:1971pn, Ketley, Egardt, Block:1965zol, 
Block:NAL, Bilenky:2013fra, Bilenky:2013iua, Akbar:2016awk, Fatima:2018gjy, Akbar:2017qsf, Graczyk:2017rti, 
Kuzmin:2003ji}. Alternatively, in the case of the reactions where $\tau$ leptons are produced, the terms proportional 
to $m_l^2/M^2$ become quite significant even in the absence of the SCC. Motivated by these considerations some 
calculations of the polarization observables of the $\tau$ leptons have been done recently in the weak processes 
induced by $\tau$ neutrinos~\cite{Kuzmin:2003ji, Kuzmin:2004ke, Graczyk:2004vg, Hagiwara:2004gs, Graczyk:2004uy, 
Valverde:2006yi}. The study of the $Q^2$ dependence of the polarization observables of the electrons and protons in the 
scattering of the polarized electrons on protons has played an important role in determining the electromagnetic form 
factors of the nucleon~\cite{Punjabi:2015bba, Punjabi:2014tna}. Similar studies of the polarization observables in the 
weak sector of the $\Delta S = 0$ and $\Delta S=1$ reactions where polarization of the muons and baryons and their 
$Q^2$ dependence made in the quasielastic and inelastic (anti)neutrino scattering from nucleons can play crucial role 
in determining the weak transition form factors of nucleons.

In the case of the weak $\Delta S = 1$ reactions corresponding to the hyperon production, i.e.
\begin{eqnarray}\label{reaction1}
\bar{\nu}_\mu + p &\longrightarrow& \mu^+ + \Lambda(\Sigma^0) \\
\label{reaction2}
\bar{\nu}_\mu + n &\longrightarrow& \mu^+ + \Sigma^-
\end{eqnarray}
the final hyperons decay predominantly into pions through two body weak decays like $\Lambda (\Sigma^0) \longrightarrow 
p \pi^-$ or $n \pi^0$, and $\Sigma^- \longrightarrow n \pi^-$ and the asymmetry in the angular distribution of the 
pions with respect to a given direction (which can be chosen to be the direction of the momentum of the hyperon, 
perpendicular to it or transverse to the reaction plane) determines the polarization component of the hyperon in that 
direction. In the case of $\Sigma^0$, which promptly decays to $\Lambda$ through the electromagnetic mode of decay, 
$\Sigma^0 \longrightarrow \Lambda \gamma$ before the weak decay of $\Sigma^0$ into pions $p \pi^-$ or $n \pi^0$, the 
pions coming from such $\Lambda$ decays can be experimentally identified to come from $\Sigma^0$~\cite{Erriquez:1978pg} 
and the asymmetries can be determined. This makes the polarization measurements on the hyperon production relatively 
easier than the polarization of the nucleon produced in $\Delta S = 0$ reactions like
\begin{eqnarray}\label{reaction3}
 \nu_\mu + n &\longrightarrow& \mu^- + p \\
 \label{reaction4}
 \bar{\nu}_\mu + p &\longrightarrow& \mu^+ + n,
\end{eqnarray}
where double scattering experiments with the final nucleons are required. Such double scattering experiments with the 
nucleons produced in the neutrino and antineutrino scattering were proposed in the days of bubble chamber experiments 
at CERN and FNAL by Block and his collaborators~\cite{Block:1965zol, Block:NAL} but were not done at that time. These 
experiments are difficult and feasibility of doing them in future could be explored with the liquid argon time 
projection chamber (LArTPC) detectors~\cite{Cao:2017hno, Brailsford:2018dzn, Adams:2018gbi,Tufanli:2017mwt}. 
 
In view of the above discussion, we have in this work considered the quasielastic production of the nucleons and 
hyperons induced by (anti)neutrinos and studied the effect of the SCC with and without TRI on the differential cross 
sections and the polarization observables of the leptons and baryons in the final state. Similar studies have been 
recently done by us in the weak quasielastic scattering induced by electrons on protons~\cite{Akbar:2017qsf, 
Fatima:2018gjy}.
 
In Section~\ref{mat_element}, we describe the structure of the matrix elements for the reactions given in 
Eqs.~(\ref{reaction1})--(\ref{reaction4}) using the SM with three flavors of quarks and discuss the form factors 
corresponding to the first and the SCC using the implications of SU(3) symmetry. The hypothesis of 
the conserved vector current (CVC) for the vector currents and the partially conserved axial vector current (PCAC) for 
the axial vector currents are assumed. In Section~\ref{cross_section}, we present the formalism for calculating the 
differential cross sections and the polarization observables of the final muons and baryons in the presence of the 
SCC with and without TRI. In Section~\ref{results}, we present and discuss our results while the 
summary and conclusions are given in Section~\ref{conclusions}.

  \begin{figure}
 \begin{center}
    \includegraphics[height=3cm,width=6cm]{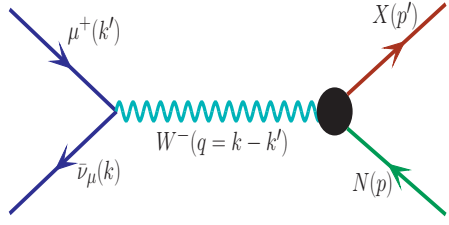}
  \caption{Feynman diagram  for the process $\bar{\nu}_\mu (k) + N (p) \rightarrow \mu^+ (k^\prime) + X (p^\prime)$, 
  where $N (=p,n)$ and $X(=n,\Lambda, \Sigma^0, \Sigma^-)$ represents the initial nucleon and the final baryon 
  respectively. The quantities in the bracket represent four momenta of the corresponding particles.}\label{fyn_hyp}
   \end{center}
 \end{figure}

\section{Matrix element and transition form factors}\label{mat_element}
\subsection{Matrix element}
The transition matrix element for the processes
\begin{eqnarray}\label{process1}
 \nu_\mu (k) + n (p) &\rightarrow& \mu^- (k^\prime) + p (p^\prime), ~\\ 
 \label{process2}
\bar{\nu}_\mu (k) + p (p) &\rightarrow& \mu^+ (k^\prime) + n (p^\prime),  \\ 
\label{process3}
 \bar{\nu}_\mu (k) + p (p) &\rightarrow& \mu^+ (k^\prime) + \Lambda (p^\prime), \\
 \label{process4}
 \bar{\nu}_\mu (k) + p (p) &\rightarrow& \mu^+ (k^\prime) + \Sigma^0 (p^\prime), \\
 \label{process5}
 \bar{\nu}_\mu (k) + n (p) &\rightarrow& \mu^+ (k^\prime) + \Sigma^- (p^\prime), 
\end{eqnarray}
 is written as
 \begin{eqnarray}
 \label{matrixelement}
 {\cal{M}} = \frac{G_F}{\sqrt{2}} a~ l^\mu {{J}}_\mu,
 \end{eqnarray}
 where the quantities in the brackets of Eqs.~(\ref{process1})$-$(\ref{process5}) represent the four momenta of the 
 particles, $G_F$ is the Fermi coupling constant, $a = \cos \theta_c$ for $\Delta S = 0$ processes i.e. 
 Eq.~(\ref{process1}) and (\ref{process2}), $a = \sin \theta_c$ for $\Delta S = 1$ processes i.e. 
 Eq.~(\ref{process3})--(\ref{process5}), and  $\theta_c~(=13.1^\circ)$ is the Cabibbo mixing angle. The leptonic 
 current $l^\mu$ is given by
 \begin{equation}\label{l}
 l^\mu = \bar{u} (k^\prime) \gamma^\mu (1 \pm \gamma_5) u (k),
\end{equation}
where $(+)-$ sign is for (anti)neutrino. The hadronic current ${J}_\mu$ is expressed as:
\begin{equation}\label{j}
 {{J}}_\mu =  \bar{u} (p^\prime) {\Gamma_\mu} u (p)
\end{equation}
with
\begin{equation}\label{gamma}
 {\Gamma_\mu} = V_\mu - A_\mu.
\end{equation}
The vector ($V_\mu$) and the axial vector ($A_\mu$) currents are given by~\cite{Akbar:2016awk, Fatima:2018gjy}:
\begin{eqnarray}\label{vx}
 \langle X(p^\prime) | V_\mu| N(p) \rangle &=& \bar{u}(p^\prime) \left[ \gamma_\mu f_1^{NX}(Q^2)+i\sigma_{\mu \nu} 
 \frac{q^\nu}{M+M^\prime} f_2^{NX}(Q^2) + \frac{2 ~q_\mu}{M+M^\prime} f_3^{NX}(Q^2) \right] u(p),
  \end{eqnarray}
and 
\begin{eqnarray}\label{vy}
  \langle X(p^\prime) | A_\mu| N(p) \rangle &=& \bar{u} (p^\prime) \left[ \gamma_\mu \gamma_5 g_1^{NX}(Q^2) + 
  i\sigma_{\mu \nu} \frac{q^\nu}{M+M^\prime} \gamma_5 g_2^{NX}(Q^2) + \frac{2 ~q_\mu} {M+M^\prime} g_3^{NX}(Q^2) 
  \gamma_5 \right] u(p), 
\end{eqnarray}
which is also expressed as
\begin{eqnarray}\label{vy1}
  \langle X(p^\prime) | A_\mu| N(p) \rangle &=& \bar{u} (p^\prime) \left[ \gamma_\mu \gamma_5 g_1^{NX}(Q^2) + 
  \left(\frac{\Delta}{M + M^\prime} \gamma_\mu \gamma_5 - \frac{p_\mu + p_\mu^\prime}{M + M^\prime} \gamma_5 \right) 
  g_2^{NX}(Q^2) \right. \nonumber \\
  &+& \left. \frac{2 ~q_\mu} {M+M^\prime} g_3^{NX}(Q^2) \gamma_5 \right] u(p), 
\end{eqnarray}
where X represents a nucleon $N (=n,p)$ or a hyperon $Y(=\Lambda, \Sigma^0 \text{ and } \Sigma^-)$, $\Delta = M^\prime 
- M$ with $M$ and $M^\prime$ being the masses of the initial and final baryons. $q_\mu (= k_\mu - k_\mu^\prime = 
p_\mu^\prime -p_\mu)$ is the four momentum transfer with $Q^2 = - q^2, Q^2 >0$. $f_1^{NX}(Q^2)$, $f_2^{NX}(Q^2)$ and 
$f_3^{NX}(Q^2)$ are the vector, weak magnetic and induced scalar form factors and $g_1^{NX}(Q^2)$,$g_2^{NX}(Q^2)$ and 
$g_3^{NX}(Q^2)$ are the axial vector, induced tensor (or weak electric) and induced pseudoscalar form factors, 
respectively. Note the difference in the coefficients containing the mass terms appearing with the form factors 
$f_3^{NX} (Q^2)$ and $g_3^{NX} (Q^2)$ from those defined in Ref.~\cite{Fatima:2018gjy}. We also define here the matrix 
elements of the electromagnetic current for protons and neutrons in terms of the electromagnetic form factors of the 
nucleons as
\begin{eqnarray}\label{vector_ff}
 \langle p(p^\prime) | J_\mu^{em}| p(p) \rangle &=& \bar{u}(p^\prime) \left[ \gamma_\mu f_1^{p}(Q^2)+i\sigma_{\mu \nu} 
 \frac{q^\nu}{2M} f_2^{p}(Q^2) \right] u(p), \\
 \label{vector_ffn}
 \langle n(p^\prime) | J_\mu^{em}| n(p) \rangle &=& \bar{u}(p^\prime) \left[ \gamma_\mu f_1^{n}(Q^2)+i\sigma_{\mu \nu} 
 \frac{q^\nu}{2M} f_2^{n}(Q^2) \right] u(p),
\end{eqnarray}
where $f_1^{p,n} (Q^2)$ and $f_2^{p,n} (Q^2)$ are, respectively, the Dirac and Pauli form factors for the proton and 
neutron.

\subsection{Weak transition form factors}
The weak transition form factors $f_i^{NX} (Q^2)$ and $g_i^{NX} (Q^2)$ for $X=N (n,p)$ or $Y=(\Lambda, \Sigma^0 
\text{ and }\Sigma^-)$ are determined using Cabibbo theory for the electroweak interactions in the SM with three quark 
flavors. In this model, only two constant couplings, $f_1 ( = 1)$ corresponding to the vector current and $g_1 (= 1)$ 
corresponding to the axial vector current occur in the matrix elements of the weak charged currents taken between 
$u(d)$ and $d(u)$ quarks which are considered as point particles. In the case of the matrix elements taken between the 
nucleon and hyperon states which are composites of three quarks bound by the strong interactions and have finite 
dimensions, the couplings are no longer constant and acquire a $Q^2$ dependence i.e. $f_1 (Q^2)$ and $g_1(Q^2)$ known 
as the form factors. Moreover, there are additional form factors which appear due to the finite dimensions of the 
nucleons and hyperons participating in the weak processes and obey the symmetries of the strong interactions 
responsible for giving the nucleons and hyperons their finite size. Using the general principles of Lorentz covariance, 
it is well known that the two additional terms $f_2 (Q^2)$ and $f_3 (Q^2)$ in the vector sector and the two additional 
terms $g_2 (Q^2)$ and $g_3 (Q^2)$ in the axial vector sector appear in Eq.~(\ref{vx}) and (\ref{vy}). 
Weinberg~\cite{Weinberg:1958ut} has introduced a classification scheme of these form factors under G-parity, a symmetry 
of the strong interaction defined as the product of the charge conjugation (C) and rotation in the isospin 
space~(I$-$space) by 180$^\circ$ around y-axis which changes neutron into proton (charge symmetry) and is defined as
\begin{equation}
 G=C e^{i \pi I_y}.
\end{equation}
The vector and axial vector currents of the SM transform  under G-parity as 
\begin{equation}\label{gparity}
G V_\mu G^{-1} = V^\mu, \qquad G A_\mu G^{-1} = - A_\mu. 
\end{equation}
The vector and axial vector currents of the SM with form factors $f_1 (Q^2)$ and $g_1 (Q^2)$ transform as in 
Eq.~(\ref{gparity}) and are termed as the first class currents. Out of the induced form factors, $f_2 (Q^2)$ and $g_3 
(Q^2)$ also transform as in Eq.~(\ref{gparity}) in their respective currents and are classified as the first class 
currents, while the form factors $f_3 (Q^2)$ and $g_2(Q^2)$ transform with opposite sign in their respective currents 
and are classified as the SCC. Since the SM has only first class currents, terms with $f_1 (Q^2)$ and $g_1 (Q^2)$ form 
factors and the induced currents comprising of $f_2 (Q^2)$ and $g_3 (Q^2)$ would contribute while the SCC form factors 
should vanish i.e. 
\begin{eqnarray}
 f_3 (Q^2) &=& 0, \\
 g_2 (Q^2) &=& 0,
\end{eqnarray}
if G-parity is a good quantum number of strong interactions. 

In the $\Delta S = 0$ sector, the violation of G-parity due to $(u-d)$ mass difference or the intrinsic charge symmetry 
violation of the strong interactions is very small, and the form factors $f_3 (Q^2)$ and $g_2 (Q^2)$ are expected to be 
very small. Moreover, in the vector sector, the charged weak vector currents $V_\mu$ along with the isovector part of 
the electromagnetic current ($J_\mu^{em}$) is assumed to form an isotriplet. Therefore, the principle of the 
conservation of the electromagnetic current, applied also to the weak currents, leads to the hypothesis of CVC and 
predicts $f_3 (Q^2) = 0$. This also relates the weak form factors $f_1 (Q^2)$ and $f_2 (Q^2)$ to the isovector 
electromagnetic form factors of the nucleons. However, in the axial vector sector there is no such constraint on the 
form factor $g_2 (Q^2)$ and it could be non-vanishing albeit small. It is because of this reason that most of the 
experiments in $\Delta S = 0$ sector are analyzed for the search of the SCC assuming $f_3 (Q^2)=0$ with a non-vanishing 
$g_2 (Q^2)$ which is found to be small~\cite{Holstein, Wilkinson:2000gx, Wilkinson:2000xq, Hardy:2004id, 
Minamisono:2011zz, Triambak:2017jpw, Akbar:2015yda, Day:2012gb}. 

Theoretically, the values of $g_2 (Q^2)$ (in the case of $p-n$ transition) calculated in various models of the nucleon 
structure are found to be very small~\cite{Jena:1991cx, Donoghue:1981uk, Barik:1985in, Shiomi:1996np, Kellett:1974kh, 
Weber:1988ca, Dominguez:1979xp, Chen:1976dq, Holstein:1976mw, Carson:1987gb, LieSvendsen:1987ai, Hwang:1988fp}. 
Experimentally they are also found to be small and consistent with zero with large uncertainties in the measurements. 
The old experimental results are summarized well by Holstein~\cite{Holstein} and Wilkinson~\cite{Wilkinson:2000gx, 
Wilkinson:2000xq}, while there are some new measurements of $g_2 (Q^2)$ in nuclear $\beta$ decay 
measurements~\cite{Minamisono:2011zz, Triambak:2017jpw}.

The general properties of the weak form factors consistent with the constraints due to the symmetry properties of the 
weak hadronic currents are summarized as~\cite{Marshak, Pais:1971er, LlewellynSmith:1971uhs}:

\begin{itemize}
 \item [(a)] T invariance implies that all the form factors $f_{1-3} (Q^2)$ and $g_{1-3} (Q^2)$ are real.
 
 \item [(b)] The assumption that the weak vector currents and its conjugate along with the isovector part of the 
 electromagnetic current form an isotriplet implies that the weak vector form factors $f_1 (Q^2)$ and $f_2 (Q^2)$ are 
 related to the isovector electromagnetic form factors of the nucleon i.e. $f_{1,2}^{np} (Q^2) = f_{1,2}^p (Q^2) - 
 f_{1,2}^n (Q^2)$.
 
 \item [(c)] The principle of CVC of the weak currents implies that $f_3 (Q^2) = 0$.
 
 \item [(d)] The principle of G-invariance implies that $f_3 (Q^2)=0$ and $g_2 (Q^2) =0$.
 
 \item [(e)] The hypothesis of PCAC relates the form factor $g_3 (Q^2)$ to the form factor $g_1 (Q^2)$ through the 
 Goldberger-Treiman~(GT) relation. 

\end{itemize}
 
 Therefore, the form factor $f_3 (Q^2) = 0$ by the CVC and G-invariance. The form factor $g_2 (Q^2) = 0$ only in the 
 presence of G-invariance. In the case of G-noninvariance, $g_2 (Q^2)$ is non-vanishing and if it is real, it preserves 
 TRI whereas if it is purely imaginary or complex, the SCC in the axial vector sector violate the TRI.
 
 In the case of $\Delta S = 1$ processes shown in Eqs.~(\ref{process3})--(\ref{process5}), while (a) above is always 
 true, whereas (b) is replaced by a more general relation between the $\Delta S = 1$ vector form factors and the 
 electromagnetic nucleon form factors in the limit of exact SU(3) symmetry which is discussed below. The other 
 properties (c)--(e) apply only in the limit of exact SU(3) symmetry. However, SU(3) symmetry is known to be only an 
 approximate symmetry due to the large mass difference between $u(d)$ and $s$ quarks and explicit dynamical breaking of 
 SU(3) symmetry leading to non-zero values of $f_3 (Q^2)$ and $g_2 (Q^2)$. While $f_3 (Q^2)$ is still constrained by 
 the CVC, there is no such constraint for $g_2 (Q^2)$ and it could be large. Indeed some theoretical calculations find 
 the value of $|g_2^{p \Lambda}/g_1^{p \Lambda}|$ to be in the range of 0.2 to 0.5~\cite{Barik:1985in, Donoghue:1981uk},
 whereas others find a larger value~\cite{Chen:1976dq, Holstein:1976mw, Garcia:1971wc, Hsueh:1988ar, Barik:1985in, 
 Carson:1987gb, Hwang:1988fp, Donoghue:1981uk}. Experimentally some earlier analysis of semileptonic hyperon 
 decays~(SHD) suggest larger values of $g_2^{p \Lambda(\Sigma)}$ and a latest experiment~\cite{AlaviHarati:2001xk} 
 reports $|g_2^{\Xi \Sigma}/f_1^{\Xi \Sigma}| = -1.77 ~\substack{+2.1 \\ -2.0} \pm 0.5$ which is consistent with zero 
 within the large systematic uncertainties of the experiment~\cite{Cabibbo:2003cu, AlaviHarati:2001xk}. It should be 
 noted that most of these analyses are done assuming a real value of $g_2 (Q^2)$. Moreover, the most extensive analysis 
 of the experimental data on $\Delta S = 1$ SHD finds the effect of SU(3) symmetry breaking to be very small provided 
 the physical masses of the nucleons and the hyperons are used in the analysis~\cite{Cabibbo:2003cu}. In view of this, 
 we calculate the form factors corresponding to the first class currents i.e. $f_1(Q^2)$, $f_2(Q^2)$, $g_1(Q^2)$ and 
 $g_3(Q^2)$ in a SU(3) symmetric analysis using physical masses for the nucleons and hyperons and take $f_3 (Q^2) = 0$. 
 For $g_2 (Q^2)$, we take a numerical value guided by some calculations done for $g_2 (Q^2)$ in the quark 
 model~\cite{Donoghue:1981uk, Hwang:1988fp, Barik:1985in, Jena:1991cx} and its value used in earlier studies done on 
 the effect of the SCC and TRI in (anti)neutrino scattering~\cite{Berman:1964zza, Fujii1, Fujii2}.
 
 \subsection{SU(3) symmetry and the weak transition form factors}
The weak vector ($V_\mu$) and the axial vector ($A_\mu$) currents corresponding to the $\Delta S=0$ and $\Delta S = 1$ 
hadronic currents whose matrix elements are defined between the states $|N\rangle$ and $|X\rangle$ in 
Eqs.~(\ref{process1})--(\ref{process5}) are assumed to belong to the octet representation of the SU(3) and are 
defined as
\begin{eqnarray}\label{su3}
V^\mu_i&=&\bar{q}F_i\gamma^\mu q\nonumber\\
A^\mu_i&=&\bar{q}F_i\gamma^\mu\gamma^5 q,
\end{eqnarray}
 where $F_i=\frac{\lambda_i}{2}$($i=1-8$) are the generators of flavor SU(3) and $\lambda_i$s are the well known 
 Gell-Mann matrices.
 The generators of the SU(3) group $F_i$ obey the following commutation and anticommutation algebra
 \begin{eqnarray}\label{algebra}
[F_i,F_j]&=&if_{ijk}F_k\nonumber\\
\{F_i,F_j\}&=&\frac{1}{3}\delta_{ij} + d_{ijk}F_k,~~ i,j,k=1-8,
\end{eqnarray}
 where $f_{ijk}$ and $d_{ijk}$ are the structure constants, and are antisymmetric and symmetric, 
 respectively, under the interchange of any two indices~\cite{Close}.

 The electromagnetic current~($J_{em}^\mu$) and the weak vector ($V_{\pm}^\mu$) and the axial vector 
 ($A_{\pm}^\mu$) charged currents are defined in terms of $V_i^\mu$ and $A_i^\mu$; $i=1-8$, as
 \begin{eqnarray}\label{jmu}
 J^\mu_{em}&=& V^\mu_3~+~ \frac{1}{\sqrt{3}} V^\mu_8, \nonumber\\
 V^\mu_{\pm}&=&\left[V^\mu_1~\pm~ i V^\mu_2\right]cos\theta_c~+~\left[V^\mu_4~\pm~ i V^\mu_5\right]sin\theta_c, 
 \nonumber \\
 A^\mu_{\pm}&=&\left[A^\mu_1~\pm~ i A^\mu_2\right]cos\theta_c~+~\left[A^\mu_4~\pm~ i A^\mu_5\right]sin\theta_c.
 \end{eqnarray}

 In the Cabibbo theory, isovector electromagnetic current $J_{em}^\mu$ along with the weak vector currents $V_{\pm}^\mu$
 are assumed to transform as an octet of vector currents under SU(3). Similarly, the axial vector currents are also 
 assumed to transform as an octet under SU(3). The form factors defined in the matrix element of an octet of the vector 
 (axial vector) currents taken between the octets of the initial and the final baryon states as defined in 
 Eqs.~(\ref{vx})--(\ref{vector_ffn}) can, therefore, be expressed in terms of the two couplings of the vector (axial 
 vector) currents corresponding to the symmetric and antisymmetric octets according to the decomposition:
\begin{equation}
 8 \times 8 = 1 + 8^S + 8^A + 10 + \overline{10} + 27
\end{equation}
and the corresponding SU(3) Clebsch-Gordan coefficients. In general, the expression for the matrix element of the 
transition between the two states of baryons (say $B_i$ and $B_k$), through the SU(3) octet ($V_j$ or $A_j$) of 
currents can be written as~\cite{Renton}:
 \begin{eqnarray}\label{bb}
< B_i | V_j | B_k > &=& if_{ijk}F^V + d_{ijk} D^V, \\
\label{b2}
< B_i | A_j | B_k > &=& if_{ijk}F^A + d_{ijk} D^A.
 \end{eqnarray}
  $F^{V}$ and $D^{V}$ are determined from the experimental data on the electromagnetic form factors, and $F^{A}$ and 
  $D^{A}$ are determined from the experimental data on the semileptonic decays of the nucleons and hyperons. Explicitly,
  the form factors defined in Eqs.~(\ref{vx})--(\ref{vy1}) can be expressed as 
\begin{eqnarray}\label{fi}
f_i (Q^2) &=& a F_i^V (Q^2) + b D_i^V  (Q^2) \qquad i=1,2,3 \\
\label{gi}
g_i (Q^2) &=& a F_i^A (Q^2) + b D_i^A  (Q^2) \qquad i =1,2,3
\end{eqnarray}
The Clebsch-Gordan coefficients $a$ and $b$ can be calculated for each transition, if we specify the quantum numbers 
($| I,~I_3,~Y\rangle$) of the initial and the final state and the current operators $V_\mu$, $A_\mu$ and $J_\mu^{em}$ 
in the octet representation. A straightforward calculation of the various Clebsch-Gordan coefficients in the case of 
weak $\Delta S=0$, $\Delta S=1$ hadronic currents and the electromagnetic currents (in the case of vector currents) 
gives the values of $a$ and $b$ which are obtained using Eqs.~(\ref{algebra}), (\ref{jmu}), (\ref{bb}) and (\ref{b2}) 
and are presented in Table~\ref{tabI}.

\begin{table*}[h]
\begin{tabular}{|c|c|c|c|}\hline
~~ Interaction ~~ & ~~ Transitions ~~ & ~~ $a$ ~~ & ~~ $b$ ~~ \\ \hline
~~ Electromagnetic ~~ & ~~ $p\rightarrow p$ ~~ & ~~ 1 ~~ & ~~ 1/3 ~~ \\
~~ Vector ~~& ~~ $n\rightarrow n$ ~~ & ~~ 0 ~~ & ~~ -2/3 ~~ \\ \hline 
& ~~ $n\rightarrow p$ ~~ & ~~1~~ & ~~1~~ \\
~~ Weak vector ~~ & ~~ $p\rightarrow \Lambda$ ~~ & ~~ $-\sqrt{\frac{3}{2}}$ ~~ & ~~ $-\frac{1}{\sqrt{6}}$ ~~\\
~~ and axial vector ~~ & ~~ $p\rightarrow \Sigma^0$ ~~ & ~~ $-\frac{1}{\sqrt{2}}$ ~~ & ~~ $\frac{1}{\sqrt{2}}$ ~~\\
& ~~ $n\rightarrow \Sigma^-$ ~~ & ~~ $-1$ ~~ & ~~ 1 ~~ \\ \hline
\end{tabular}
\caption{Values of the coefficients $a$ and $b$ given in Eqs.~(\ref{fi})$-$(\ref{gi}).}
\label{tabI}
\end{table*}

\subsubsection{Vector form factors}
 
 In the case of the vector currents, the following equations are obtained for the electromagnetic form factors using 
 the values of $a$ and $b$ from Table~\ref{tabI}:
 \begin{eqnarray}\label{fipp}
f_i^{p \rightarrow p} (Q^2) &=& f_i^p (Q^2) = F_i^V (Q^2) + \frac{1}{3} D_i^V  (Q^2) \qquad i=1,2 \\
\label{finn}
f_i^{n \rightarrow n} (Q^2) &=& f_i^n (Q^2) = - \frac{2}{3} D_i^V  (Q^2) \qquad  \qquad \qquad i =1,2
 \end{eqnarray}
These equations are solved to determine $F_i^V (Q^2)$ and $D_i^V (Q^2)$ in terms of the electromagnetic form factors 
$f_{1,2}^p (Q^2)$ and $f_{1,2}^n (Q^2)$ for the protons and neutrons. Once the functions $F_i^V (Q^2)$ and $D_i^V(Q^2)$ 
are determined, all the form factors $f_{1,2}^{NX} (Q^2)$ for the transitions under consideration are determined with 
the help of the coefficients $a$ and $b$ given in Table~\ref{tabI}. The expressions for the vector form factors in 
terms of the electromagnetic form factors $f_{1,2}^p (Q^2)$ and $f_{1,2}^n (Q^2)$ for the various processes given in 
Eqs.~(\ref{process1})--(\ref{process5}) are given as 
\begin{eqnarray}
 f_{1,2}^{np}(Q^2)&=&f^p_{1,2}(Q^2)-f^n_{1,2}(Q^2), \\
 f_{1,2}^{p \Lambda}(Q^2)&=& -\sqrt{\frac{3}{2}}~f_{1,2}^p(Q^2), \\
 f_{1,2}^{n \Sigma^-}(Q^2)&=& -\left[f_{1,2}^p(Q^2) + 2 f_{1,2}^n(Q^2) \right],  \\
 f_{1,2}^{p \Sigma^0}(Q^2)&=& -\frac{1}{\sqrt2}\left[f_{1,2}^p(Q^2) + 2 f_{1,2}^n(Q^2) \right].
\end{eqnarray}
The electromagnetic form factors $f_{1,2}^p (Q^2)$ and $f_{1,2}^n (Q^2)$ are expressed in terms of the Sachs electric 
and magnetic form factors $G_E^{p,n} (Q^2)$ and $G_M^{p,n} (Q^2)$ of the nucleons as
\begin{eqnarray}\label{f1pn}
f_1^{p,n}(Q^2)&=&\left(1+\frac{Q^2}{4M^2}\right)^{-1}~\left[G_E^{p,n}(Q^2)+\frac{Q^2}{4M^2}~G_M^{p,n}(Q^2)\right],\\
\label{f2pn}
f_2^{p,n}(Q^2)&=&\left(1+\frac{Q^2}{4M^2}\right)^{-1}~\left[G_M^{p,n}(Q^2)-G_E^{p,n}(Q^2)\right].
\end{eqnarray}
For $G_E^{p,n}(Q^2)$ and $G_M^{p,n}(Q^2)$ various parameterizations are available in the literature and in our 
numerical calculations, we have used the parameterization given by Bradford et al.~\cite{Bradford:2006yz}.
 
\subsubsection{Axial vector form factors}
 
 The axial vector form factors $g_i^{NX} (Q^2) (i=1,2,3)$ are expressed in terms of the two functions $F_i^{A} (Q^2)$ 
 and $D_i^{A} (Q^2)$ corresponding to the asymmetric and symmetric couplings of the two octets. But we express the 
 form factors $g_i^{NX} (Q^2)$ in terms of $g_i (Q^2)$ and $x_i (Q^2)$ which are defined as 
 \begin{eqnarray}\label{gnp}
  g_i (Q^2) &=& F_i^A (Q^2) + D_i^A (Q^2) = g_i^{np} (Q^2) \\
  x_i (Q^2) &=& \frac{F_i^A (Q^2)}{F_i^A (Q^2) + D_i^A (Q^2)}; \qquad i=1-3
 \end{eqnarray}
 and the expressions for the axial vector transition form factors for the various processes given in 
 Eq.~(\ref{process1})--(\ref{process5}) are given as:
 \begin{eqnarray}\label{gnp}
 g_{1,2,3}^{np}(Q^2)&=& g_{A,2,3}^{np}(Q^2), \\
 \label{gplam}
 g_{1,2,3}^{p \Lambda}(Q^2)&=& -\frac{1}{\sqrt{6}}(1+2x_{1,2,3}) g_{A,2,3}^{np} (Q^2), \\
 \label{gnsig}
 g_{1,2,3}^{n \Sigma^-}(Q^2)&=& (1-2x_{1,2,3})g_{A,2,3}^{np}(Q^2),  \\
 \label{gpsig}
 g_{1,2,3}^{p \Sigma^0}(Q^2)&=& \frac{1}{\sqrt2}(1-2x_{1,2,3})g_{A,2,3}^{np}(Q^2).
\end{eqnarray}
 In the following we describe the explicit forms of the axial vector form factors used for calculating the numerical 
 results.

 \begin{itemize}
  \item [(a)] \underline{Axial vector form factor $g_1^{NX} (Q^2)$:}  
  We note from Eq.~(\ref{gnp}), that $g_1^{np} (Q^2)$ is the axial vector form factor for $n \rightarrow p$ 
  transition which is determined experimentally from the quasielastic (anti)neutrino scattering from the nucleons and 
  is parameterized as
  \begin{equation}\label{ga}
  g_1^{np} (Q^2) = g_A^{np} (Q^2) = \frac{g_A (0)}{\left( 1 + \frac{Q^2}{M_A^2} \right)^2},
  \end{equation}
  with $g_A(0) = 1.267$~\cite{Cabibbo:2003cu} and $M_A = 1.026$ GeV~\cite{Bernard:2001rs}. The parameter $x_1 (Q^2)$ 
  occurring in Eqs.~(\ref{gplam})--(\ref{gpsig}) for $g_1^{NY} (Q^2)$ ($Y = \Lambda, \Sigma^0, \Sigma^-$) is determined 
  at low $Q^2$ from the analysis of SHD and is found to be $x_1 (Q^2 \approx 0) = 0.364$~\cite{Cabibbo:2003cu}. There 
  is no experimental information about the $Q^2$ dependence of $x_1 (Q^2)$, therefore, we assume it to be constant i.e.
  $x_1 (Q^2) \approx x_1(0) = 0.364$ for convenience.
  
    \item [(b)] \underline{Second class current form factor $g_2^{NX} (Q^2)$:}
  The expression for $g_2^{NY} (Q^2)$ for $Y (= \Lambda, \Sigma^-, \Sigma^0)$ are given in 
  Eqs.~(\ref{gplam})--(\ref{gpsig}) in terms of $g_2^{np} (Q^2)$ and $x_2 (Q^2)$, where $g_2^{np} (Q^2)$ is 
  parameterized as 
  \begin{equation}\label{g2}
  g_2^{np} (Q^2) = \frac{g_2^{np} (0)}{\left( 1 + \frac{Q^2}{M_2^2} \right)^2},
  \end{equation} 
  in analogy with $g_1^{np} (Q^2) = g_A (Q^2)$. There is some information on $g_2^{np} (Q^2)$ from neutrino and 
  antineutrino scattering off the nucleons~\cite{Ahrens:1988rr, Belikov:1983kg, Baker:1981su}. It is shown that the 
  value of $g_2^{np} (0)$ is correlated with the value of $M_2$ used in the analysis. There exists theoretical 
  calculations for the $Re~g_2^{np} (0)$ and $Re~g_2^{NY} (0)$ for $Y=\Lambda, \Sigma^-, \Sigma^0$ using quark 
  models~\cite{Donoghue:1981uk, Hwang:1988fp, Barik:1985in, Jena:1991cx}. There are also some older calculations of 
  T violating effects in weak processes where, phenomenologically, the values of $Im~g_2^{N X} (Q^2)$ have been used 
  in a large range of $1<Im~g_2^{pn}(0)<10$, like the works of Fearing et al. ($Im~g_2^{pn}(0) = 1, 5, 
  10$)~\cite{Fearing:1969nr}, Berman and Veltman ($Im~g_2^{pn} (0) = 3.71, 6$)~\cite{Berman:1964zza} and Fujii and 
  Yamaguchi ($Im~g_2^{p \Lambda}(0) = 1.92$)~\cite{Fujii1}. However, there is no information about $x_2(Q^2)$. In view 
  of this we vary the value of $Re~g_2^{np}(0)$ and $Im~g_2^{np}(0)$ in the range of $0-3$ to study the effect of the 
  SCC in (anti)neutrino scattering~\cite{Berman:1964zza, Fujii1, Fujii2} and use $M_2 = M_A$. For the $Q^2$ dependence 
  of the form factor i.e. $g_2^{NY}(Q^2)$, we use the SU(3) symmetric expressions for $g_2^{NY} (Q^2)$ taken to be of 
  dipole form given in Eq.~(\ref{g2}) for the various transitions given in Eqs.~(\ref{gplam})--(\ref{gpsig}), in terms 
  of $g_2^{np} (Q^2)$ and $x_2 (Q^2)$, treating $x_2 (Q^2)$ to be constant and take $x_2 = x_1$ for simplicity.
 
  \item [(c)] \underline{The induced pseudoscalar form factor $g_3^{NX} (Q^2)$:}
  We see from Eqs.~(\ref{gplam})--(\ref{gpsig}), that $g_3^{NY} (Q^2)$ is determined in terms of $g_3^{np} (Q^2)$ and 
  $x_3 (Q^2)$ for $Y = \Lambda$, $\Sigma^0$ and $\Sigma^-$. In general, the contribution of $g_3^{np} (Q^2)$ to the 
  (anti)neutrino scattering cross sections is proportional to $m_l^2$, where $m_l$ is the mass of the corresponding 
  charged lepton, and is very small. There is very little experimental information available about $g_3^{np} (Q^2)$ at 
  very small $Q^2$ from the muon capture experiments in the nucleons and nuclei~\cite{Bernard:1998rs}. This information 
  is consistent with the prediction of $g_3^{np} (Q^2)$ obtained using the hypothesis of PCAC and GT 
  relation~\cite{Goldberger:1958vp} in the $\Delta S = 0$ sector which is given as:
  \begin{equation}\label{g3}
  g_3^{np}(Q^2)=\frac{2M^2 g_A^{np}(Q^2)}{m_\pi^2+Q^2}.
  \end{equation}
  
  There is no information about $x_3(Q^2)$ from the analysis of the SHD due to the small contribution of 
  $g_3^{NY} (Q^2)$. In the absence of any information about $x_3 (Q^2)$, we do not use SU(3) symmetric expressions for 
  $g_3^{NY} (Q^2)$ given in Eqs.~(\ref{gplam})--(\ref{gpsig}). Instead we use the expression for $g_3^{NY} (Q^2)$ using 
  the generalized PCAC and GT relation~\cite{Goldberger:1958vp} for the $\Delta S = 1$ currents given by 
  Nambu~\cite{Nambu:1960xd}, i.e.
  \begin{equation}\label{g3_Nambu}
   g_3^{NY}(Q^2)=\frac{(M + M_Y)^2}{2(m_K^2+Q^2)}g_1^{NY}(Q^2),
  \end{equation}
  where $m_K$ 
  is the mass of kaon and $g_1^{NY} (Q^2)$ is given in Eqs.~(\ref{gplam})--(\ref{gpsig}), for $Y = \Lambda, \Sigma^-, 
  \Sigma^0$.
  
 \end{itemize}

\section{Cross section and the polarization observables}\label{cross_section}
\subsection{Cross section}
The general expression of the differential cross section for the processes~(\ref{process1})$-$(\ref{process5}), in the 
laboratory frame, is written as
 \begin{eqnarray}
 \label{crosv.eq}
 d\sigma&=&\frac{1}{(2\pi)^2}\frac{1}{4 M E_{\nu}}\delta^4(k+p-k^\prime-p^\prime) 
 \frac{d^3k^\prime}{2E_{k^\prime}} \frac{d^3p^\prime}{2E_{p^\prime}} \overline{\sum} \sum |{\cal{M}}|^2,
 \end{eqnarray}
where $E_\nu$ = ($E_{{\bar{\nu}_{\mu}}}$)$E_{\nu_\mu}$ is the incoming (anti)neutrino energy. The transition matrix 
element squared is defined as:
\begin{equation}\label{matrix}
  \overline{\sum} \sum |{\cal{M}}|^2 = \frac{G_F^2 a^2}{2} \cal{J}^{\mu \nu} \cal{L}_{\mu \nu},
\end{equation} 

where the hadronic and the leptonic tensors are obtained using Eqs.~(\ref{l}) and (\ref{j}) as
\begin{eqnarray}\label{J}
\cal{J}_{\mu \nu} &=& \overline{\sum} \sum J_\mu J_\nu^\dagger = \frac{1}{2} \mathrm{Tr}\left[\Lambda({p^\prime}) 
\Gamma_{\mu} \Lambda({p}) \tilde{\Gamma}_{\nu} \right], \\
  \label{L}
\cal{L}^{\mu \nu} &=& \overline{\sum} \sum l_\mu l_\nu^\dagger = \mathrm{Tr}\left[\gamma^{\mu}(1 \pm \gamma_{5}) 
\Lambda({k^\prime})\gamma^{\nu}(1 \pm \gamma_{5}) \Lambda({k})~\right],
\end{eqnarray} 
with $\tilde{\Gamma}_{\nu} =\gamma^0 \Gamma^{\dagger}_{\nu} \gamma^0$ and the expression for $\Gamma_\nu$ is given in 
Eq.~(\ref{gamma}). The spin $\frac{1}{2}$ projection operator $\Lambda (P)$ for momentum $P=k, k^\prime,p,p^\prime$ 
corresponding to the initial and the final baryons and the leptons are given by
\begin{equation}\label{lam}
 \Lambda(P)=(P\!\!\!\!/+M_P),
\end{equation}
where $M_P$ is the mass of the particle with momentum $P$.
 
 Following the above definitions, the differential scattering cross section $d\sigma/dQ^2$ for the processes given in 
 Eq.~(\ref{process1})--(\ref{process5}) is written as
\begin{equation}\label{dsig}
 \frac{d\sigma}{dQ^2}=\frac{G_F^2 a^2}{8 \pi {M}^2 {E^2_\nu}} N(Q^2),
\end{equation}
where $N(Q^2) = \cal{J}^{\mu \nu} \cal{L}_{\mu \nu}$ and the expression of $N(Q^2)$ is given in the Appendix-I. The 
expression of $N(Q^2)$ is consistent with the expression given by Llewellyn Smith~\cite{LlewellynSmith:1971uhs} in the 
limit $M^\prime = M$ and $g_2 (Q^2) = 0$. In this limit, it is also consistent with Bilenky and 
Christova~\cite{Bilenky:2013fra}.

\subsection{Polarization of the final baryon}
Using the covariant density matrix formalism, the polarization 4-vector($\xi^\tau$) of the baryon produced in the 
final state in reactions~(\ref{process1})$-$(\ref{process5}) is written as~\cite{Bilenky}:
\begin{equation}\label{polar}
\xi^{\tau}=\frac{\mathrm{Tr}[\gamma^{\tau}\gamma_{5}~\rho_{f}(p^\prime)]}
{\mathrm{Tr}[\rho_{f}(p^\prime)]},
\end{equation}
where the spin density matrix $\rho_f(p^\prime)$ corresponding to the final baryon of momentum $p^\prime$ is 
given by 
\begin{equation}\label{polar1}
 \rho_{f}(p^\prime)= {\cal L}^{\alpha \beta} ~\mathrm{Tr}[\Lambda(p')\Gamma_{\alpha} \Lambda(p)\tilde{\Gamma}_{\beta} 
 \Lambda(p')].
\end{equation} 
In the above expression, $\Lambda (p)$ and $\Gamma_\alpha$ are given in Eqs. (\ref{lam}) and (\ref{gamma}) respectively.
Using the following relations:
\begin{equation}\label{polar3}
\Lambda(p')\gamma^{\tau}\gamma_{5}\Lambda(p')=2M^\prime \left(g^{\tau\sigma}-\frac{p'^{\tau}p'^{\sigma}}{M^{\prime 2}}
\right)\Lambda(p')\gamma_{\sigma}\gamma_{5}
\end{equation}
and
\begin{equation}\label{polar31}
 \Lambda(p^\prime)\Lambda(p^\prime) = 2M^\prime \Lambda(p^\prime),
\end{equation}
 $\xi^\tau$ defined in Eq.~(\ref{polar}) may be rewritten after some algebra as
\begin{equation}\label{polar4}
\xi^{\tau}=\left( g^{\tau\sigma}-\frac{p'^{\tau}p'^{\sigma}}{M^{\prime 2}}\right) 
\frac{  {\cal L}^{\alpha \beta}  \mathrm{Tr}
\left[\gamma_{\sigma}\gamma_{5}\Lambda(p')\Gamma_{\alpha} \Lambda(p)\tilde{\Gamma}_{\beta} \right]}
{ {\cal L}^{\alpha \beta} \mathrm{Tr}\left[\Lambda(p')\Gamma_{\alpha} \Lambda(p)\tilde{\Gamma}_{\beta} \right]}.
\end{equation}
Note that in Eq.~(\ref{polar4}), $\xi^\tau$ is manifestly orthogonal to $p^{\prime \tau}$, \textit{i.e.} $p^\prime 
\cdot \xi=0$. Moreover, the denominator is directly related to the differential cross section given in Eq.~(\ref{dsig}).

With ${\cal J}^{\alpha \beta}$ and ${\cal L}_{\alpha \beta}$ given in Eqs.~(\ref{J}) and (\ref{L}), respectively, an 
expression for $\xi^\tau$ is obtained. In the laboratory frame where the initial nucleon is at rest, the polarization 
vector $\vec{\xi}$ for the final baryon is calculated to be a function of 3-momenta of incoming antineutrino 
$({\vec{k}})$ and outgoing baryon $({\vec{p}}\,^{\prime})$, and is given as  
\begin{equation}\label{3pol}
 \vec{\xi} =\left[{A^h(Q^2)\vec{ k}} + B^h(Q^2){\vec{p}}\,^{\prime} + C^h(Q^2) M (\vec{k} \times {\vec{p}}\,^{\prime} ) 
 \right], 
\end{equation}
where the expressions of $A^h(Q^2)$, $B^h(Q^2)$ and $C^h(Q^2)$ are given in the Appendix-I.

\begin{figure}
 \begin{center}  
        \includegraphics[height=6cm,width=11cm]{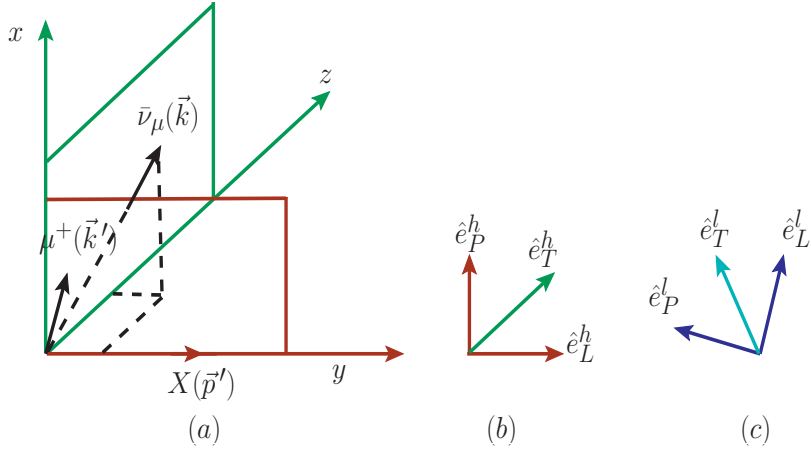}
  \caption{(a) Momentum and polarization directions of the final baryon and the lepton. $\hat{e}_{L}^{h,l}$, 
  $\hat{e}_{P}^{h,l}$ and $\hat{e}_{T}^{h,l}$ represent the orthogonal unit vectors corresponding to the longitudinal, 
  perpendicular and transverse directions with respect to the momentum of the final baryon in (b) and the final lepton 
  in (c).}\label{TRI}
   \end{center}
 \end{figure}

We now expand the polarization vector $\vec{\xi}$ along the orthogonal directions, ${\hat{e}}_L^h$, ${\hat{e}}_P^h$ 
and ${\hat{e}_T^h}$ in the reaction plane corresponding to the longitudinal, perpendicular and transverse directions, 
defined as
\begin{equation}\label{vectors}
\hat{e}_{L}^h=\frac{\vec{ p}^{\, \prime}}{|\vec{ p}^{\,\prime}|},\qquad
\hat{e}_{P}^h=\hat{e}_{L}^h \times \hat{e}_T^h,\qquad   {\rm where}~~~~~ 
\hat{e}_T^h=\frac{\vec{ p}^{\,\prime}\times \vec{ k}}{|\vec{ p}^{\,\prime}\times \vec{ k}|},
\end{equation}
and have depicted in Fig.~\ref{TRI}. We then write $\vec{\xi}$ as:
 \begin{equation}\label{polarLab}
\vec{\xi}=\xi_{L} \hat{e}_{L}^h+\xi_{P} \hat{e}_{P}^h + \xi_{T} \hat{e}_{T}^h,
\end{equation}
such that the longitudinal, perpendicular and transverse components of the polarization vector ($\vec{\xi}$) in the 
laboratory frame are given by
\begin{equation}\label{PL}
 \xi_L(Q^2)=\vec{\xi} \cdot \hat{e}_L^h,\qquad \xi_P(Q^2)= \vec{\xi} \cdot \hat{e}_P^h, \qquad \xi_T(Q^2) = \vec{\xi}
 \cdot \hat{e}_T^h.
\end{equation}
From Eq.~(\ref{PL}), the longitudinal $P_L^h(Q^2)$, perpendicular $P_P^h(Q^2)$ and transverse $P_T^h(Q^2)$ components 
of the polarization vector defined in the rest frame of the initial nucleon are given by 
\begin{equation}\label{PL1}
 P_L^h(Q^2)=\frac{M^\prime}{E_{p^\prime}} \xi_L(Q^2), \qquad P_P^h(Q^2)=\xi_P(Q^2), \qquad P_T^h(Q^2)=\xi_T(Q^2),
\end{equation}
where $\frac{M^\prime}{E_{p^\prime}}$ is the Lorentz boost factor along ${\vec p}\, ^\prime$.
With the help of Eqs.~(\ref{3pol}), (\ref{vectors}), (\ref{PL}) and (\ref{PL1}), the longitudinal $P_L^h(Q^2)$, 
perpendicular $P_P^h (Q^2)$ and transverse $P_T^h (Q^2)$ components are calculated to be
\begin{eqnarray}
  P_L^h (Q^2) &=& \frac{M^\prime}{E_{p^{\prime}}} \frac{A^h(Q^2) \vec{k}.\vec{p}^{\,\prime} + B^h (Q^2) 
  |\vec{p}^{\,\prime}|^2}{N(Q^2)~|\vec{p}^{\,\prime}|},\label{Pl} \\
 P_P^h (Q^2) &=& \frac{A^h(Q^2) [(\vec{k}.\vec{p}^{\,\prime})^2 - |\vec{k}|^2 |\vec{p}^{\,\prime}|^2]}{N(Q^2)~
 |\vec{p}^{\,\prime}| ~ |\vec{p}^{\,\prime}\times \vec{k}|},\label{Pp} \\
  P_T^h (Q^2) &=& \frac{C^h(Q^2) M [(\vec{k}.\vec{p}^{\,\prime})^2 - |\vec{k}|^2 |\vec{p}^{\,\prime}|^2]}{N(Q^2)~
  |\vec{p}^{\,\prime} \times \vec{k}|}, \label{Pt}
\end{eqnarray}
which are consistent with the expressions of the components of polarization given by Erriquez {\it et 
al.}~\cite{Erriquez:1978pg} except for a change in sign in one of the terms in $P_P^h (Q^2)$. The above expressions for 
$P_L^h(Q^2)$ and $P_P^h(Q^2)$ are also consistent with the expressions given by Bilenky and 
Christova~\cite{Bilenky:2013fra} in the limit $M^\prime = M$ and $g_2(0) = g_3(Q^2) = 0$.

If the T invariance is assumed then all the vector and the axial vector form factors are real and the expression for 
$C^h(Q^2)$ vanishes which implies that the transverse component of the polarization perpendicular to the production 
plane, $ P_T^h (Q^2)$ vanishes.

\subsection{Polarization of the final lepton}
Instead of the final baryon polarization if one assumes the final lepton to be polarized, then the polarization 
4-vector($\zeta^\tau$) in reaction~(\ref{process1})--(\ref{process5}) is written as
\begin{equation}\label{polarl}
\zeta^{\tau}=\frac{\mathrm{Tr}[\gamma^{\tau}\gamma_{5}~\rho_{f}(k^\prime)]}
{\mathrm{Tr}[\rho_{f}(k^\prime)]},
\end{equation}
and the spin density matrix for the final lepton $\rho_f(k^\prime)$ is given by 
\begin{equation}\label{polar1l}
 \rho_{f}(k^\prime)= {\cal J}^{\alpha \beta}  \text{ Tr}[\Lambda(k') \gamma_\alpha (1 \pm \gamma_5) \Lambda(k) 
 \tilde\gamma_ {\beta} (1 \pm \tilde\gamma_5)\Lambda(k')], 
\end{equation} 
with $\tilde{\gamma}_{\alpha} =\gamma^0 \gamma^{\dagger}_{\alpha} \gamma^0$ and $\tilde{\gamma}_{5} =\gamma^0 
\gamma^{\dagger}_{5} \gamma^0$.

Using Eqs.~(\ref{polar3}) and (\ref{polar31}), $\zeta^\tau$ defined in Eq.~(\ref{polarl}) may also be rewritten as
\begin{equation}\label{polar4l}
\zeta^{\tau}=\left( g^{\tau\sigma}-\frac{k'^{\tau}k'^{\sigma}}{m_\mu^2}\right)
\frac{  {\cal J}^{\alpha \beta}  \mathrm{Tr}
\left[\gamma_{\sigma}\gamma_{5}\Lambda(k') \gamma_\alpha (1 \pm \gamma_5) \Lambda(k) \tilde\gamma_ {\beta} (1 \pm 
\tilde\gamma_5) \right]}
{ {\cal J}^{\alpha \beta} \mathrm{Tr}\left[\Lambda(k') \gamma_\alpha (1 \pm \gamma_5) \Lambda(k) \tilde\gamma_ {\beta} 
(1 \pm \tilde\gamma_5) \right]},
\end{equation}
where $m_\mu$ is the mass of the muon. In Eq.~(\ref{polar4l}), the denominator is directly related to the differential 
cross section given in Eq.~(\ref{dsig}).

With ${\cal J}^{\alpha \beta}$ and ${\cal L}_{\alpha \beta}$ given in Eqs.~(\ref{J}) and (\ref{L}), respectively, an 
expression for $\zeta^\tau$ is obtained. In the laboratory frame where the initial nucleon is at rest, the polarization 
vector $\vec{\zeta}$ is calculated to be a function of 3-momenta of incoming antineutrino $({\vec{k}})$ and outgoing 
lepton $({\vec{k}}\,^{\prime})$, and is given as  
\begin{equation}\label{3poll}
 \vec{\zeta} =\left[{A^l(Q^2)\vec{ k}} + B^l(Q^2){\vec{k}}\,^{\prime} + C^l(Q^2) M (\vec{k} \times 
 {\vec{k}}\,^{\prime}) \right], 
\end{equation}
where the expressions of $A^l(Q^2)$, $B^l(Q^2)$ and $C^l(Q^2)$ are given in the Appendix-II.

One may expand the polarization vector $\vec{\zeta}$ along the orthogonal directions, ${\hat{e}}_L^l$, ${\hat{e}}_P^l$ 
and ${\hat{e}_T^l}$ in the reaction plane corresponding to the longitudinal, perpendicular and transverse directions, 
defined as
\begin{equation}\label{vectorsl}
\hat{e}_{L}^l=\frac{\vec{ k}^{\, \prime}}{|\vec{ k}^{\,\prime}|},\qquad
\hat{e}_{P}^l=\hat{e}_{L}^l \times \hat{e}_T^l,\qquad   {\rm where}~~~~~ 
\hat{e}_T^l=\frac{\vec{ k}\times \vec{ k}^{\,\prime}}{|\vec{ k}\times \vec{ k}^{\,\prime}|},
\end{equation}
and depicted in Fig.~\ref{TRI}. We then write $\vec{\zeta}$ as:
 \begin{equation}\label{polarLabl}
\vec{\zeta}=\zeta_{L} \hat{e}_{L}^l+\zeta_{P} \hat{e}_{P}^l + \zeta_{T} \hat{e}_{T}^l,
\end{equation}
such that the longitudinal, perpendicular and transverse components of the $\vec{\zeta}$ in the laboratory frame are 
given by
\begin{equation}\label{PLl}
 \zeta_L(Q^2)=\vec{\zeta} \cdot \hat{e}_L^l,\qquad \zeta_P(Q^2)= \vec{\zeta} \cdot \hat{e}_P^l, 
 \qquad \zeta_T(Q^2) = \vec{\zeta} \cdot \hat{e}_T^l.
\end{equation}
From Eq.~(\ref{PLl}), the longitudinal $P_L^l(Q^2)$, perpendicular $P_P^l(Q^2)$ and transverse $P_T^l(Q^2)$ 
components of the polarization vector defined in the rest frame of the initial nucleon are given by 
\begin{equation}\label{PL1l}
 P_L^l(Q^2)=\frac{m_\mu}{E_{k^\prime}} \zeta_L(Q^2), \qquad P_P^l(Q^2)=\zeta_P(Q^2), \qquad P_T^l(Q^2)=\zeta_T(Q^2),
\end{equation}
where $\frac{m_\mu}{E_{k^\prime}}$ is the Lorentz boost factor along ${\vec k}\, ^\prime$. Using Eqs.~(\ref{3poll}), 
(\ref{vectorsl}) and (\ref{PLl}) in Eq. (\ref{PL1l}), the longitudinal $P_L^l(Q^2)$, perpendicular 
$P_P^l (Q^2)$ and transverse $P_T^l (Q^2)$ components are calculated to be
\begin{eqnarray}
  P_L^l (Q^2) &=& \frac{m_\mu}{E_{k^{\prime}}} \frac{A^l(Q^2) \vec{k}.\vec{k}^{\,\prime} + B^l (Q^2) 
  |\vec{k}^{\,\prime}|^2}{N(Q^2)~|\vec{k}^{\,\prime}|},\label{Pll} \\
 P_P^l (Q^2) &=& \frac{A^l(Q^2) [|\vec{k}|^2 |\vec{k}^{\,\prime}|^2 - (\vec{k}.\vec{k}^{\,\prime})^2]}{N(Q^2)~
 |\vec{k}^{\,\prime}| ~ |\vec{k}\times \vec{k}^{\,\prime}|},\label{Ppl} \\
  P_T^l (Q^2) &=& \frac{C^l(Q^2) M [(\vec{k}.\vec{k}^{\,\prime})^2 - |\vec{k}|^2 |\vec{k}^{\,\prime}|^2]}{N(Q^2)~
  |\vec{k} \times \vec{k}^{\,\prime} |}. \label{Ptl}
\end{eqnarray}

\section{Results and discussion}\label{results}

We have used Eq.~(\ref{dsig}) to numerically evaluate the differential cross section $d \sigma/d Q^2$, 
Eqs.~(\ref{Pl}), (\ref{Pp}) and (\ref{Pt}) to evaluate the longitudinal $P_L^h(Q^2)$, perpendicular $P_P^h(Q^2)$ and 
transverse $P_T^h(Q^2)$ components of the polarization of the final baryon and Eqs.~(\ref{Pll}), (\ref{Ppl}) and 
(\ref{Ptl}) to evaluate the longitudinal $P_L^l(Q^2)$, perpendicular $P_P^l(Q^2)$ and transverse $P_T^l(Q^2)$ 
components of polarization of the final lepton. The Dirac and Pauli form factors $f_{1,2}^{N} (Q^2); ~ (N=p,n)$ are 
expressed in terms of the electric and magnetic Sachs' form factors, for which the parameterization given by Bradford 
et al.~\cite{Bradford:2006yz} have been used. Using the SU(3) symmetry, the axial vector form factors, $g_1^{N X} 
(Q^2)$ and $g_2^{N X}(Q^2)$ are expressed in terms of $g_A^{np}(Q^2)$ and $g_2^{np}(Q^2)$. For $g_A^{np}(Q^2)$ and 
$g_2^{np}(Q^2)$, dipole parameterizations have been used as written in Eqs. (\ref{ga}) and (\ref{g2}) with $M_A = $ 
1.026 GeV and $M_2 = M_A$. For $g_3^{np} (Q^2)$, PCAC along with GT relation has been used and for $g_3^{NY} (Q^2)$, 
the generalized PCAC along with GT relation given by Nambu~\cite{Nambu:1960xd} has been used to relate them, 
respectively, with $g_1^{np} (Q^2)$ and $g_1^{NY} (Q^2)$ as shown in Eqs.~(\ref{g3}) and (\ref{g3_Nambu}). To study 
the dependence of the cross section $\sigma(E_{\nu_\mu (\bar{\nu}_\mu)})$ and the polarization observables 
$\overline{P}_L^{h,l}(E_{\nu_\mu(\bar{\nu}_\mu)})$, $\overline{P}_P^{h,l} (E_{\nu_\mu(\bar{\nu}_\mu)})$ and 
$\overline{P}_T^{h,l} (E_{\nu_\mu(\bar{\nu}_\mu)})$ of the hadrons $(h = n, p, \Lambda, \Sigma^0, \Sigma^-)$ and the 
leptons ($l=\mu^\pm$) on the (anti)neutrino's energy $E_{\nu_\mu(\bar{\nu}_\mu)}$, we have integrated $d\sigma/dQ^2$, 
$P_L^{h,l} (Q^2),~P_P^{h,l} (Q^2)$ and $P_T^{h,l} (Q^2)$ over $Q^2$ and obtained the expressions for $\sigma 
(E_{\nu_\mu(\bar{\nu}_\mu)})$ and $\overline{P}_{L,P,T}^{h,l} (E_{\nu_\mu(\bar{\nu}_\mu)})$ as:
\begin{eqnarray}\label{total_sig}
\sigma (E_{\nu_\mu(\bar{\nu}_\mu)}) &=& \int_{Q^2_{min}}^{Q^2_{max}} \frac{d\sigma}{dQ^2} dQ^2, 
\end{eqnarray}
and 
\begin{eqnarray}\label{average_Plpt}
 \overline{P}_{L,P,T}^{h,l} (E_{\nu_\mu(\bar{\nu}_\mu)}) &=& \frac{\int_{Q^2_{min}}^{Q^2_{max}} P_{L,P,T}^{h,l} (Q^2) 
 \frac{d\sigma}{dQ^2} dQ^2}{\int_{Q^2_{min}}^{Q^2_{max}} \frac{d\sigma}{dQ^2} dQ^2}.
\end{eqnarray}

\begin{figure}
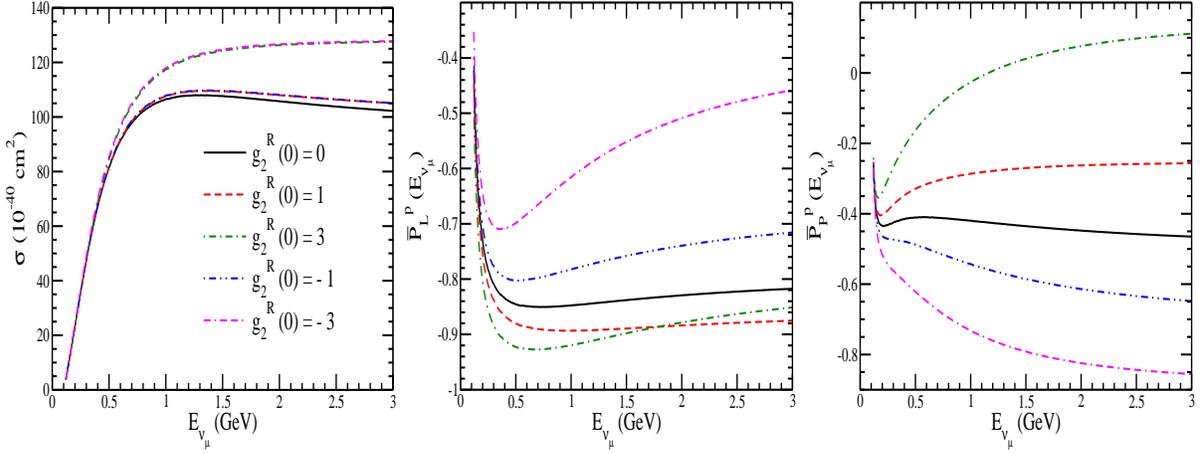

 \includegraphics[height=6cm,width=5.2cm]{total_sigma_real_g2_variation_with_Fp_GT_Ma_1026GeV_proton_polarized.eps}
 \includegraphics[height=6cm,width=5.2cm]{Pl_enu_real_g2_variation_with_Fp_GT_Ma_1026GeV_proton_polarized.eps}
 \includegraphics[height=6cm,width=5.2cm]{Pp_enu_real_g2_variation_with_Fp_GT_Ma_1026GeV_proton_polarized.eps}
\caption{$\sigma ~vs.~ E_{\nu_{\mu}}$ (left panel) for the process ${\nu_\mu + n \rightarrow \mu^- + p}$, 
$\overline{P}_L^p (E_{\nu_{\mu}}) ~vs.~ E_{\nu_{\mu}}$ (middle panel) and $\overline{P}_P^p (E_{\nu_{\mu}}) ~vs.~ 
E_{\nu_{\mu}}$ (right panel), for the polarized proton in the final state, at the different values of $g_2^R (0)$ viz. 
$g_2^R (0) = $ 0~(solid line), 1~(dashed line), 3~(dashed-dotted line), $-1$~(double-dotted-dashed line) and 
$-3$~(double-dashed-dotted line).}\label{fig1}
\end{figure}

\begin{figure}
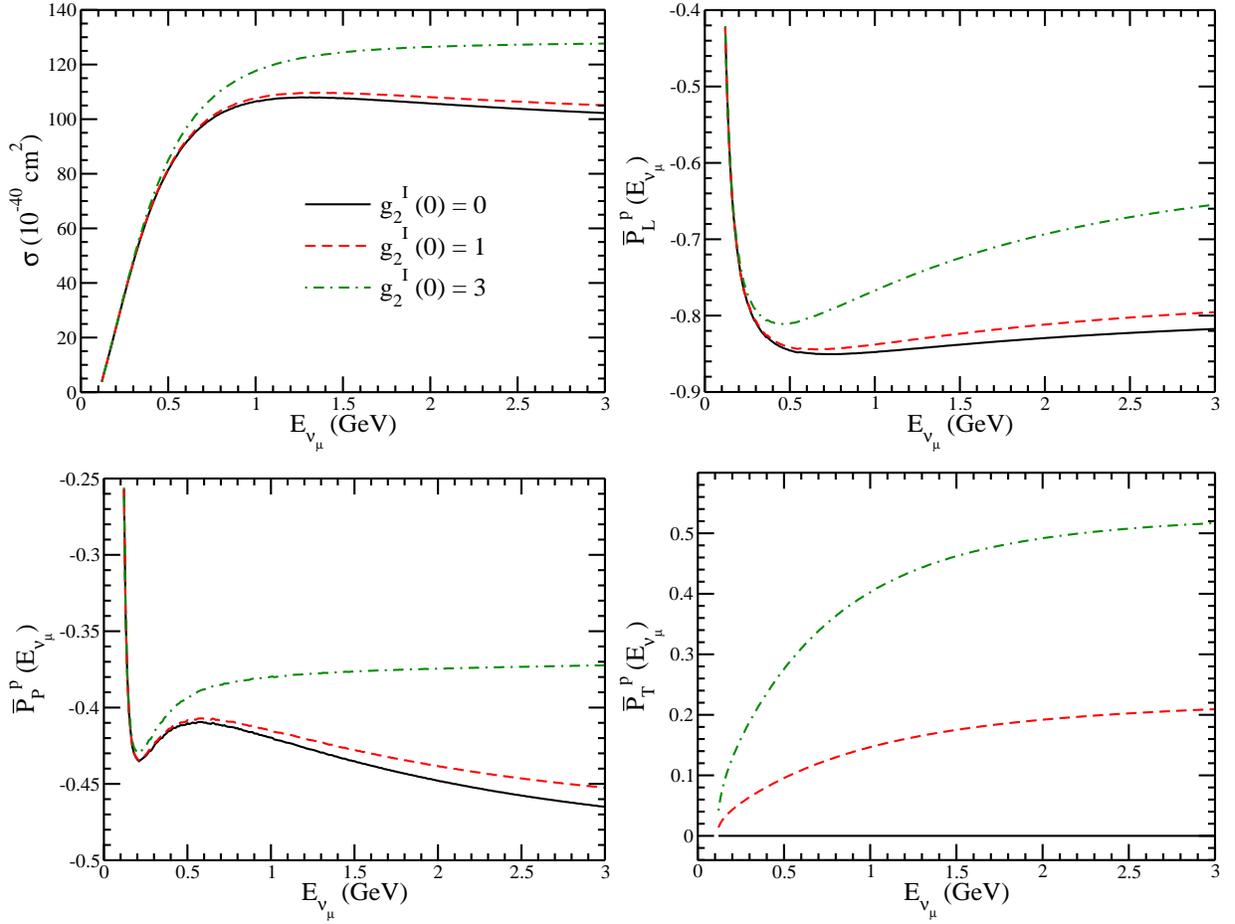

 \includegraphics[height=6cm,width=8cm]{total_sigma_imaginary_g2_variation_with_Fp_GT_Ma_1026GeV_proton_polarized.eps}
 \includegraphics[height=6cm,width=8cm]{Pl_enu_imaginary_g2_variation_with_Fp_GT_Ma_1026GeV_proton_polarized.eps} \\
 \vspace{2mm}
 \includegraphics[height=6cm,width=8cm]{Pp_enu_imaginary_g2_variation_with_Fp_GT_Ma_1026GeV_proton_polarized.eps}
 \includegraphics[height=6cm,width=8cm]{Pt_enu_imaginary_g2_variation_with_Fp_GT_Ma_1026GeV_proton_polarized.eps}
\caption{$\sigma ~vs.~ E_{\nu_{\mu}}$ (upper left panel) for the process ${\nu_\mu + n \rightarrow \mu^- + p}$,
$\overline{P}_L^p (E_{\nu_{\mu}}) ~vs.~ E_{\nu_{\mu}}$ (upper right panel), $\overline{P}_P^p (E_{\nu_{\mu}}) ~vs.~ 
E_{\nu_{\mu}}$ (lower left panel) and $\overline{P}_T^p (E_{\nu_{\mu}}) ~vs.~ E_{\nu_{\mu}}$ (lower right panel), for 
the polarized proton in the final state, at the different values of $g_2^I (0)$ viz. $g_2^I (0) = $ 0~(solid line), 
1~(dashed line) and 3~(dashed-dotted line).}
\label{fig2}
\end{figure}

\begin{figure}
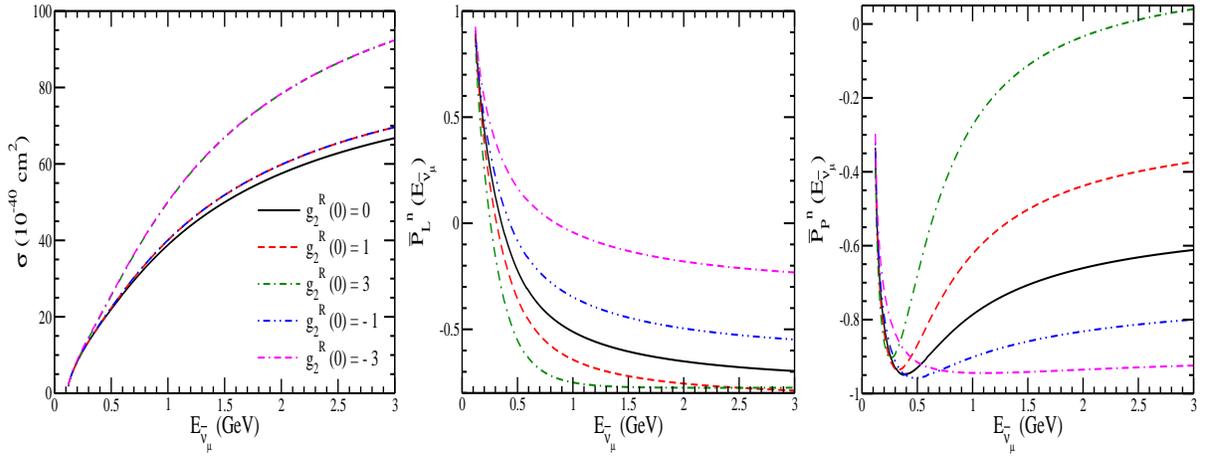

 \includegraphics[height=6cm,width=5.2cm]{total_sigma_real_g2_variation_with_Fp_GT_Ma_1026GeV_neutron_polarized.eps}
 \includegraphics[height=6cm,width=5.2cm]{Pl_enu_real_g2_variation_with_Fp_GT_Ma_1026GeV_neutron_polarized.eps}
 \includegraphics[height=6cm,width=5.2cm]{Pp_enu_real_g2_variation_with_Fp_GT_Ma_1026GeV_neutron_polarized.eps}
\caption{$\sigma ~vs.~ E_{\bar{\nu}_{\mu}}$ (left panel) for the process ${\bar{\nu}_\mu + p \rightarrow \mu^+ + n}$, 
$\overline{P}_L^n (E_{\bar{\nu}_{\mu}}) ~vs.~ E_{\bar{\nu}_{\mu}}$ (middle panel) and $\overline{P}_P^n 
(E_{\bar{\nu}_{\mu}}) ~vs.~ E_{\bar{\nu}_{\mu}}$ (right panel), for the polarized neutron in the final state. Lines and 
points have the same meaning as in Fig.~\ref{fig1}.}\label{fig3}
\end{figure}

\begin{figure}
 \includegraphics[height=6cm,width=8cm]{total_sigma_imaginary_g2_variation_with_Fp_GT_Ma_1026GeV_neutron_polarized.eps}
 \includegraphics[height=6cm,width=8cm]{Pl_enu_imaginary_g2_variation_with_Fp_GT_Ma_1026GeV_neutron_polarized.eps} \\
 \includegraphics[height=6cm,width=8cm]{Pp_enu_imaginary_g2_variation_with_Fp_GT_Ma_1026GeV_neutron_polarized.eps}
 \includegraphics[height=6cm,width=8cm]{Pt_enu_imaginary_g2_variation_with_Fp_GT_Ma_1026GeV_neutron_polarized.eps}
\caption{$\sigma ~vs.~ E_{\bar{\nu}_{\mu}}$ (upper left panel) for the process ${\bar{\nu}_\mu + p \rightarrow \mu^+ 
+ n}$, $\overline{P}_L^n (E_{\bar{\nu}_{\mu}}) ~vs.~ E_{\bar{\nu}_{\mu}}$ (upper right panel), $\overline{P}_P^n 
(E_{\bar{\nu}_{\mu}}) ~vs.~ E_{\bar{\nu}_{\mu}}$ (lower left panel) and $\overline{P}_T^n (E_{\bar{\nu}_{\mu}}) ~vs.~ 
E_{\bar{\nu}_{\mu}}$ (lower right panel), for the polarized neutron in the final state. Lines and points have the same 
meaning as in Fig.~\ref{fig2}.}
\label{fig4}
\end{figure}

\begin{figure}
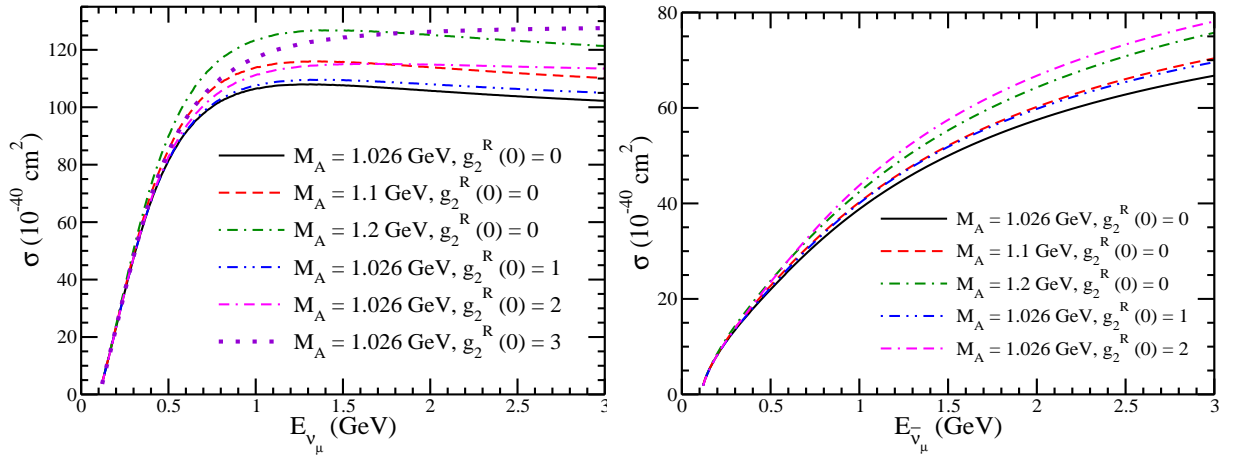

 \includegraphics[height=6cm,width=8cm]{total_sigma_compare_g2_Ma_proton_real.eps}
 \includegraphics[height=6cm,width=8cm]{total_sigma_compare_g2_Ma_neutron_real.eps} 
\caption{$\sigma ~vs.~ E_{{\nu}_{\mu}(\bar{\nu}_\mu)}$ for the process ${{\nu}_\mu + n \rightarrow \mu^- + p}$~(left 
panel) and ${\bar{\nu}_\mu + p \rightarrow \mu^+ + n}$~(right panel) for the different combinations of $M_A$, and 
${{g_2^R (0)}}$ viz. $M_A = 1.026$ GeV and ${{g_2^{R} (0)}} = 0$~(solid line), $M_A = 1.1$ GeV and ${{g_2^{R} (0)}} = 
0$~(dashed line), $M_A = 1.2$ GeV and ${{g_2^{R} (0)}} = 0$~(dashed-dotted line), $M_A = 1.026$ GeV and ${{g_2^{R} (0)
}} = 1$~(double-dotted-dashed line), $M_A = 1.026$ GeV and ${{g_2^{R} (0)}} = 2$~(double-dashed-dotted line) and $M_A = 
1.026$ GeV and ${{g_2^R (0)}} = 3$~(dotted line).} \label{figb}
\end{figure}

\begin{figure}
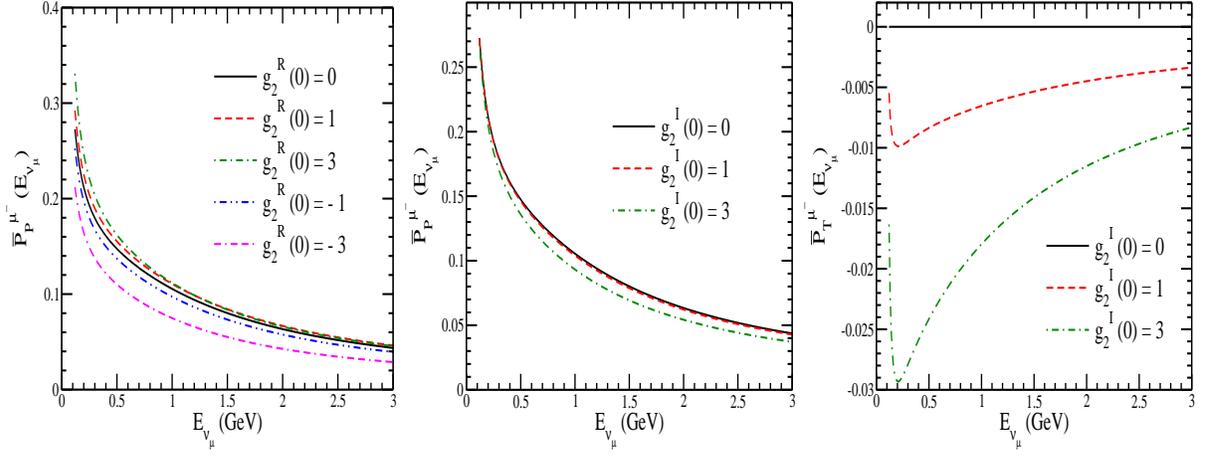

\includegraphics[height=6cm,width=5.2cm]{Pp_enu_real_g2_variation_proton_muon_polarized.eps}
\includegraphics[height=6cm,width=5.2cm]{Pp_enu_imaginary_g2_variation_proton_muon_polarized.eps}
\includegraphics[height=6cm,width=5.2cm]{Pt_enu_imaginary_g2_variation_proton_muon_polarized.eps}
\caption{$\overline{P}_P^{\mu^-} (E_{{\nu}_{\mu}}) ~vs.~ E_{{\nu}_{\mu}}$ for the real values of $g_2^{np} (0)$ (left 
panel), $\overline{P}_P^{\mu^-} (E_{{\nu}_{\mu}}) ~vs.~ E_{{\nu}_{\mu}}$ for the imaginary values of $g_2^{np} (0)$ 
(middle panel) and $\overline{P}_T^{\mu^-} (E_{{\nu}_{\mu}}) ~vs.~ E_{{\nu}_{\mu}}$ (right panel) for the process 
${{\nu}_\mu + n \rightarrow {\mu}^- + p}$ at the different values of ${{g_2^{R,I} (0)}}$ viz. 
${{g_2^{R,I} (0)}} = 0$~(solid line), 1~(dashed line), 3~(dashed-dotted line), $-1$~(double-dotted-dashed line) and 
$-3$~(double-dashed-dotted line).}\label{figmup}
\end{figure}

\begin{figure}
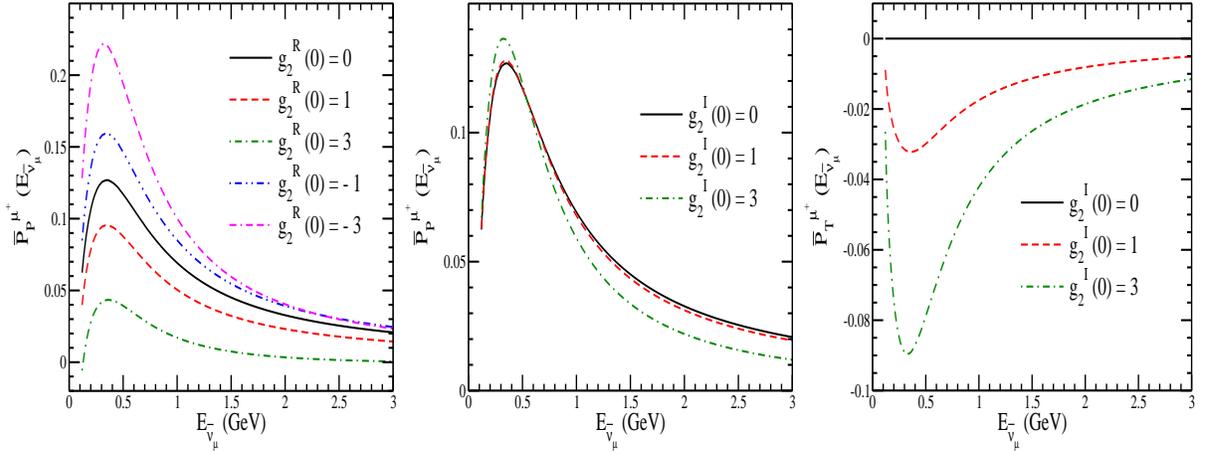

\includegraphics[height=6cm,width=5.2cm]{Pp_enu_real_g2_variation_neutron_muon_polarized.eps}
\includegraphics[height=6cm,width=5.2cm]{Pp_enu_imaginary_g2_variation_neutron_muon_polarized.eps}
\includegraphics[height=6cm,width=5.2cm]{Pt_enu_imaginary_g2_variation_neutron_muon_polarized.eps}
\caption{$\overline{P}_P^{\mu^+} (E_{\bar{\nu}_{\mu}}) ~vs.~ E_{\bar{\nu}_{\mu}}$ for the real values of $g_2^{np} (0)$ 
(left panel), $\overline{P}_P^{\mu^+} (E_{\bar{\nu}_{\mu}}) ~vs.~ E_{\bar{\nu}_{\mu}}$ for the imaginary values of 
$g_2^{np} (0)$ (middle panel) and $\overline{P}_T^{\mu^+} (E_{\bar{\nu}_{\mu}}) ~vs.~ E_{\bar{\nu}_{\mu}}$ (right 
panel) for the process ${\bar{\nu}_\mu + p \rightarrow {\mu}^+ + n}$. Lines and points have the same meaning as 
in Fig.~\ref{figmup}.}
\label{figmu}
\end{figure}

\begin{figure}
 \includegraphics[height=6cm,width=5.2cm]{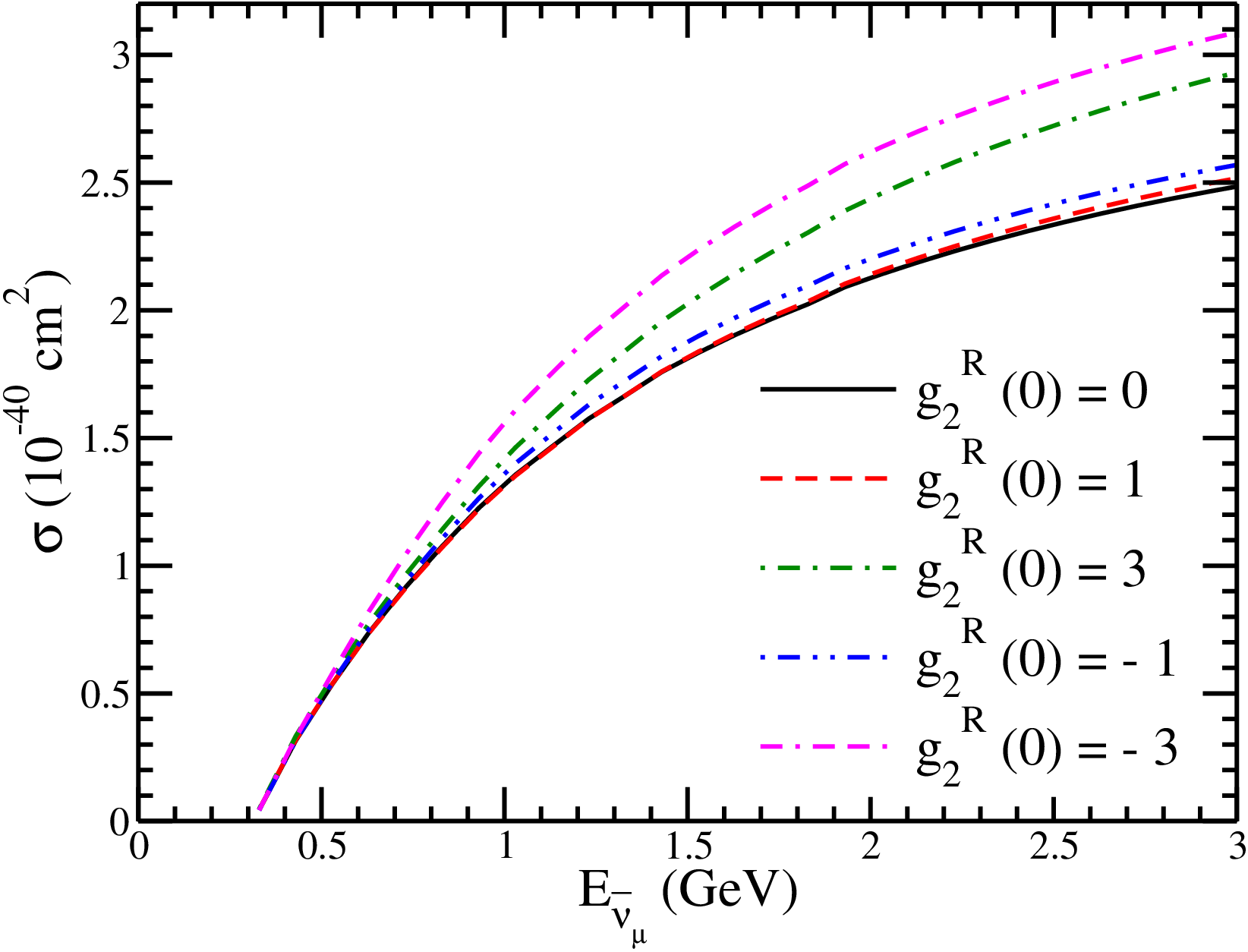}
 \includegraphics[height=6cm,width=5.2cm]{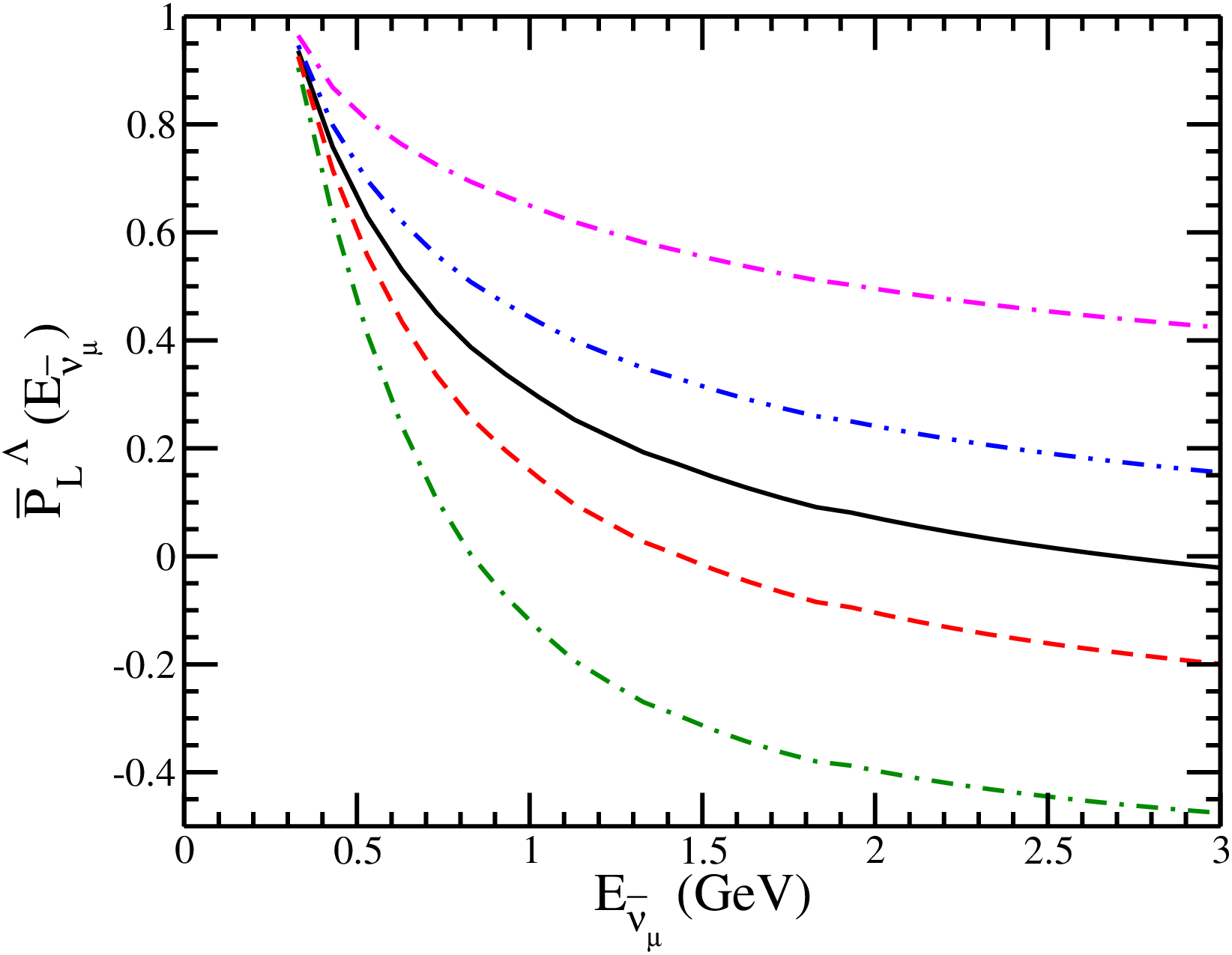}
 \includegraphics[height=6cm,width=5.2cm]{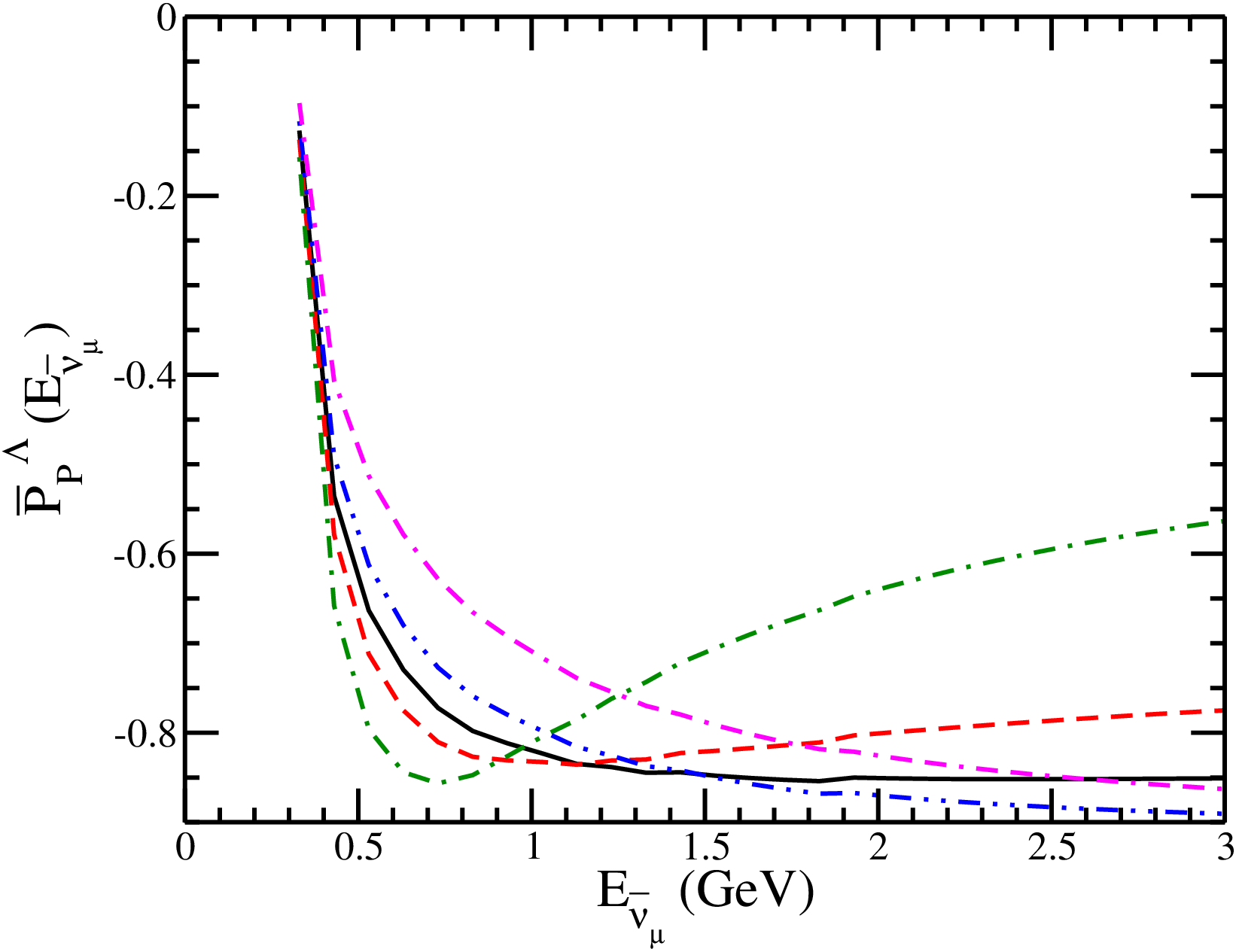}
\caption{$\sigma ~vs.~ E_{\bar{\nu}_{\mu}}$ (left panel) for the process ${\bar{\nu}_\mu + p \rightarrow \mu^+ + 
\Lambda}$, $\overline{P}_L^{\Lambda} (E_{\bar{\nu}_{\mu}}) ~vs.~ E_{\bar{\nu}_{\mu}}$ (middle panel) and 
$\overline{P}_P^{\Lambda} (E_{\bar{\nu}_{\mu}}) ~vs.~ E_{\bar{\nu}_{\mu}}$ (right panel), for the polarized lambda in 
the final state. Lines and points have the same meaning as in Fig.~\ref{fig1}.}
\label{fig11}
\end{figure}

\begin{figure}
 \includegraphics[height=6cm,width=8cm]{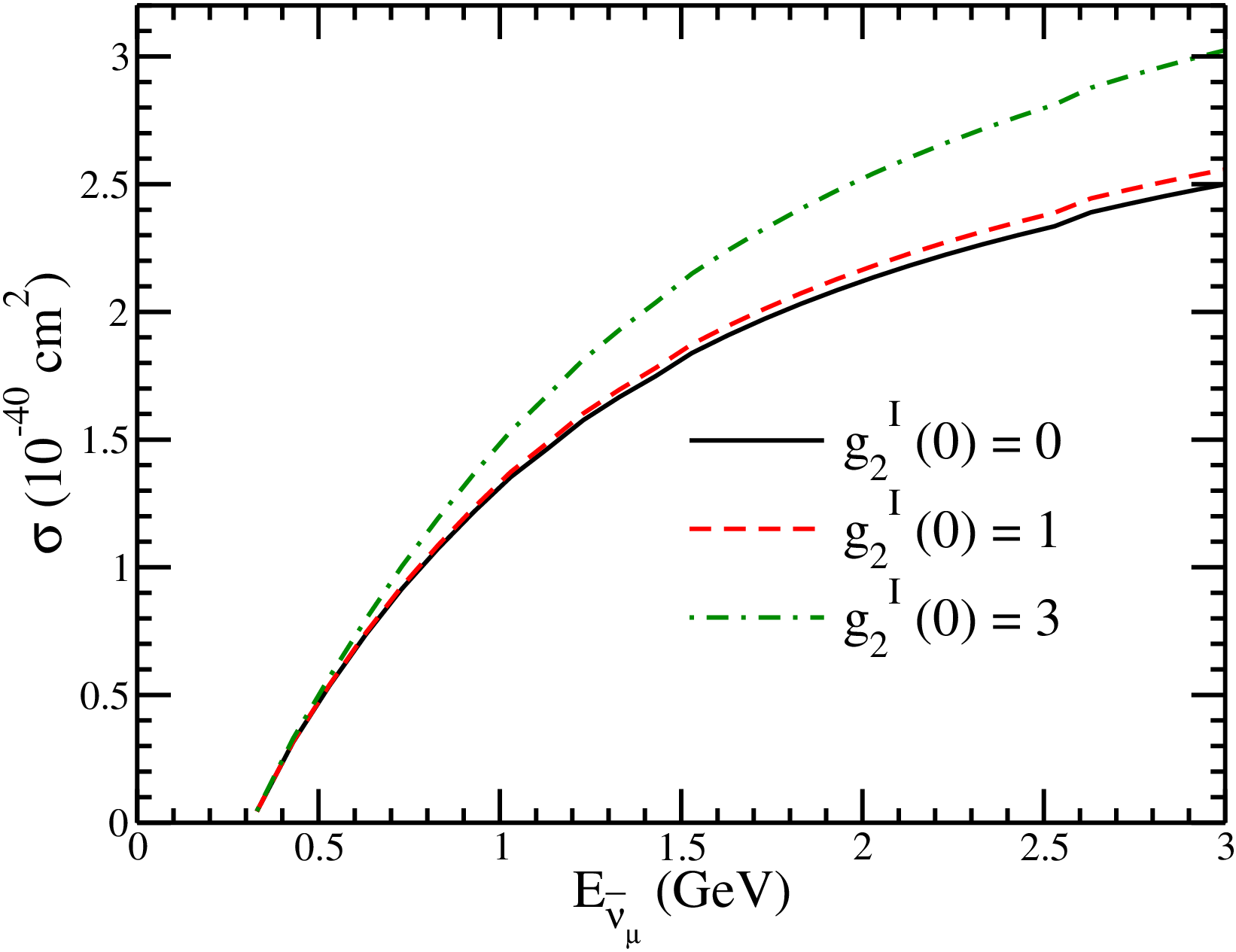}
 \includegraphics[height=6cm,width=8cm]{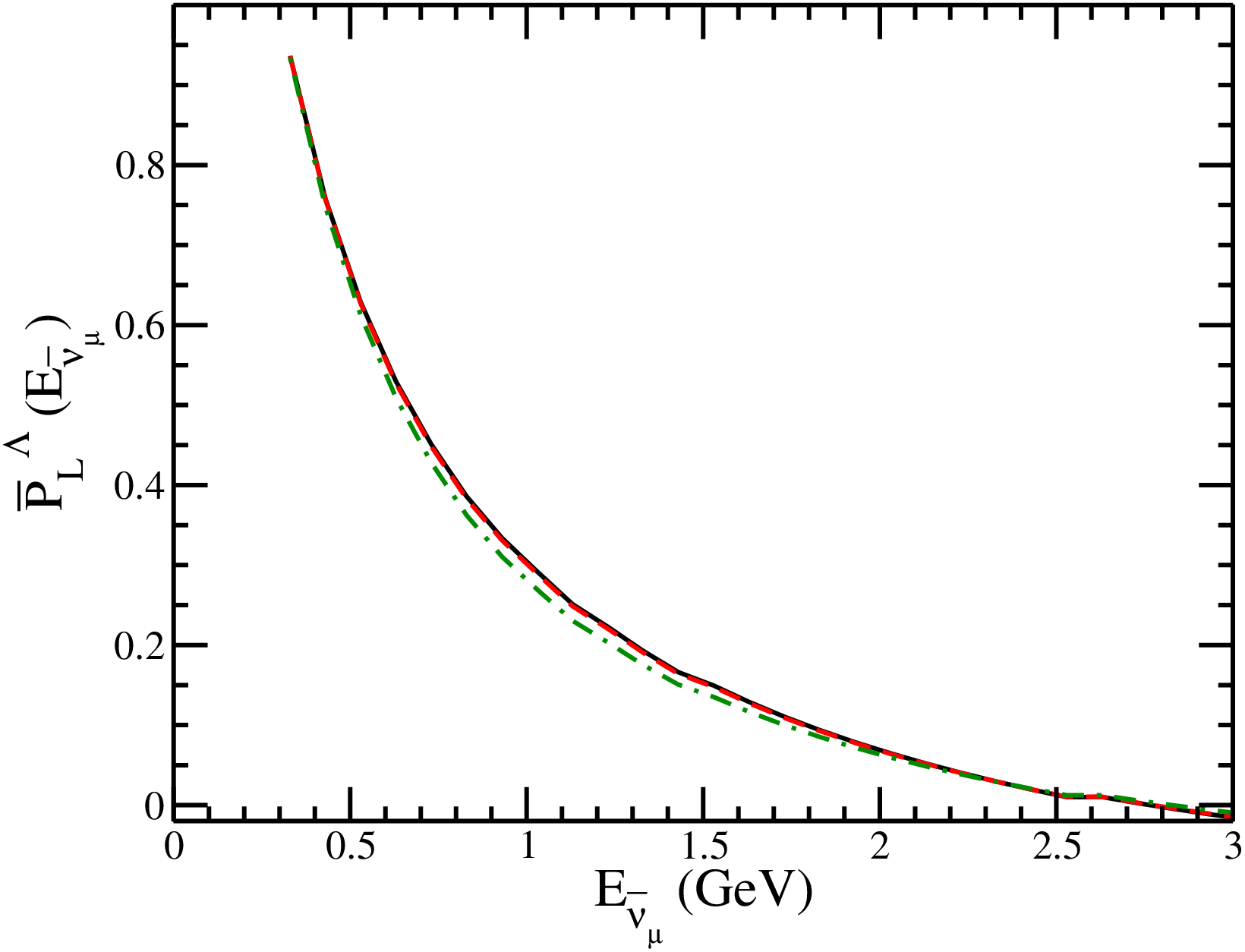} \\
 \includegraphics[height=6cm,width=8cm]{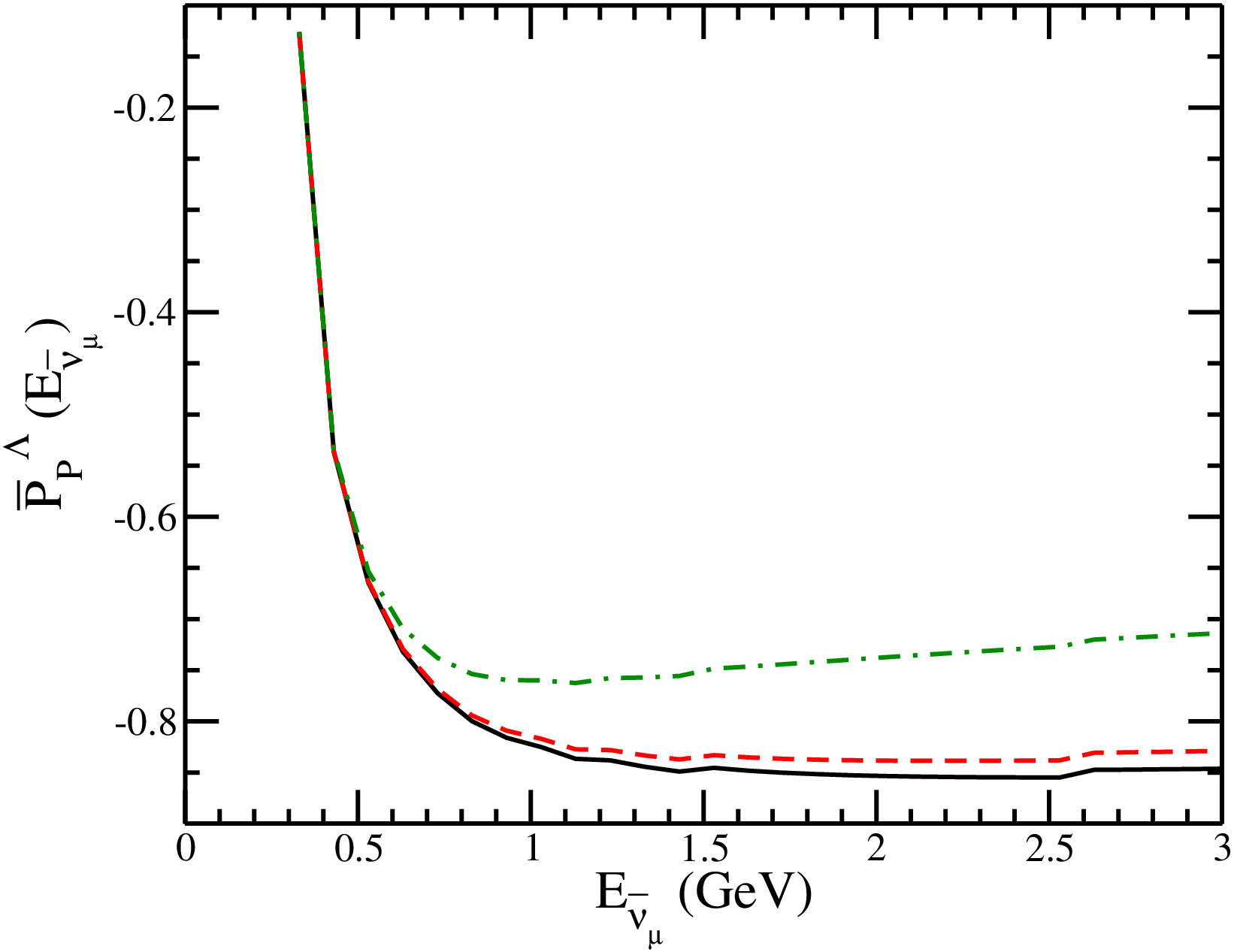}
 \includegraphics[height=6cm,width=8cm]{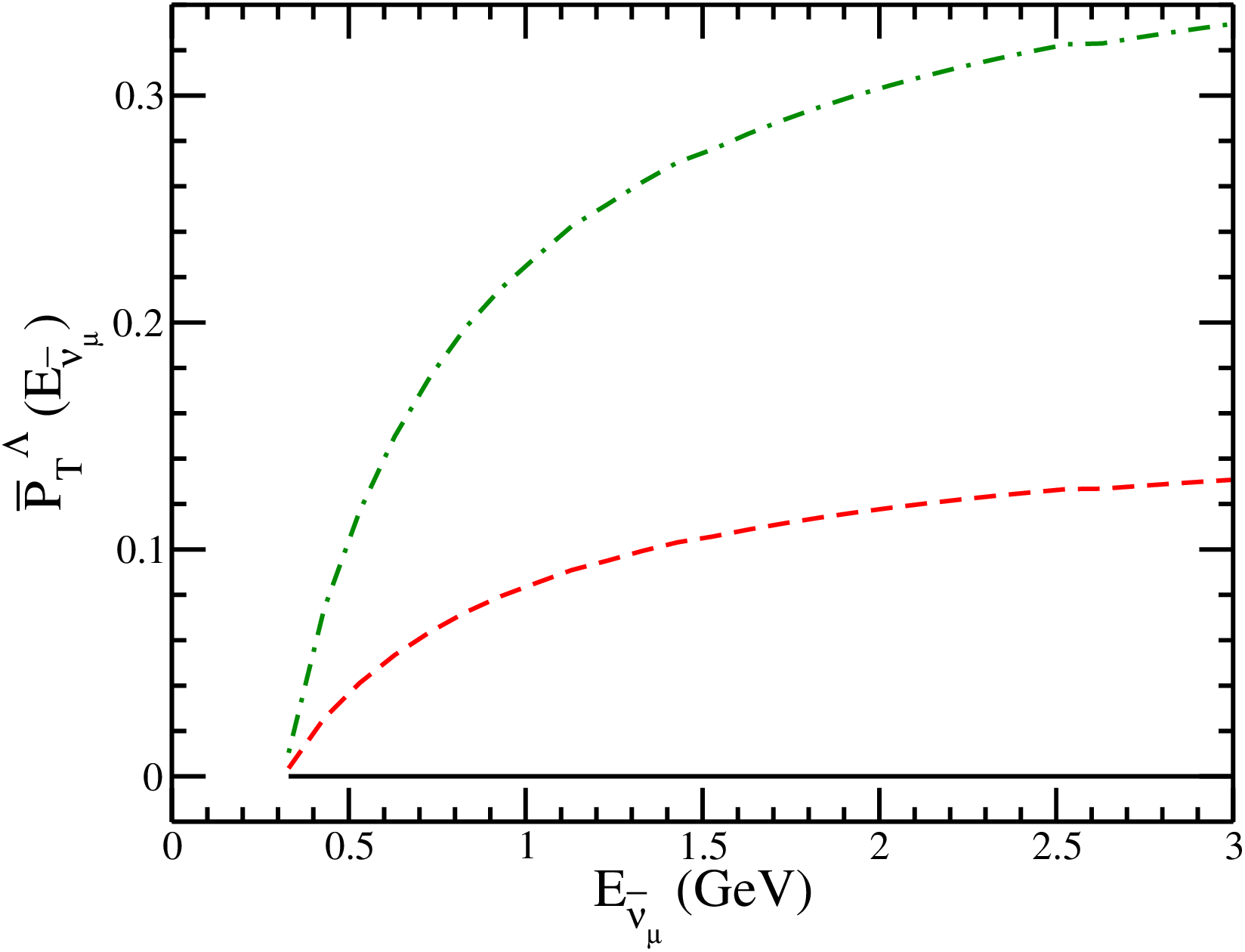}
\caption{$\sigma ~vs.~ E_{\bar{\nu}_{\mu}}$ (upper left panel) for the process ${\bar{\nu}_\mu + p \rightarrow \mu^+ 
+ \Lambda}$, $\overline{P}_L^{\Lambda} (E_{\bar{\nu}_{\mu}}) ~vs.~ E_{\bar{\nu}_{\mu}}$ (upper right panel), 
$\overline{P}_P^{\Lambda} (E_{\bar{\nu}_{\mu}}) ~vs.~ E_{\bar{\nu}_{\mu}}$ (lower left panel) and 
$\overline{P}_T^{\Lambda}(E_{\bar{\nu}_{\mu}}) ~vs.~ E_{\bar{\nu}_{\mu}}$ (lower right panel), for the polarized lambda 
in the final state. Lines and points have the same meaning as in Fig.~\ref{fig2}.}
\label{fig12}
\end{figure}

\begin{figure}
 \includegraphics[height=6cm,width=5.2cm]{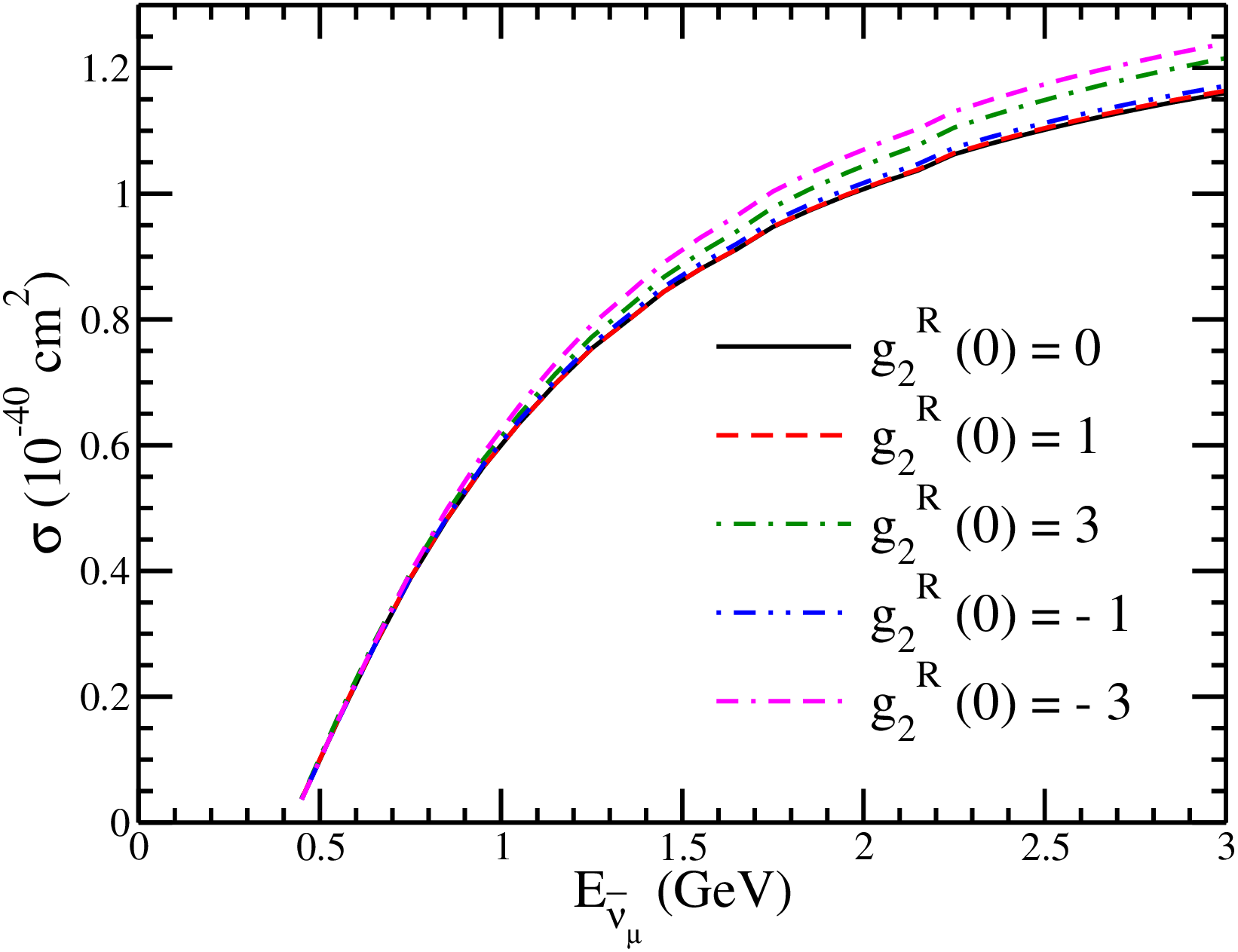}
 \includegraphics[height=6cm,width=5.2cm]{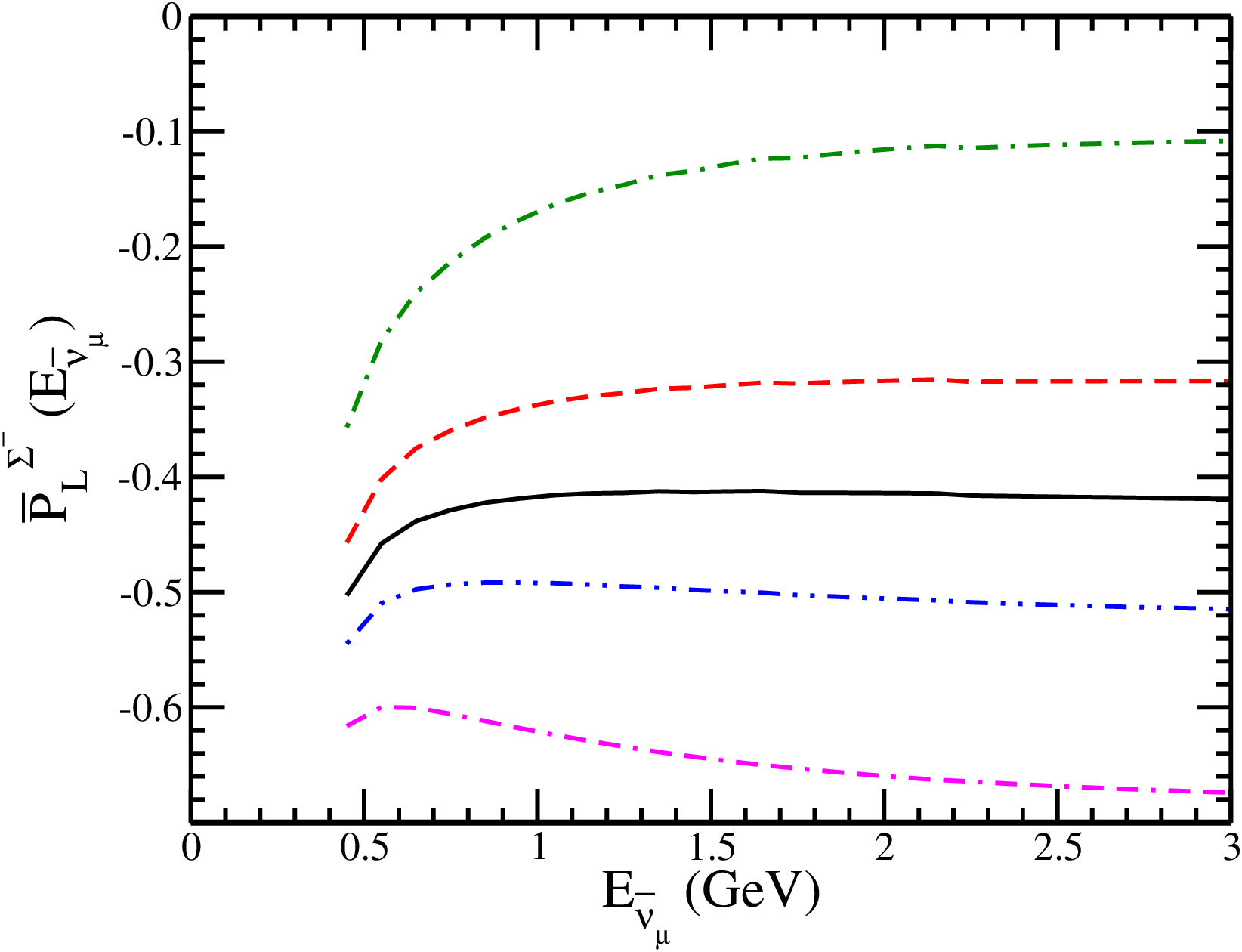}
 \includegraphics[height=6cm,width=5.2cm]{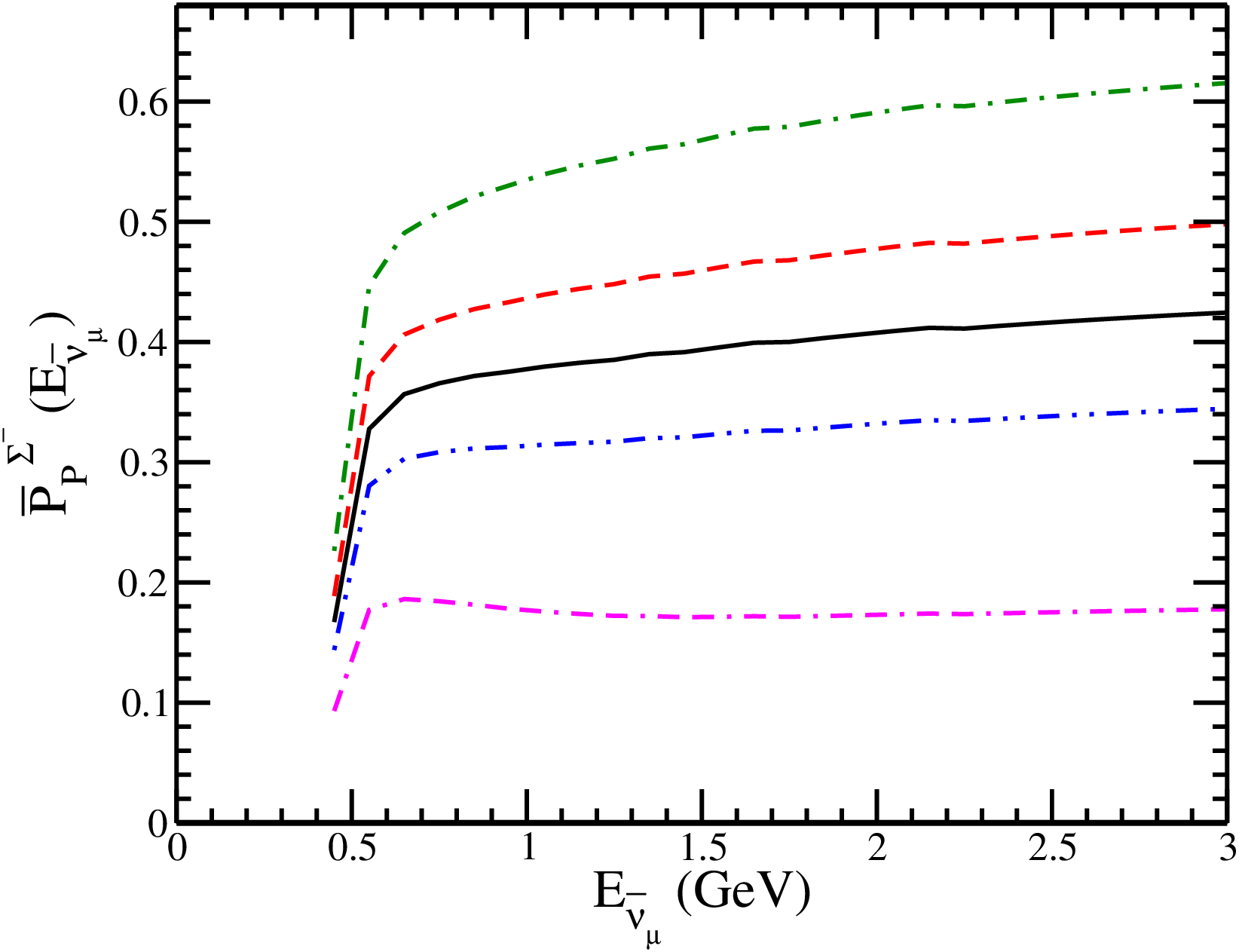}
\caption{$\sigma ~vs.~ E_{\bar{\nu}_{\mu}}$ (left panel) for the process ${\bar{\nu}_\mu + n \rightarrow \mu^+ + 
\Sigma^-}$, $\overline{P}_L^{\Sigma^-} (E_{\bar{\nu}_{\mu}}) ~vs.~ E_{\bar{\nu}_{\mu}}$ (middle panel) and 
$\overline{P}_P^{\Sigma^-} (E_{\bar{\nu}_{\mu}}) ~vs.~ E_{\bar{\nu}_{\mu}}$ (right panel), for a polarized sigma in the 
final state. Lines and points have the same meaning as in Fig.~\ref{fig1}.}
\label{fig16}
\end{figure}

\begin{figure}
 \includegraphics[height=6cm,width=8cm]{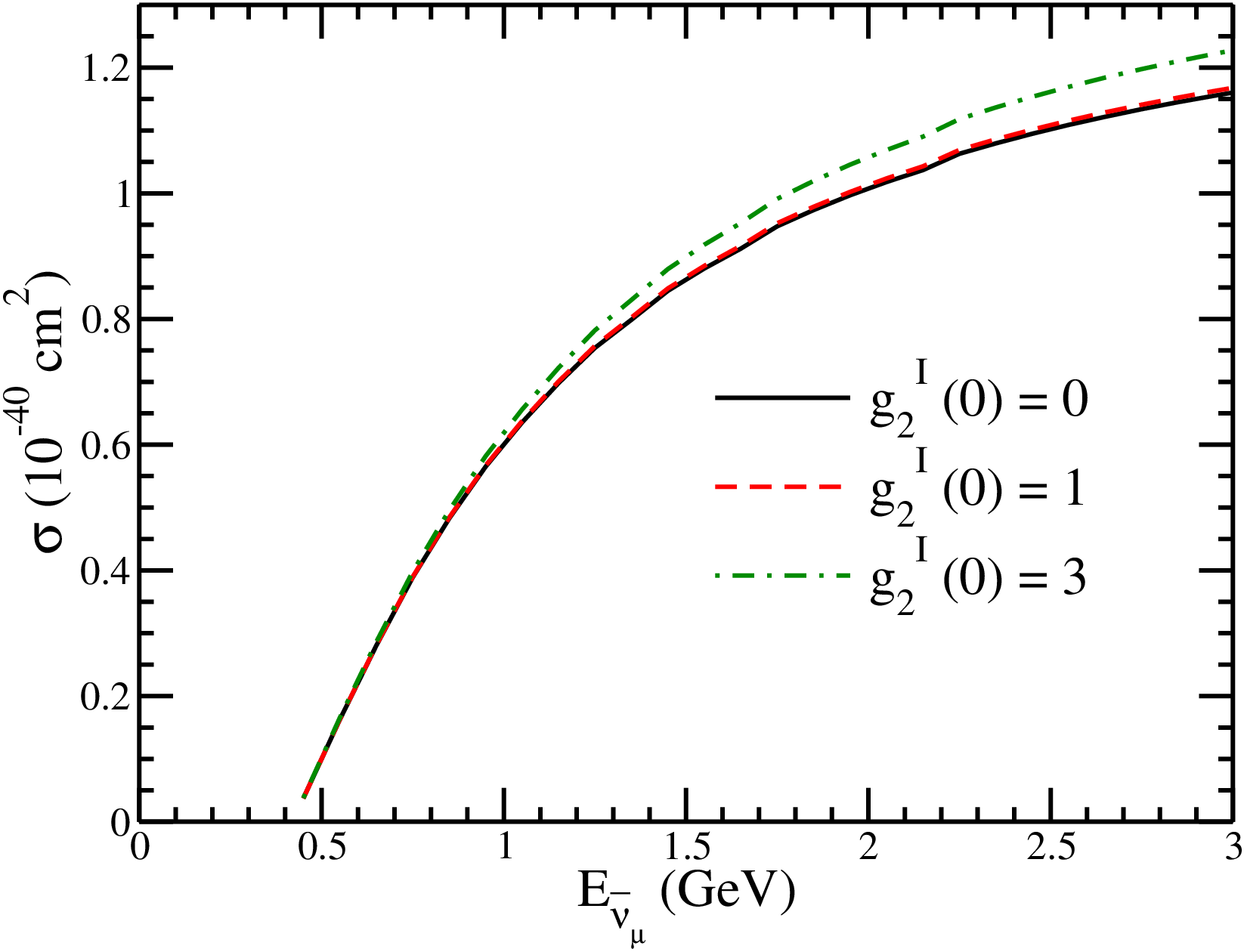}
 \includegraphics[height=6cm,width=8cm]{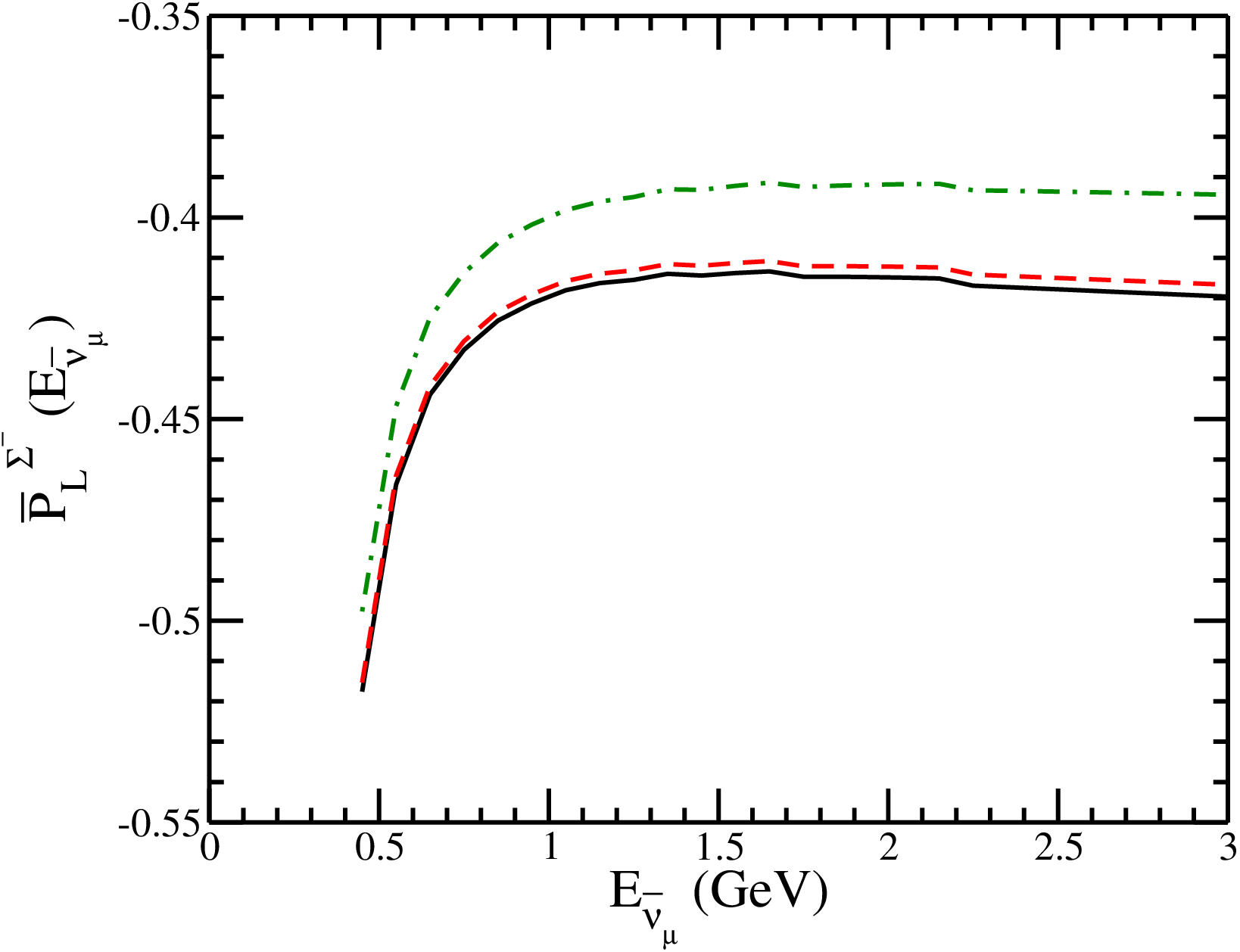} \\
 \includegraphics[height=6cm,width=8cm]{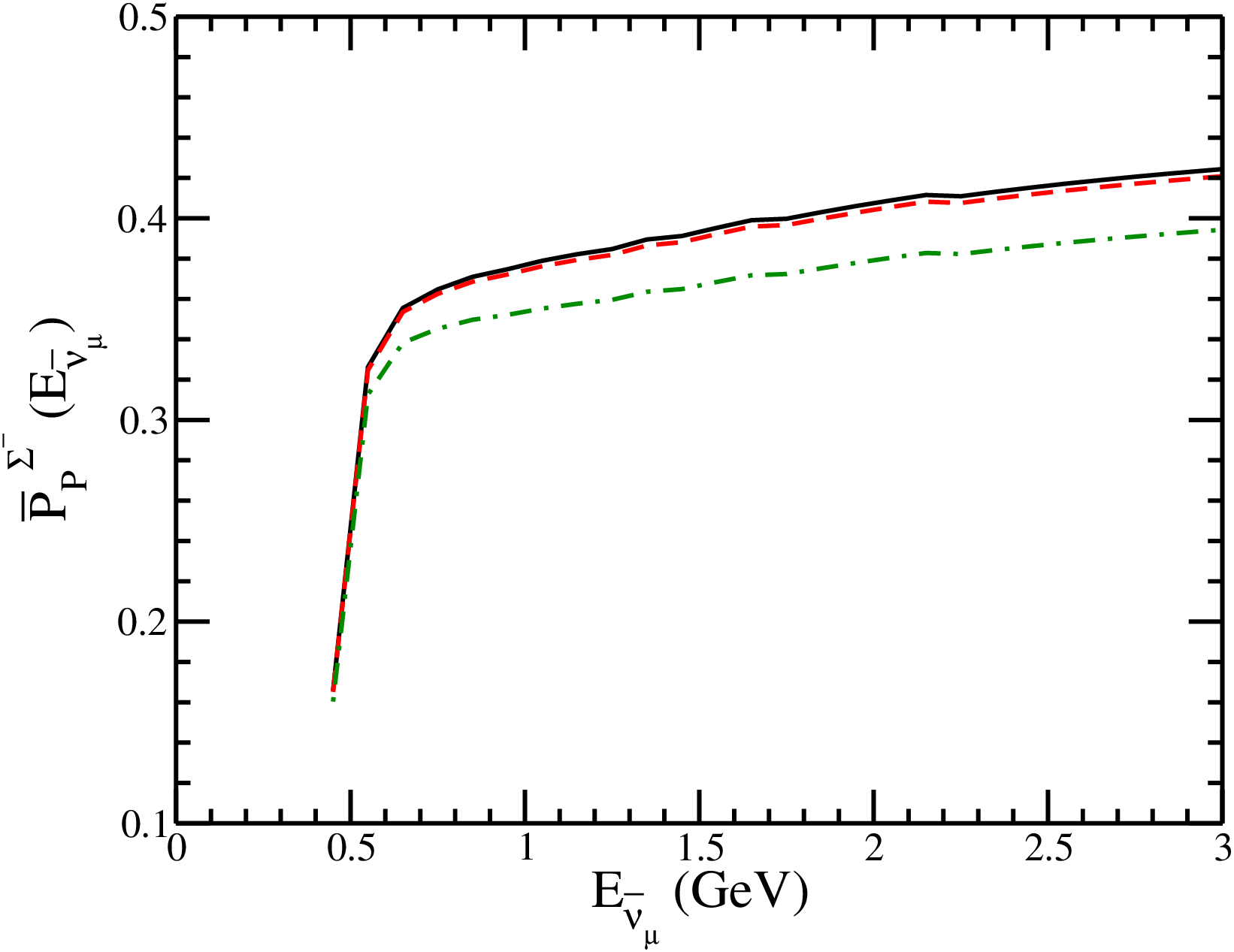}
 \includegraphics[height=6cm,width=8cm]{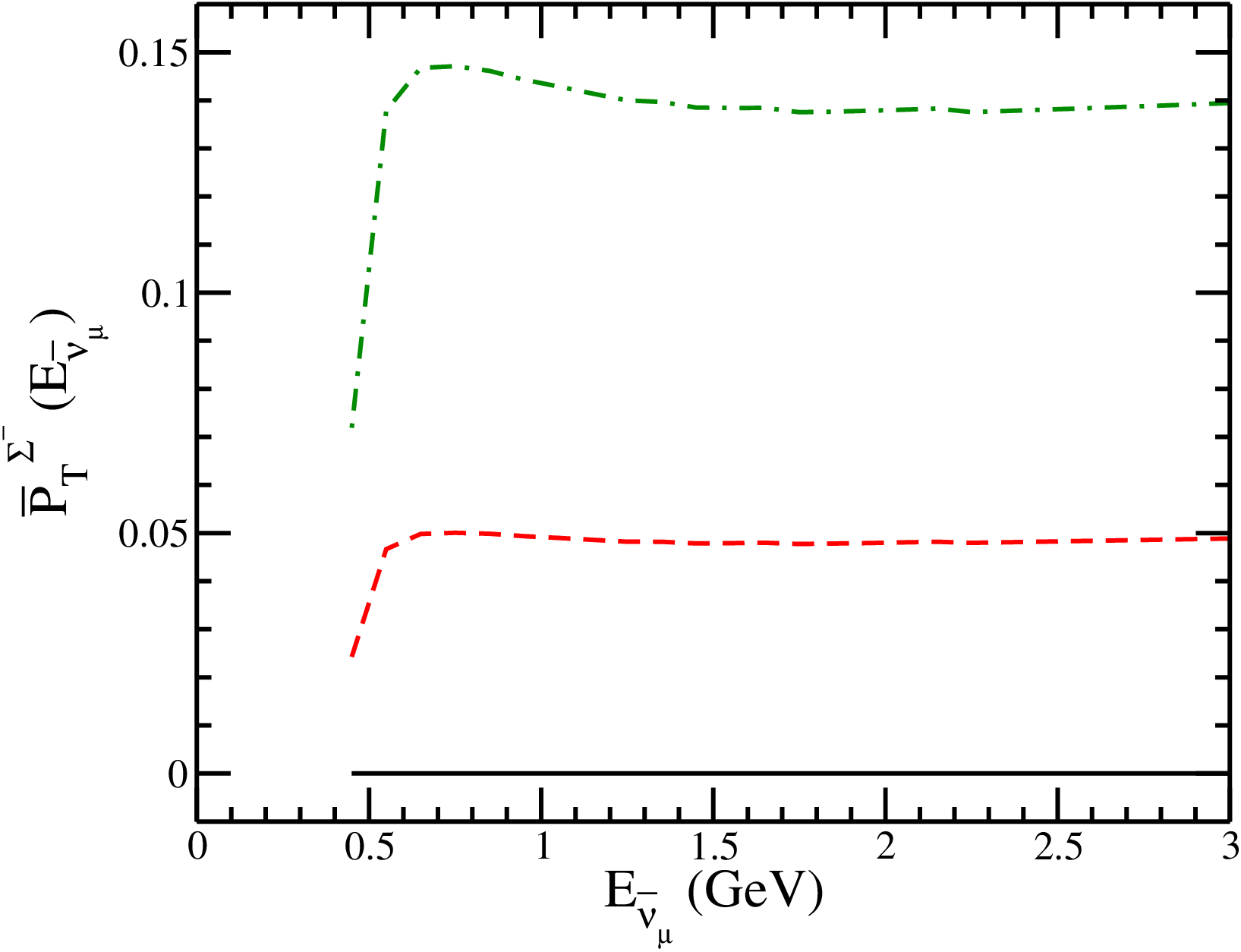}
\caption{$\sigma ~vs.~ E_{\bar{\nu}_{\mu}}$ (upper left panel) for the process ${\bar{\nu}_\mu + n \rightarrow \mu^+ 
+ \Sigma^-}$, $\overline{P}_L^{\Sigma^-} (E_{\bar{\nu}_{\mu}}) ~vs.~ E_{\bar{\nu}_{\mu}}$ (upper right panel), 
$\overline{P}_P^{\Sigma^-} (E_{\bar{\nu}_{\mu}}) ~vs.~ E_{\bar{\nu}_{\mu}}$ (lower left panel) and 
$\overline{P}_T^{\Sigma^-} (E_{\bar{\nu}_{\mu}}) ~vs.~ E_{\bar{\nu}_{\mu}}$ (lower right panel), for a polarized sigma 
in the final state. Lines and points have the same meaning as in Fig.~\ref{fig2}.}
\label{fig17}
\end{figure}

\begin{figure}
\includegraphics[height=6cm,width=5.2cm]{Pp_enu_real_g2_variation_lambda_with_Fp_Nambu_muon_polarized.eps}
\includegraphics[height=6cm,width=5.2cm]{Pp_enu_imaginary_g2_variation_lambda_with_Fp_Nambu_muon_polarized.eps}
 \includegraphics[height=6cm,width=5.2cm]{Pt_enu_imaginary_g2_variation_lambda_with_Fp_Nambu_muon_polarized.eps}
\caption{$\overline{P}_P^{\mu^+} (E_{\bar{\nu}_{\mu}}) ~vs.~ E_{\bar{\nu}_{\mu}}$ for the real values of $g_2^{np} (0)$ 
(left panel), $\overline{P}_P^l (E_{\bar{\nu}_{\mu}}) ~vs.~ E_{\bar{\nu}_{\mu}}$ for the imaginary values of $g_2^{np} 
(0)$ (middle panel) and $\overline{P}_T^l (E_{\bar{\nu}_{\mu}}) ~vs.~ E_{\bar{\nu}_{\mu}}$ (right panel) for the 
process ${\bar{\nu}_\mu + p \rightarrow {\mu}^+ + {\Lambda}}$. Lines and points have the same meaning as in 
Fig.~\ref{figmup}.}
\label{fig10}
\end{figure}

\begin{figure}
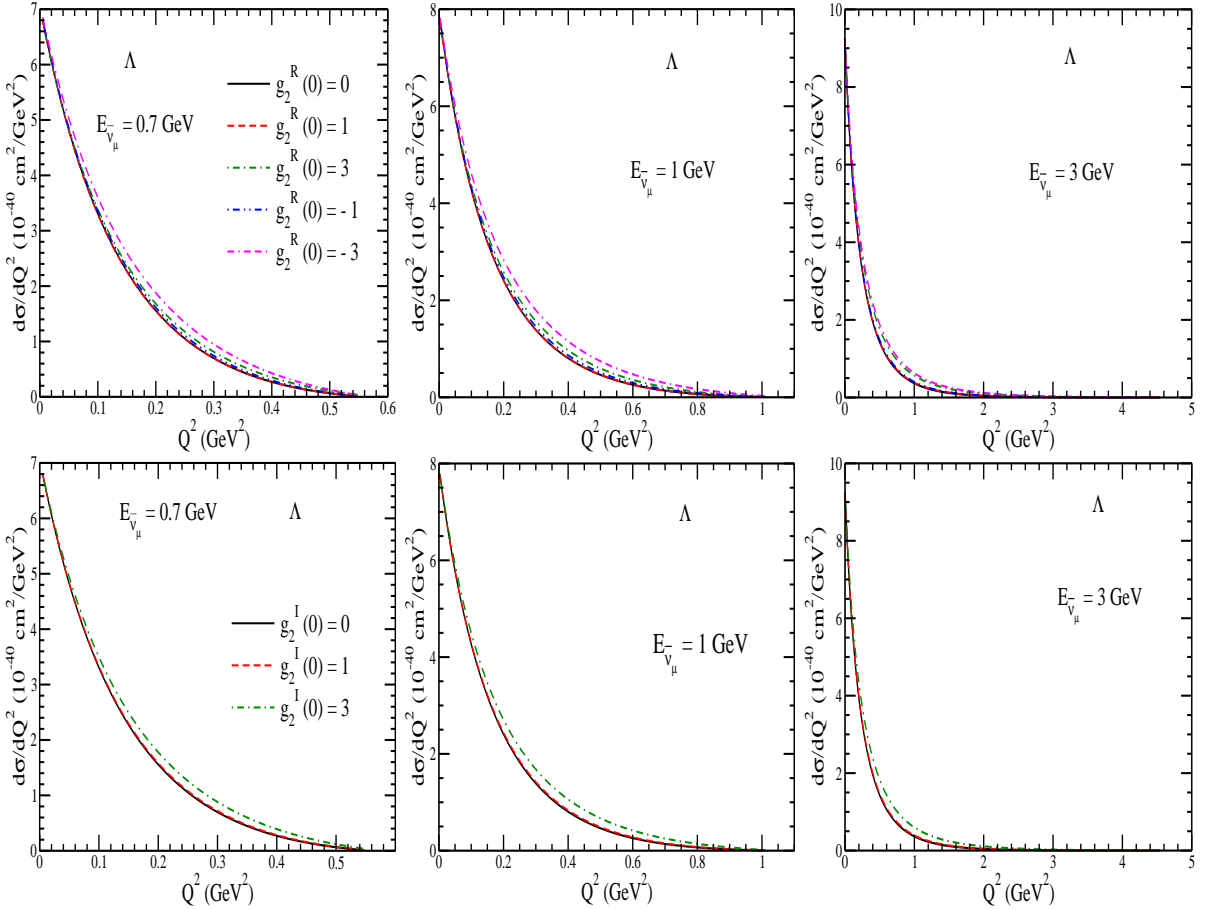

 \includegraphics[height=6cm,width=5.2cm]{dsigma_dq2_g2_variation_with_Fp_Nambu_real_g2_enu_700MeV.eps}  
 \includegraphics[height=6cm,width=5.2cm]{dsigma_dq2_g2_variation_with_Fp_Nambu_real_g2_enu_1GeV.eps} 
 \includegraphics[height=6cm,width=5.2cm]{dsigma_dq2_g2_variation_with_Fp_Nambu_real_g2_enu_3GeV.eps} \\ 
 \includegraphics[height=6cm,width=5.2cm]{dsigma_dq2_imaginary_g2_variation_with_Fp_Nambu_enu_700MeV_lambda.eps}
 \includegraphics[height=6cm,width=5.2cm]{dsigma_dq2_imaginary_g2_variation_with_Fp_Nambu_enu_1GeV_lambda.eps}
 \includegraphics[height=6cm,width=5.2cm]{dsigma_dq2_imaginary_g2_variation_with_Fp_Nambu_enu_3GeV_lambda.eps}
\caption{$d \sigma/d Q^2$ for the process ${\bar{\nu}_\mu + p \rightarrow \mu^+ + \Lambda}$ at the different values of 
the incoming antineutrino energy, $E_{\bar{\nu}_{\mu}} =$ 0.7~GeV~(left panel), 1~GeV~(middle panel) and 3~GeV~(right 
panel). The upper panel represents the results with ${{g_2^R (0)}}$ = 0~(solid line), 1~(dashed line), 3~(dashed-dotted 
line), $-1$~(double-dotted-dashed line) and $-3$~(double-dashed-dotted line) and the lower panel represents the results 
with ${{g_2^I (0)}}$ = 0~(solid line), 1~(dashed line) and 3~(dashed-dotted line).}\label{fig15}
\end{figure}

\begin{figure}
 \includegraphics[height=6cm,width=5.2cm]{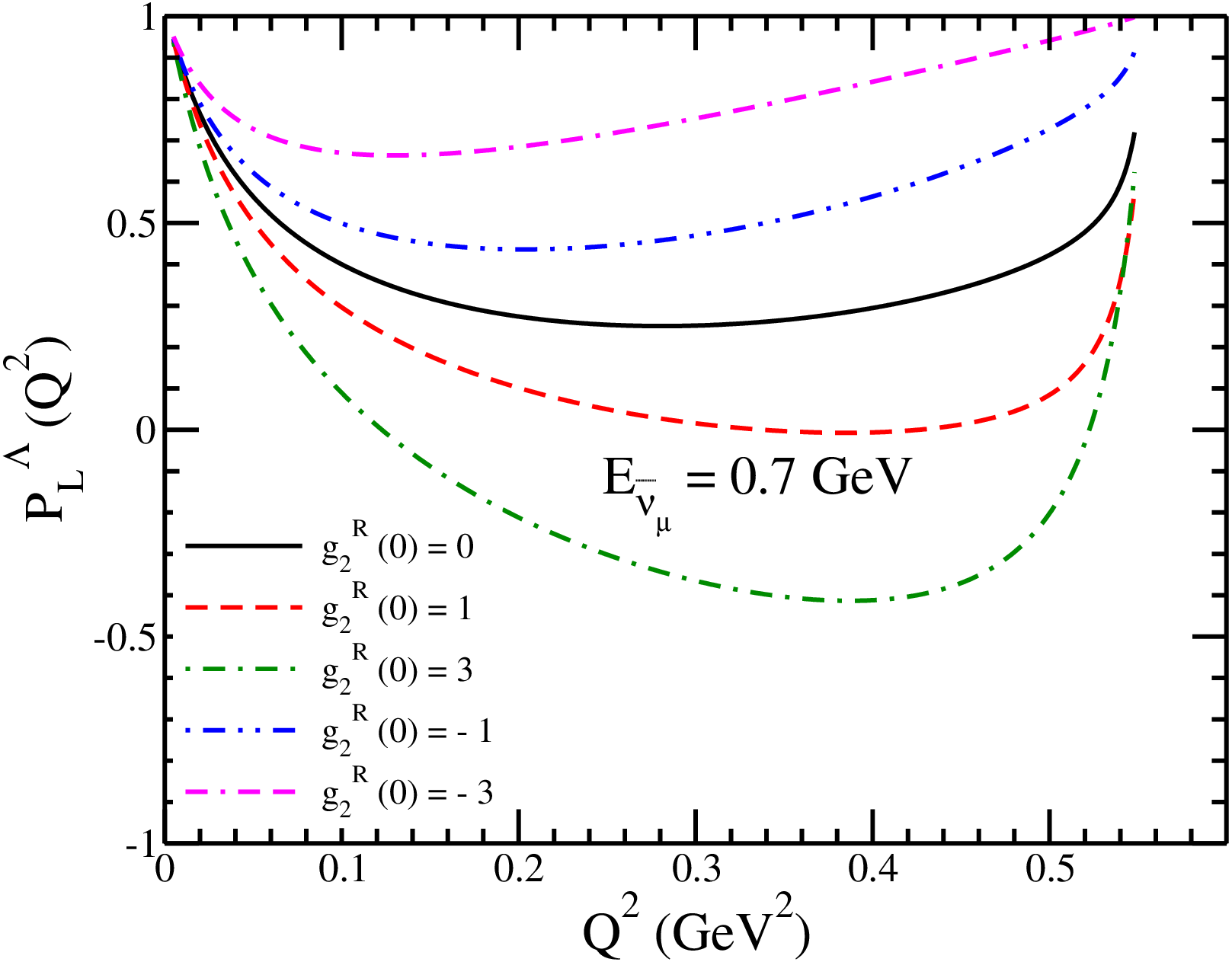} 
 \includegraphics[height=6cm,width=5.2cm]{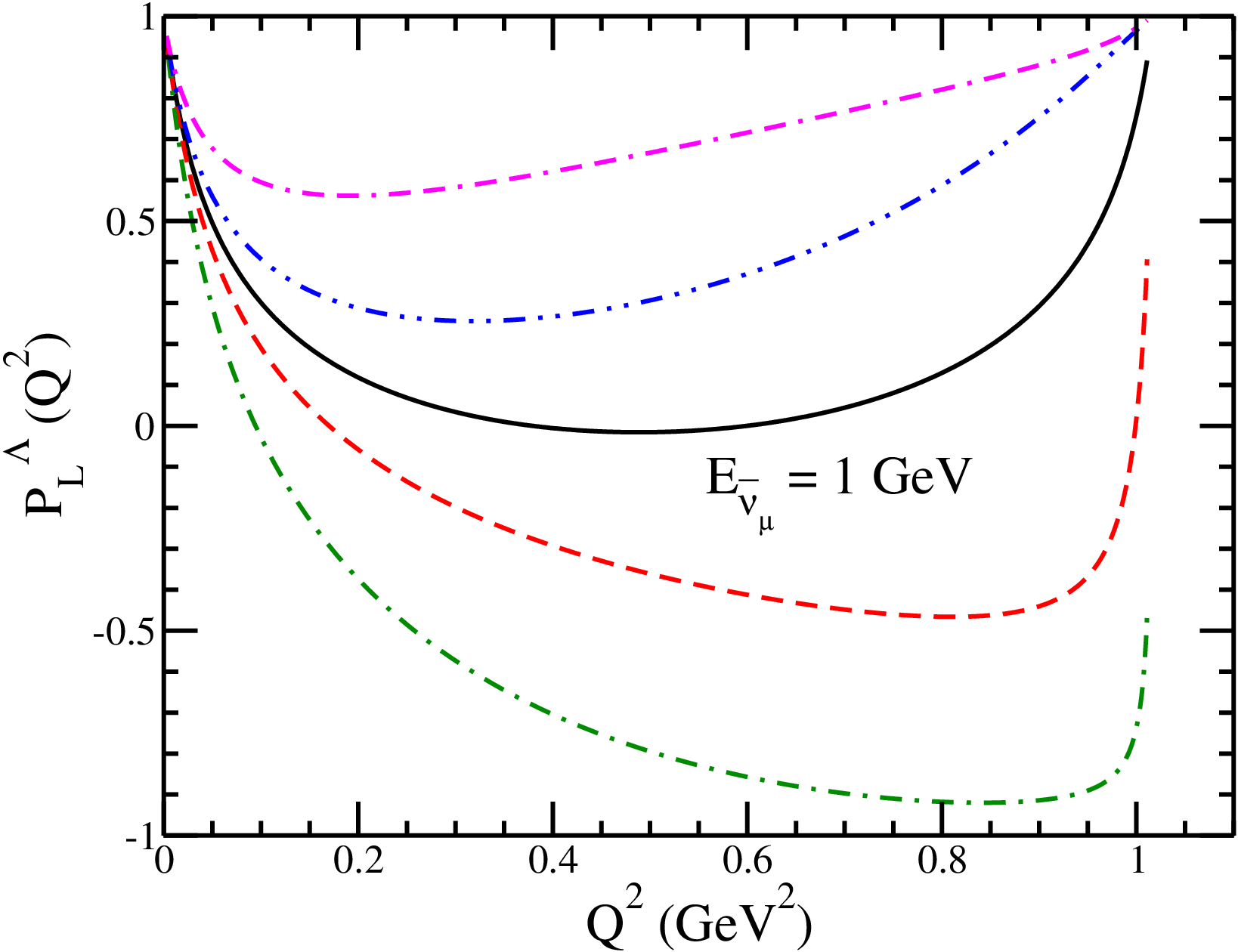} 
 \includegraphics[height=6cm,width=5.2cm]{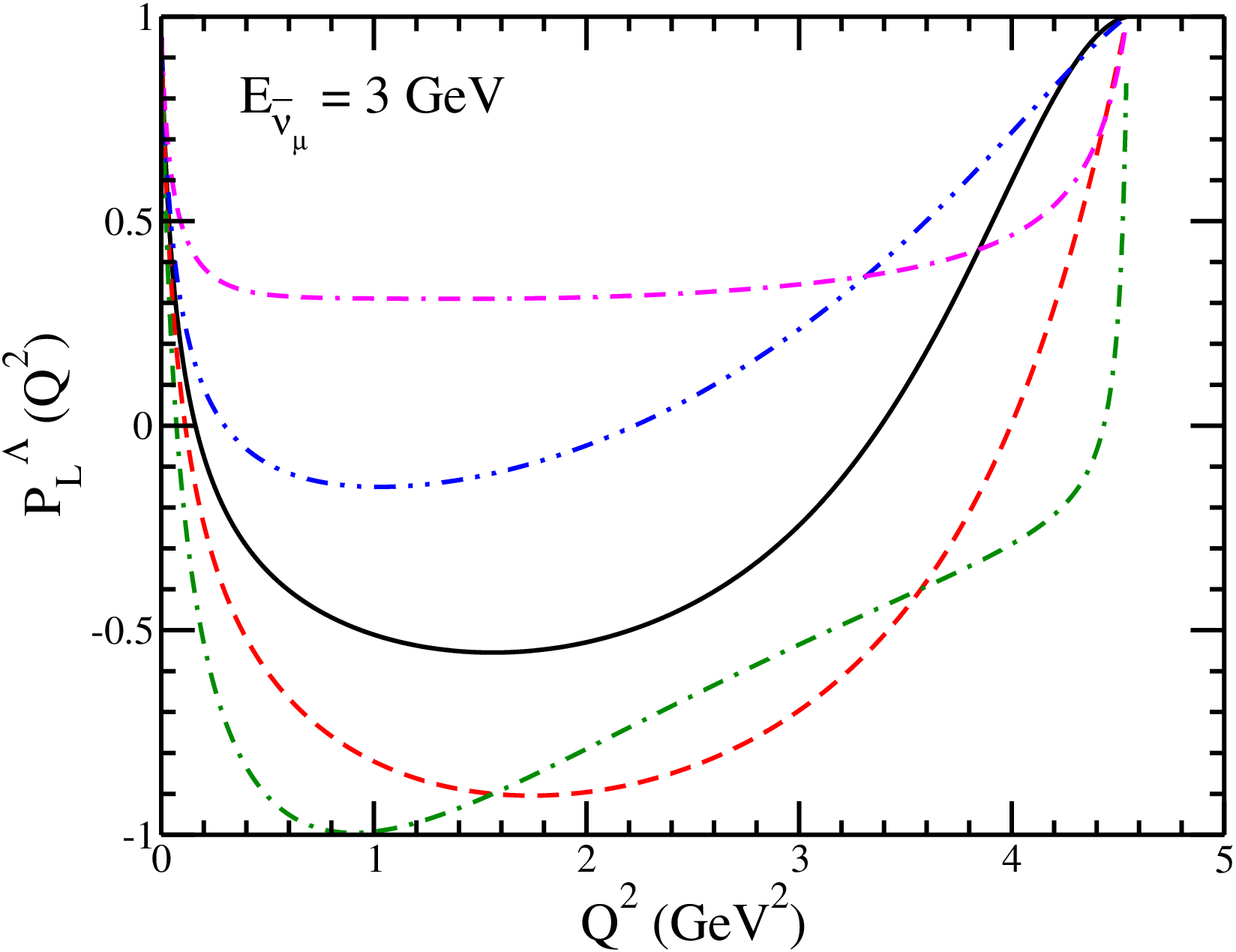} \\
 \includegraphics[height=6cm,width=5.2cm]{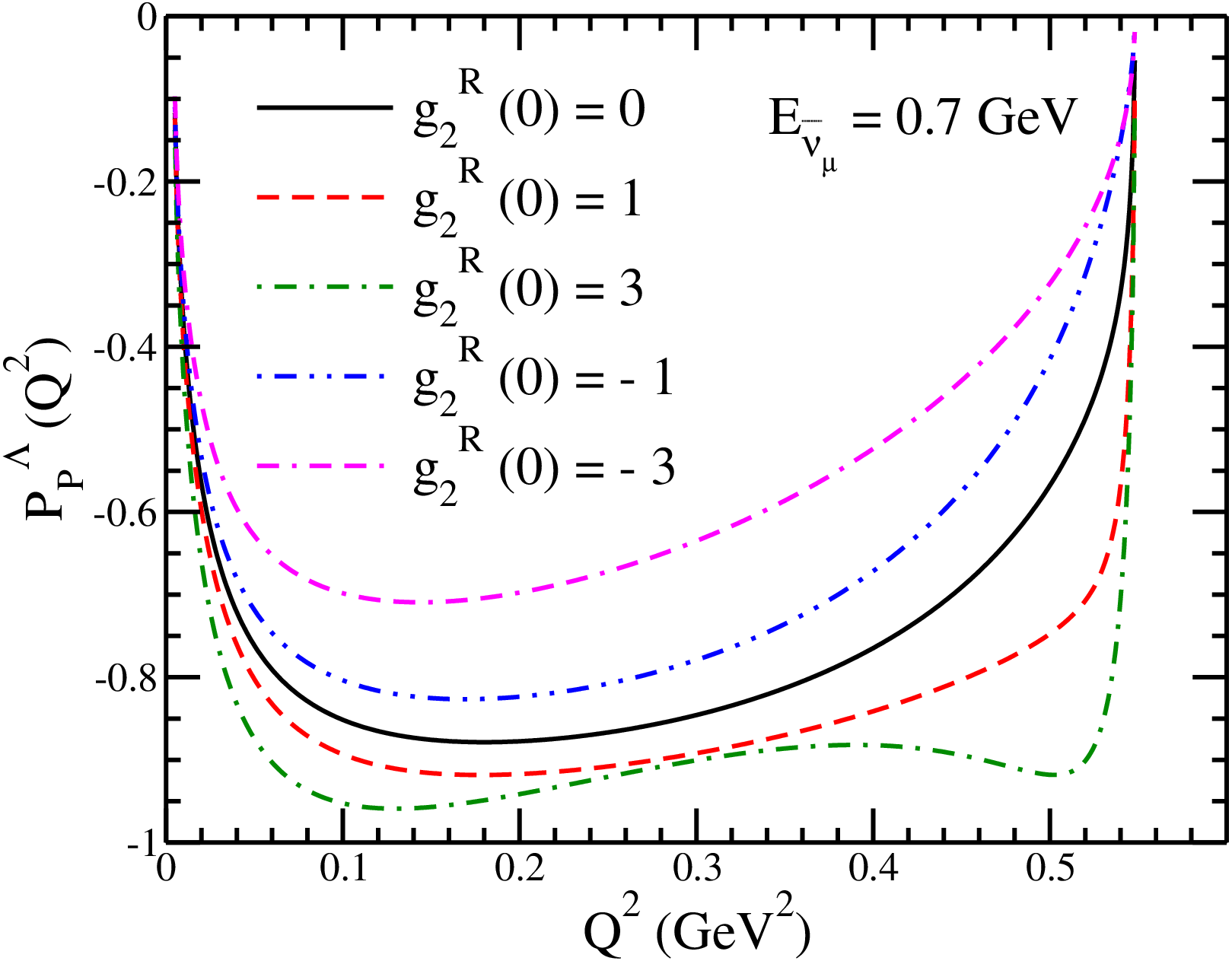}
 \includegraphics[height=6cm,width=5.2cm]{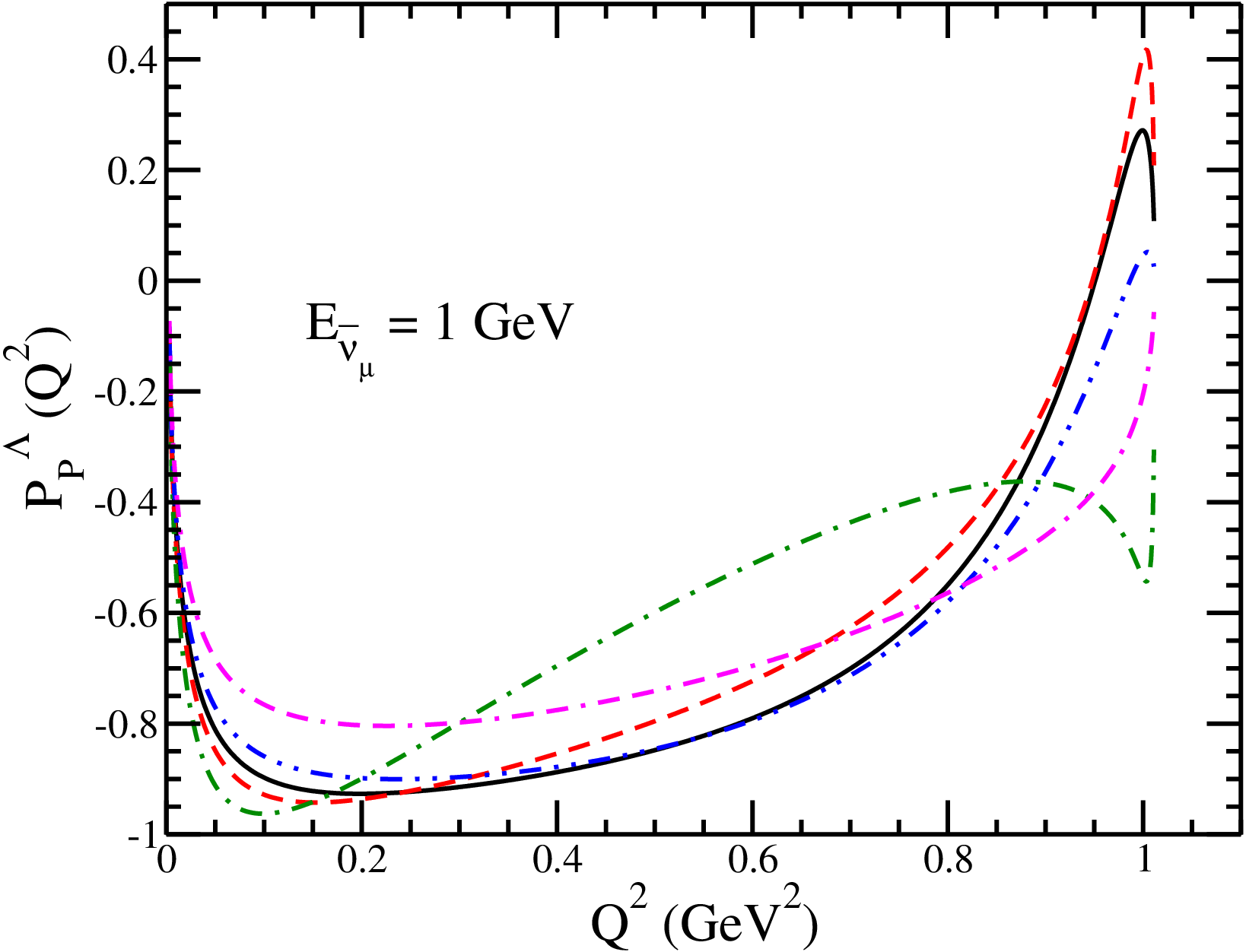} 
 \includegraphics[height=6cm,width=5.2cm]{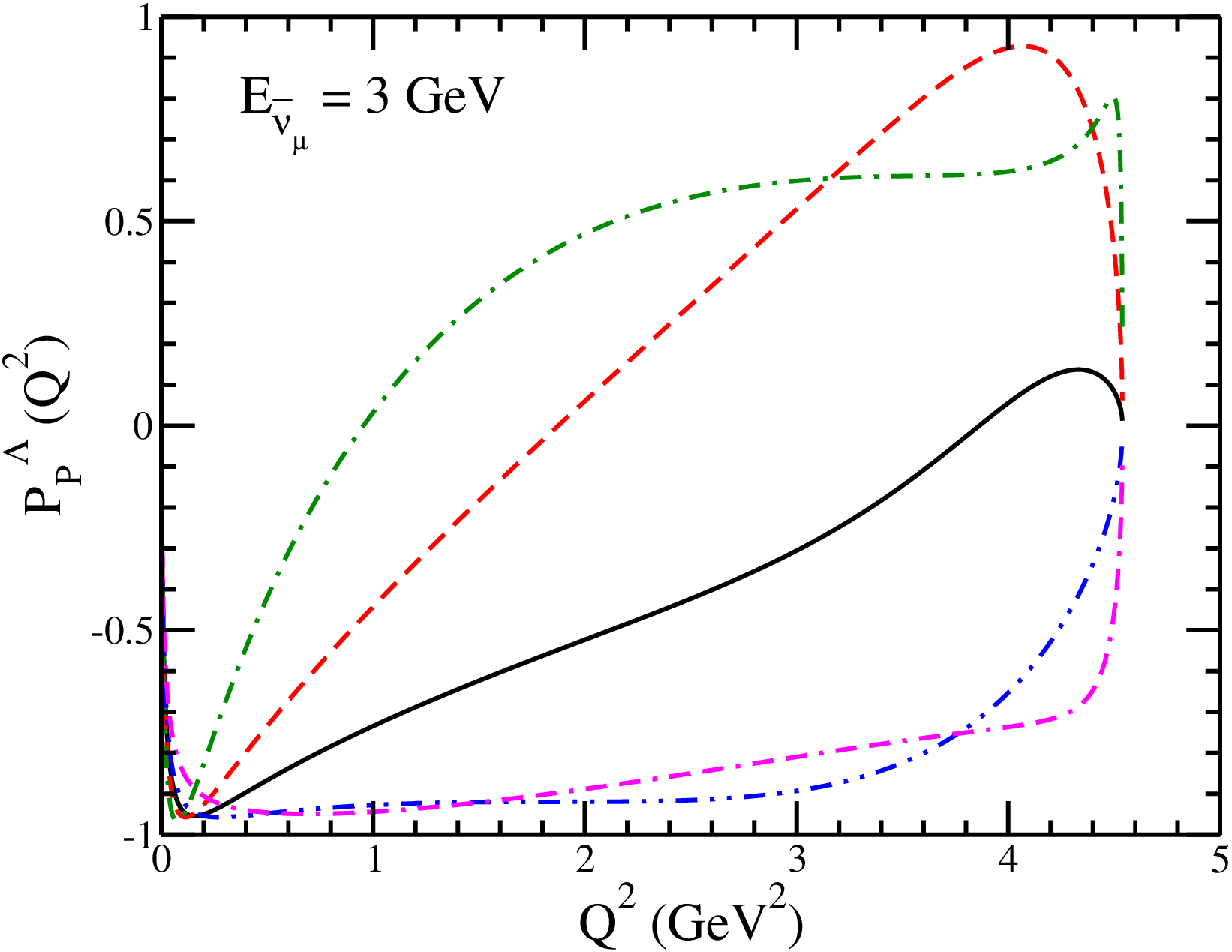}  
\caption{$P_L^{\Lambda} (Q^2)~vs.~ Q^2$ (upper panel) and $P_P^{\Lambda} (Q^2)~vs.~ Q^2$ (lower panel) for the process 
${\bar{\nu}_\mu + p \rightarrow \mu^+ + {\Lambda}}$ at the different values of the incoming antineutrino energy, 
$E_{\bar{\nu}_{\mu}} =$ 0.7~GeV~(left panel), 1~GeV~(middle panel) and 3~GeV~(right panel). Lines and points have the 
same meaning as in Fig.~\ref{fig1}.}\label{fig5}
\end{figure}

\begin{figure}
 \includegraphics[height=6cm,width=5.2cm]{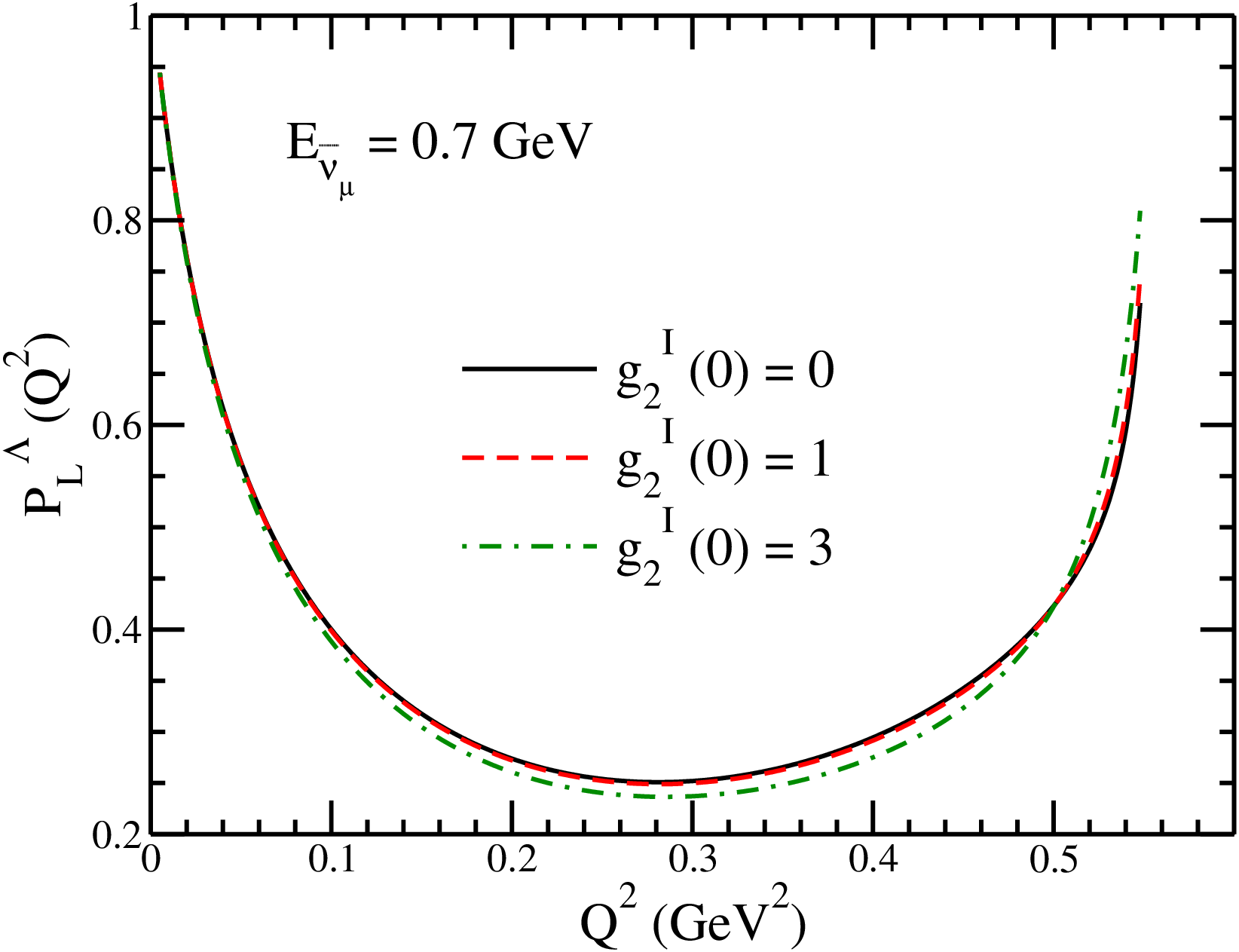}
 \includegraphics[height=6cm,width=5.2cm]{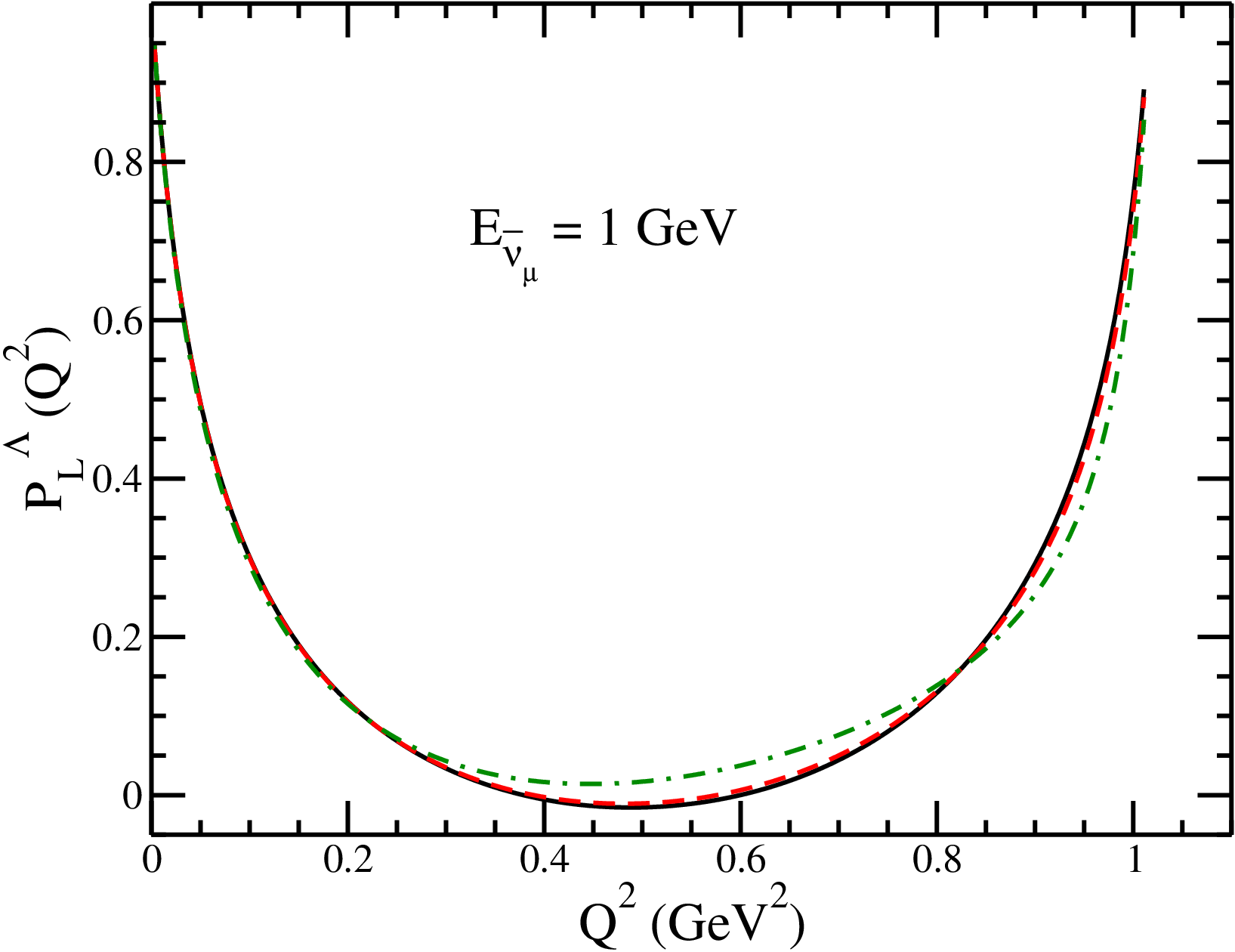}
 \includegraphics[height=6cm,width=5.2cm]{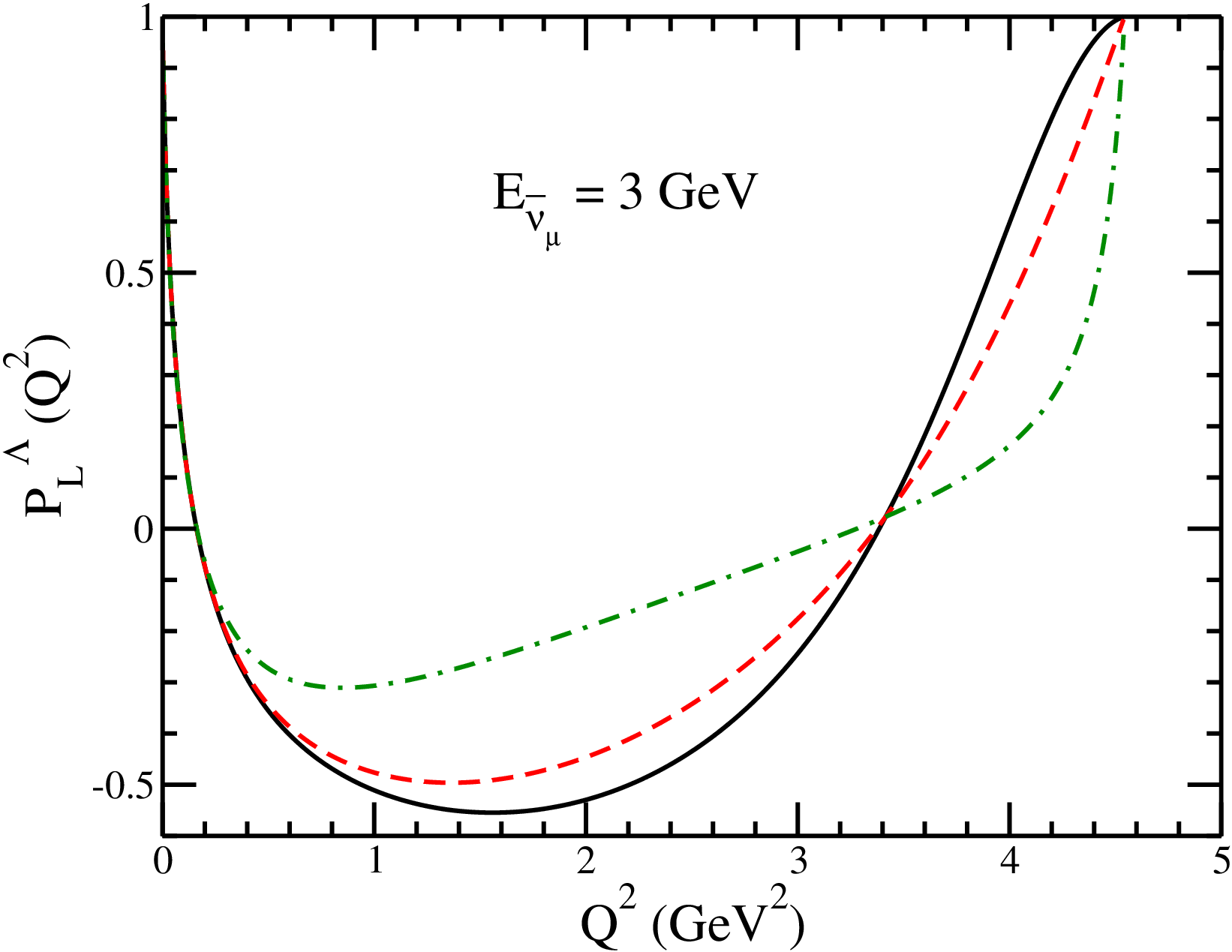} \\
 \includegraphics[height=6cm,width=5.2cm]{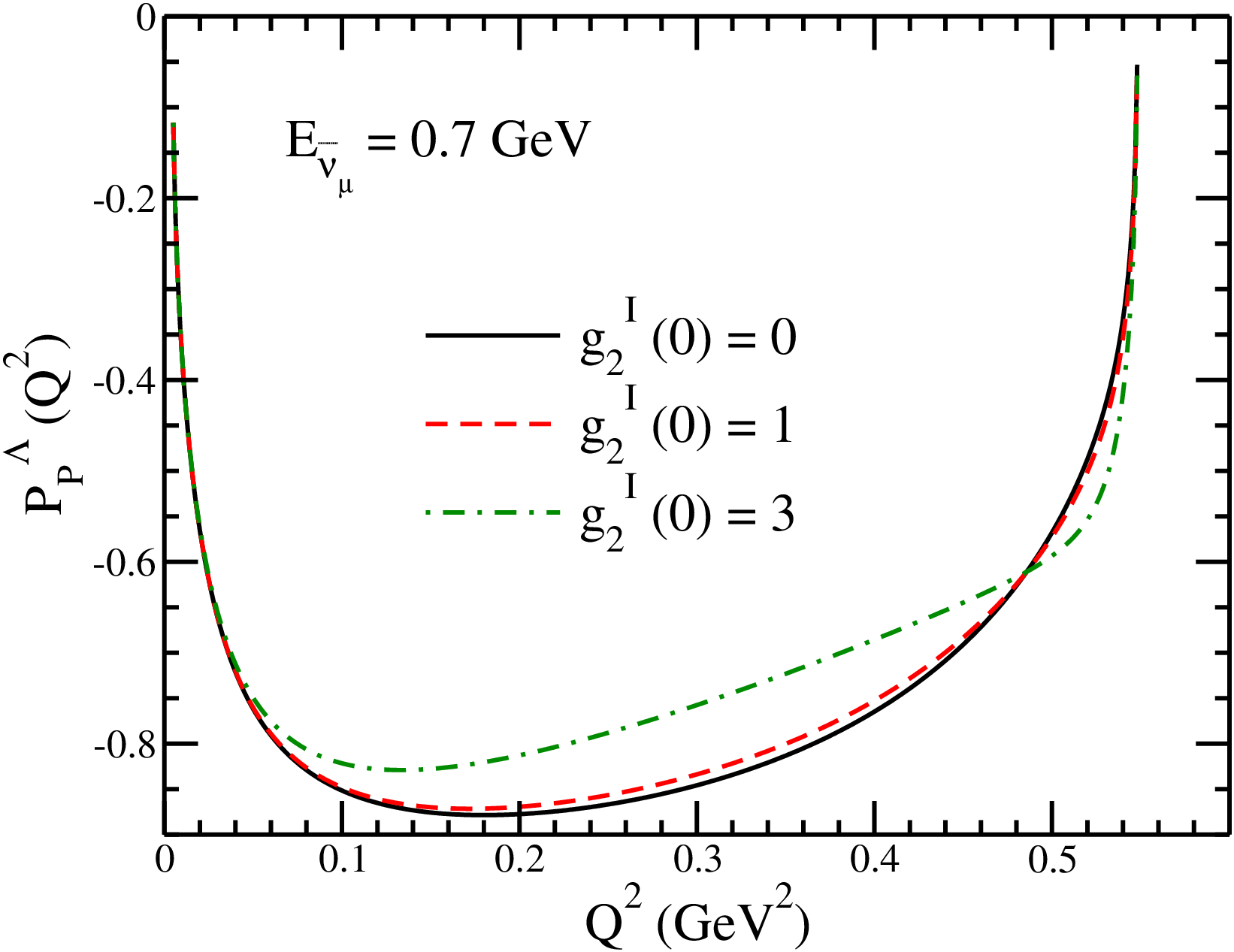}
 \includegraphics[height=6cm,width=5.2cm]{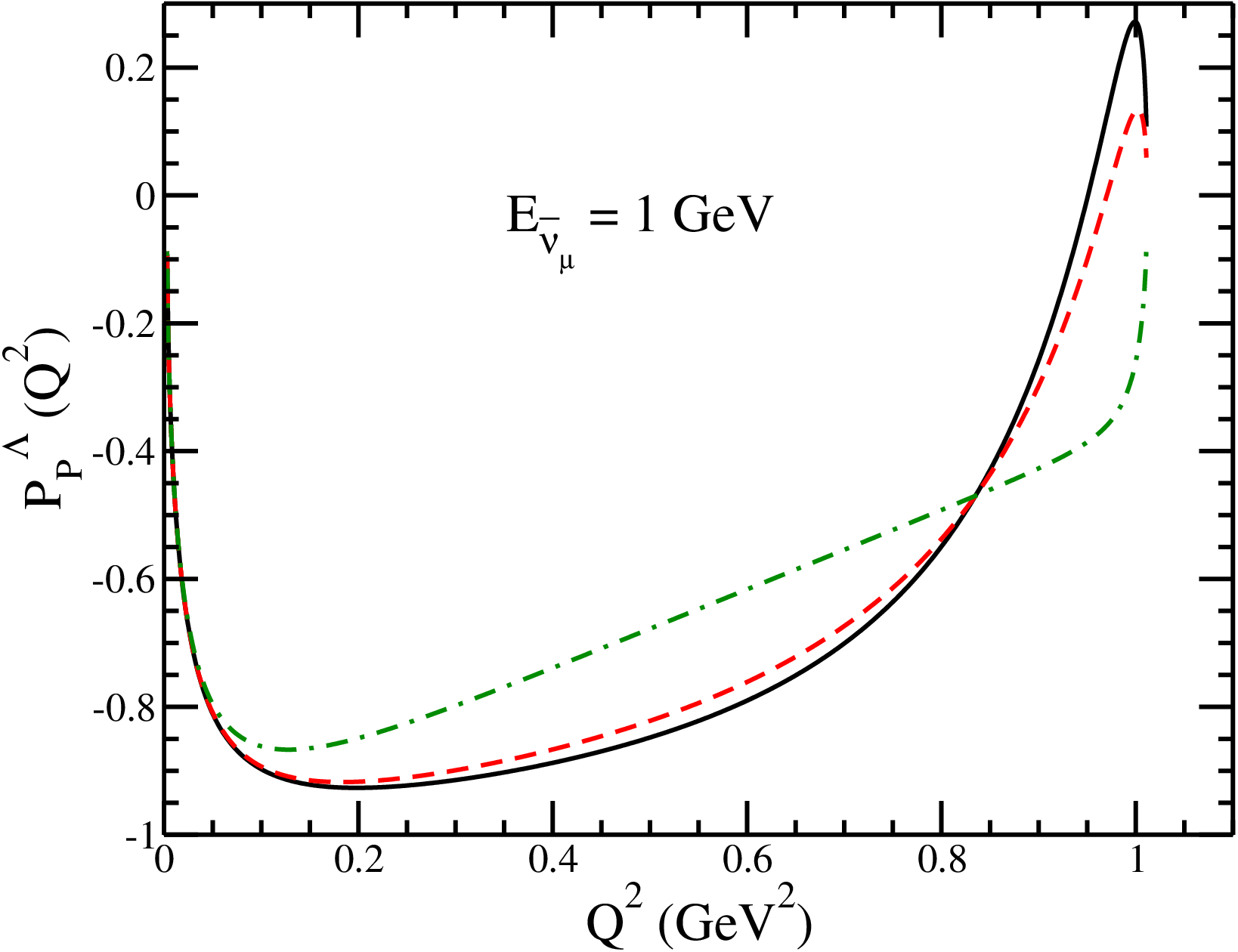} 
 \includegraphics[height=6cm,width=5.2cm]{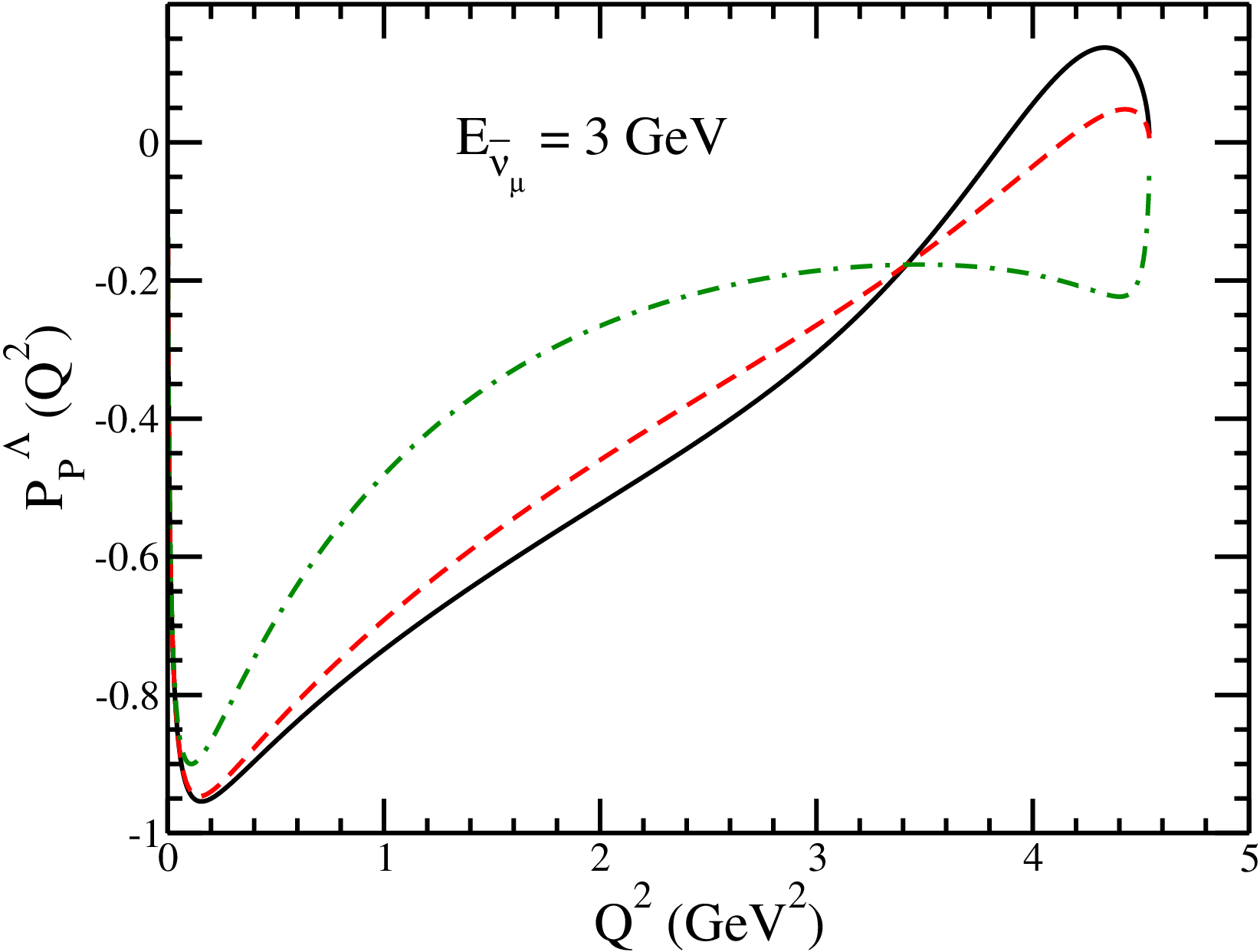}\\
 \includegraphics[height=6cm,width=5.2cm]{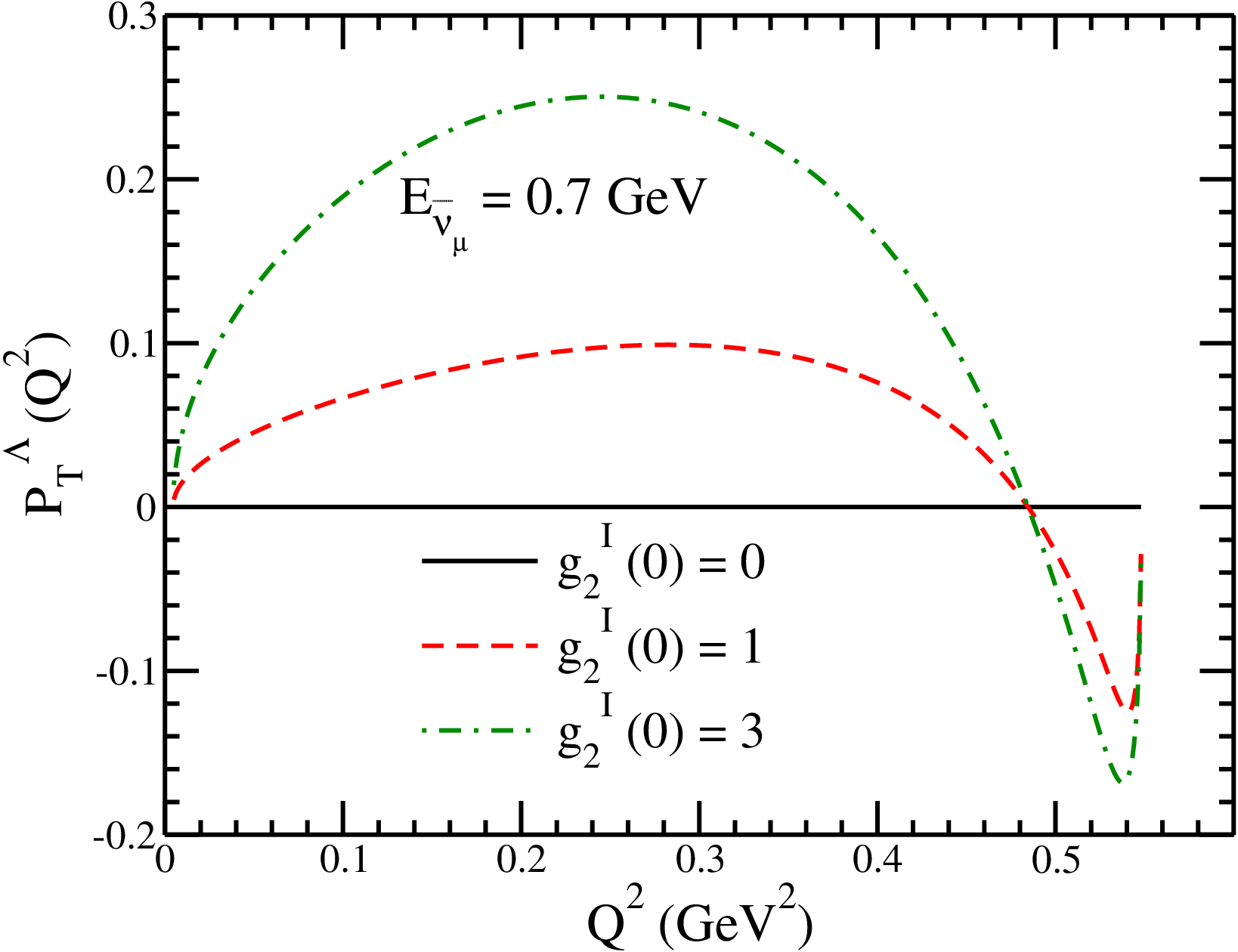}
 \includegraphics[height=6cm,width=5.2cm]{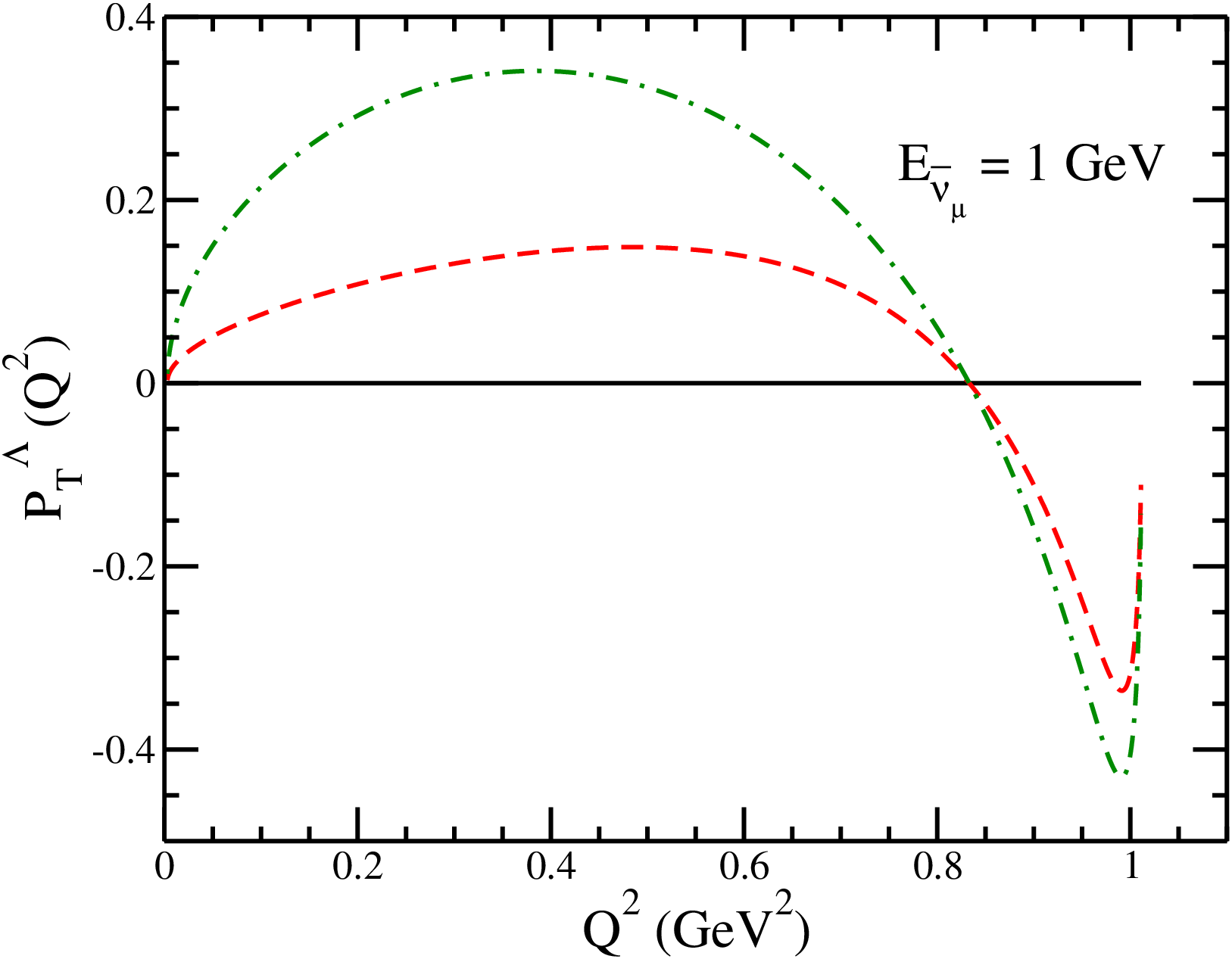}
 \includegraphics[height=6cm,width=5.2cm]{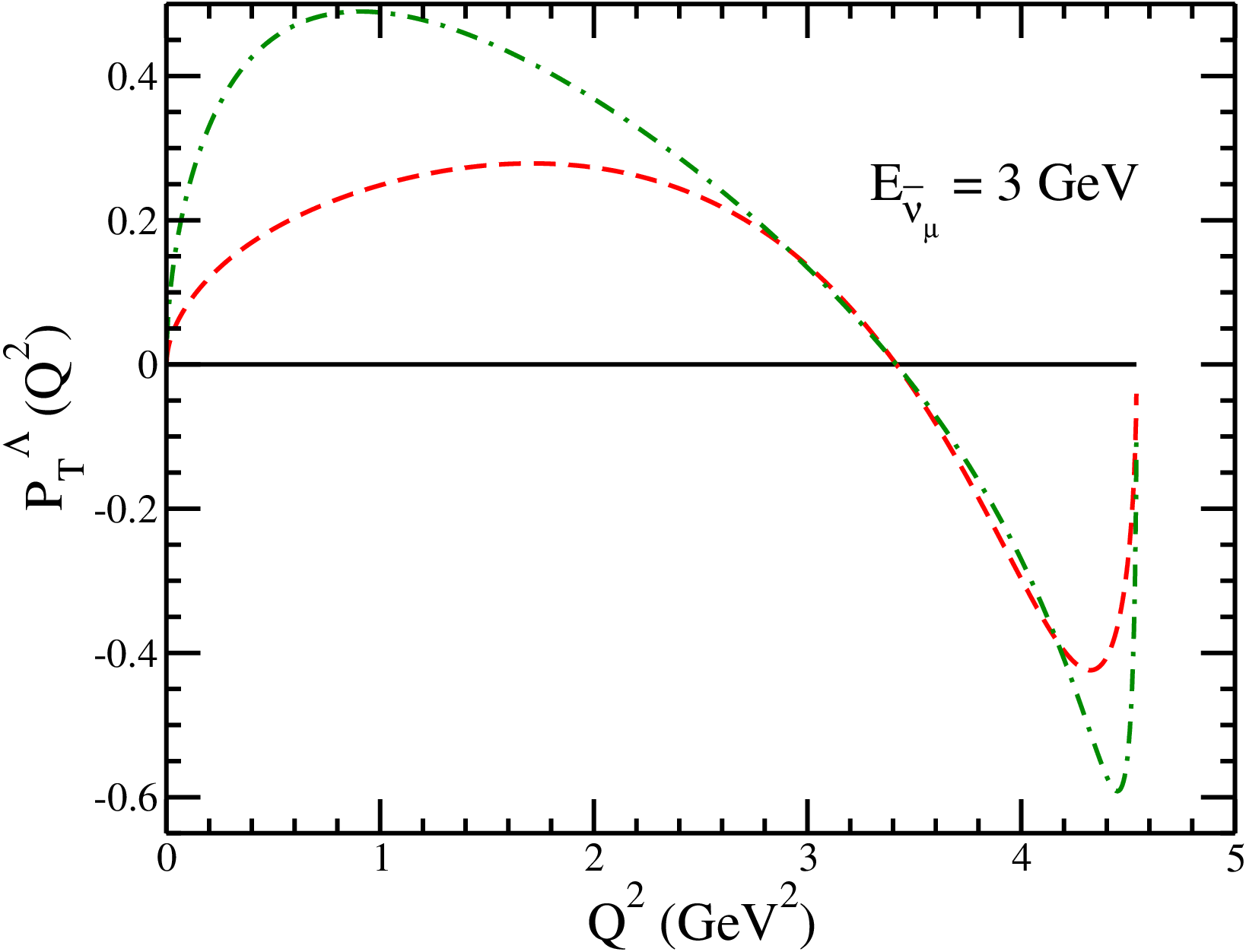}
\caption{$P_L^{\Lambda} (Q^2)~vs.~ Q^2$ (upper panel), $P_P^{\Lambda} (Q^2)~vs.~ Q^2$ (middle panel) and $P_T^{\Lambda} 
(Q^2)~vs.~ Q^2$~(lower panel) for the process ${\bar{\nu}_\mu + p \rightarrow \mu^+ + {\Lambda}}$ at the 
different values of the incoming antineutrino energy, $E_{\bar{\nu}_{\mu}} =$ 0.7~GeV~(left panel), 1~GeV~(middle panel)
and 3~GeV~(right panel). Lines and points have the same meaning as in Fig.~\ref{fig2}.}\label{fig6}
\end{figure}

 In Fig.~\ref{fig1}, the numerical results are presented for the total cross sections $\sigma (E_{\nu_\mu})$ for the 
 process $\nu_\mu + n \rightarrow \mu^- + p$ and the average polarizations $\overline{P}_L^p (E_{\nu_{\mu}})$ and 
 $\overline{P}_P^p (E_{\nu_{\mu}})$ for the outgoing proton as a function of neutrino energy for the various values of 
 $g_2^{np} (0)$ taking it to be real i.e. $g_2^{np} (0) = |g_2^R (0)| =$ 0, 1 and 3. We find that $\sigma (E_{\nu_\mu})$
 increases with the neutrino energy for all values of $g_2^R (0)$. In the case of $g_2^R (0) \ne 0$, there is a further 
 enhancement in the cross section which increases with $g_2^R (0)$ and the increase is significant only when $|g_2^R (0)
 | \ge 1$. The increase in $\sigma(E_{\nu_\mu})$ is almost the same for the positive as well as the negative values of 
 $g_2^R (0)$. However, the average values of the longitudinal and perpendicular components of the polarization 
 $\bar{P}_L^p (E_{\nu_\mu})$ and $\bar{P}_P^p (E_{\nu_\mu})$ are negative for all the values of $|g_2^R (0)|$ in the 
 range $0-3$ in the entire energy range of $E_{\nu_\mu}$ considered here. The absolute value of $\overline{P}_L^{p} 
 (E_{\nu_\mu})$ shows similar trend for the positive as well as the negative values of $g_2^R (0)$, i.e. it decreases 
 with the increase in energy. The absolute value of $\overline{P}_P^p (E_{\nu_\mu})$ decreases with energy for 
 $g_2^R (0) > 0$ and increases for $g_2^R (0) \le 0$.
 
 In Fig.~\ref{fig2}, we present the results for $\sigma (E_{\nu_\mu})$ and $\overline{P}_{L,P,T} (E_{\nu_{\mu}})$ as a 
 function of neutrino energy $E_{\nu_{\mu}}$ for the various values of $g_2^{np} (0)$ taken to be purely imaginary i.e. 
 $g_2^{np} (0) = g_2^I (0)$= 0, 1 and 3. We find that the value of $\sigma(E_{\nu_\mu})$ increases with energy for all 
 values of $g_2^I (0)$. The effect of $g_2^I (0) \ne 0$ is to further increase the cross section and the increase is 
 significant only for $|g_2^I (0)| \ge 1$. The longitudinal $\overline{P}_L^p (E_{\nu_\mu})$ and perpendicular 
 $\overline{P}_P^p (E_{\nu_\mu})$ components of the proton polarization are negative and their absolute values decrease 
 with the increase in $g_2^I (0) \ge 0$. The results for $\overline{P}_L^p (E_{\nu_\mu})$ and $\overline{P}_P^p 
 (E_{\nu_\mu})$ remain unchanged when the negative values of $g_2^I (0)$ are taken as they depend on $Re~g_2 (Q^2)$ and 
 $|g_2 (Q^2)|^2$ (see Eq.~(\ref{Ah}) and(\ref{Bh})). The transverse component of the proton polarization which violates 
 TRI is now non-vanishing and is positive for $g_2^I (0)\ge 0$ and increases with energy. Even for $g_2^I (0) = 1$, it 
 could be $\sim 10\%$ for $E_{\nu_\mu} \sim 1$ GeV and becomes larger for $g_2^I (0) \ge 1$. In the case of $g_2^I (0) 
 < 0$, the transverse component of the polarization $\overline{P}_T^p (E_{\nu_\mu})$ just changes sign but the magnitude
 remains the same, and therefore, the results have not been depicted in the figures.
  
 The results for the total cross section for the reaction $\bar{\nu}_\mu + p \rightarrow \mu^+ + n$, and the various 
 components of the polarization for the neutron are presented in Fig. \ref{fig3} for $g_2^R (0)$ by taking the different
 values of $g_2^R (0)$ viz. 0, $\pm$1 and $\pm$3. We see that with the increase in the value of $|g_2^R (0)|$, the 
 total cross section $\sigma (E_{\bar{\nu}_\mu})$ increases with energy for all values of $g_2^R (0)$. In the case of 
 $g_2^R (0) \ne 0$, there is further increase in the cross section which is almost the same for $g_2^R (0) > 0$ and 
 $g_2^R (0) < 0$. The longitudinal component of the polarization $\overline{P}_{L}^n (E_{\bar{\nu}_\mu})$ is positive 
 for the smaller values of energy ($E_{\bar{\nu}_\mu} <0.3$ GeV) and then becomes negative for all values of $g_2^R(0)$.
 In the case of non-zero values of $g_2^R (0)$, the absolute value of $\overline{P}_{L}^n (E_{\bar{\nu}_\mu})$ 
 increases with the energy and the increase is more (less) for the positive (negative) values of $g_2^R (0)$ as 
 compared to $g_2^R (0) = 0$. In the case of the perpendicular component of the polarization, $\overline{P}_P^n 
 (E_{\bar{\nu}_\mu})$ is always negative and its absolute value increases with energy at low energies $E_{\bar{\nu}_\mu}
 \le 0.3$ GeV and then decreases with energy for all $g_2^R (0)$. In the case of non-zero values of $g_2^R (0)$, the 
 absolute value of $\overline{P}_P^n (E_{\bar{\nu}_\mu})$ still decreases but the decrease is more (less) for positive 
 (negative) values of $g_2^R (0)$ as compared to $g_2^R (0) = 0$.
 
 In Fig.~\ref{fig4}, we have shown the results for $\sigma(E_{\bar{\nu}_\mu})$ and $\overline{P}_{L,P,T}^n 
 (E_{\bar{\nu}_\mu})$ for various values of $g_2^{np} (0)$ taken to be purely imaginary i.e. $g_2^I (0) =$ 0, 1 and 3. 
 The results for $\sigma(E_{\bar{\nu}_\mu})$, $\overline{P}_L^n (E_{\bar{\nu}_\mu})$ and $\overline{P}_P^n 
 (E_{\bar{\nu}_\mu})$ are qualitatively similar to Fig.~\ref{fig3}. The average value of the transverse component of 
 the polarization $\overline{P}_T^n (E_{\bar{\nu}_\mu})$ is non-zero and qualitatively similar to the case of 
 $\overline{P}_T^p (E_{\nu_\mu})$ for proton shown in Fig.~\ref{fig2}. 
  
 We notice from Figs.~\ref{fig1}--\ref{fig4} that the results for $\sigma (E_{\nu_\mu(\bar{\nu}_\mu)})$ increases in 
 the presence of the SCC with (without) TRI i.e. $g_2^R(0) (g_2^I(0)) \ne 0$. The increase is almost the same for the 
 real as well as the imaginary values of $g_2 (0)$. It is also well known that with the increase in the value of the 
 axial dipole mass $M_A$, the cross section increases. For example, with $20\%$ increase in the value of $M_A$, i.e. 
 from 1 GeV to 1.2 GeV, the cross section increases by about $20\%$.  A higher value of $M_A$ from the world average 
 value~(i.e. $M_A = 1.026$ GeV) is suggested from (anti)neutrino scattering experiments in the higher energy 
 region~\cite{Lyubushkin:2008pe, Formaggio:2013kya}. It is, therefore, possible that a non-zero value of $g_2^R (0)$ 
 and/or $g_2^I (0)$ i.e. the existence of the SCC with or without TRI may lead to a smaller value of $M_A$. Keeping 
 this in mind, in Fig.~\ref{figb}, we have studied the dependence of the cross section on $M_A$ with or without the 
 presence of $g_2^R (0)$ or $g_2^I (0)$. It may be observed from the figure that, in the case of neutrino induced 
 process i.e. ${{\nu}_\mu + n \rightarrow \mu^- + p}$, the results obtained by taking $M_A = 1.1$ GeV and $g_2^R(0) = 0$
 are comparable to the results obtained with $M_A = 1.026$ GeV and $g_2^R(0) = 2$, whereas the results obtained by 
 taking $M_A = 1.2$ GeV and $g_2^R(0) = 0$ are comparable to the results obtained using $M_A = 1.026$ GeV and $g_2^R(0) 
 = 3$. While in the case of antineutrino induced process i.e. ${\bar{\nu}_\mu + p \rightarrow \mu^+ + n}$, the results 
 obtained by taking $M_A = 1.1$ GeV and $g_2^R(0) = 0$ are comparable to the results obtained with $M_A = 1.026$ GeV and
 $g_2^R(0) = 1$ whereas the results obtained by taking $M_A = 1.2$ GeV and $g_2^R(0) = 0$ are slightly lower than the 
 results obtained using $M_A =1.026$ GeV and $g_2^R(0) = 2$. Thus, a higher value of $\sigma (E_{\bar{\nu}_\mu})$ may 
 be obtained by either taking a non-zero value of $g_2^{np} (0)$ or increasing the value of $M_A$. Furthermore, the 
 cross section measurements may give information only about the non-zero value of $g_2^{np} (0)$ irrespective of the 
 nature of the SCC current i.e. with or without TRI. One may obtain the nature of the SCC by measuring the polarization 
 observables which gives different results with real and imaginary values of $g_2^{np} (0)$ corresponding to the SCC 
 with or without TRI.
 
 In Fig.~\ref{figmup}, we have presented the results for the perpendicular and the transverse components of the 
 polarization of the muon produced in the reaction ${{\nu}_\mu + n \rightarrow {\mu}^- + p}$ for the real as well as 
 the imaginary value of $g_2^{np} (0)$. Since the relativistic muons are known to be longitudinal, therefore, we have 
 presented the results only for $\overline{P}_P^{\mu} (E_{{\nu}_\mu})$ and $\overline{P}_T^{\mu} (E_{{\nu}_\mu})$. We 
 find that $\overline{P}_P^{\mu} (E_{{\nu}_\mu})$ is less sensitive to the values of $g_2^R (0)$ as well $g_2^I (0)$ 
 unless $|g_2^R (0)|>1$. The transverse component of the polarization $\overline{P}_T^{\mu} (E_{{\nu}_\mu})$ is mildly 
 sensitive to $g_2^I (0)$, and the sensitivity increases with the increase in $g_2^I (0)$.
 
 Similarly in Fig.~\ref{figmu}, we have presented the results for the perpendicular and the transverse components of 
 polarization of the muon produced in the reaction ${\bar{\nu}_\mu + p \rightarrow {\mu}^+ + n}$ for the real as well 
 as the imaginary values of $g_2^{np} (0)$. We find that in the case of real $g_2^{np}(0)$, $\overline{P}_P^{\mu} 
 (E_{\bar{\nu}_\mu})$ shows appreciable sensitivity towards $g_2^R(0)$. The results for the positive (negative) values 
 of $g_2^R (0)$ are less (more) as compared to the results  obtained with $g_2^R(0) = 0$. In the case of imaginary 
 $g_2^{np}(0)$, the perpendicular component of polarization is not very sensitive to $g_2^I(0)$, whereas the transverse 
 component of polarization is sensitive to $g_2^I(0)$,  and the sensitivity increases with the increase in $g_2^I (0)$.

 In Fig.~\ref{fig11}, we have presented the results for the total cross section $\sigma (E_{\bar{\nu}_\mu})$ for the 
 process ${\bar{\nu}_\mu + p \rightarrow \mu^+ + \Lambda}$, and the average polarizations $\overline{P}_L^{\Lambda} 
 (E_{\bar{\nu}_{\mu}})$ and $\overline{P}_P^{\Lambda}(E_{\bar{\nu}_{\mu}})$ as a function of $E_{\bar{\nu}_{\mu}}$ 
 for the polarized $\Lambda$ at the different values of $g_2^R (0) =$ 0, $\pm$1 and $\pm$3 and in Fig.~\ref{fig12}, the 
 results for the total cross section $\sigma(E_{\bar{\nu}_\mu})$, and the average polarizations 
 $\overline{P}_L^{\Lambda} (E_{\bar{\nu}_{\mu}})$, $\overline{P}_P^{\Lambda} (E_{\bar{\nu}_{\mu}})$ and 
 $\overline{P}_T^{\Lambda} (E_{\bar{\nu}_{\mu}})$ are presented for the different values of $g_2^I (0) =$ 0, 1 and 3. 
 We see that the results for $\overline{P}_{L,P,T}^{\Lambda} (E_{\bar{\nu}_{\mu}})$ are qualitatively similar to the 
 results for neutron in the reaction ${\bar{\nu}_\mu + p \rightarrow \mu^+ + n}$ but quantitatively they are smaller. 
 However, in the case of the total cross section, unlike the neutron production cross section in the reaction 
 $\bar{\nu}_\mu + p \longrightarrow \mu^+ + n$, we find that the positive and negative values of $g_2^R (0)$ give 
 different values of the total cross section $\sigma (E_{\bar{\nu}_\mu})$. The results with $g_2^R (0) < 0$ are higher 
 than the results obtained with $g_2^R (0)>0$.

 Similarly, in Figs.~\ref{fig16} and \ref{fig17}, we present the results for $\sigma (E_{\bar{\nu}_\mu})$ for the 
 reaction ${\bar{\nu}_\mu + p \rightarrow \mu^+ + \Sigma^-}$ and $\overline{P}_{L,P,T}^{\Sigma^-} (E_{\bar{\nu}_{\mu}})
 $ for the polarized $\Sigma^-$ in the final state for $|g_2^R (0)|$ and $g_2^I (0)$ in the range $0-3$. We see that 
 while the results for $\sigma (E_{\bar{\nu}_\mu})$ are qualitatively similar to the results for $\Lambda$ production 
 in the reaction ${\bar{\nu}_\mu + p \rightarrow \mu^+ + \Lambda}$, the results for $\overline{P}_{L,P,T}^{\Sigma^-} 
 (E_{\bar{\nu}_{\mu}})$ are qualitatively different from the results for $\overline{P}_{L,P,T}^{\Lambda} 
 (E_{\bar{\nu}_{\mu}})$. It should be noted that results for $\sigma^{\Sigma^0} (E_{\bar{\nu}_\mu})$ and 
 $\overline{P}_{L,P,T}^{\Sigma^0} (E_{\bar{\nu}_{\mu}})$ in the reaction ${\bar{\nu}_\mu + p \rightarrow \mu^+ + 
 {\Sigma}^0}$ will be similar to $\Sigma^-$ production due to the SU(3) Clebsch Gordan coefficients shown in 
 Table~\ref{tabI}, and the numerical results for this reaction are not presented here.
 
 In Fig.~\ref{fig10}, we present the results for the perpendicular and transverse polarization of muons for the various 
 values of $g_2^R (0)$ and $g_2^I (0)$ in the case of $\Lambda$ production. We see that while $\overline{P}_P^{\mu} 
 (E_{\bar{\nu}_\mu})$ is not very sensitive to the value of $g_2^R (0)$ and $g_2^I(0)$ unless the value of $g_2^R (0)$ 
 is much greater than one i.e. $g_2^R (0) \ge 1$. However, the transverse component of the polarization 
 $\overline{P}_T^{\mu} (E_{\bar{\nu}_\mu})$ is sensitive to the numerical value of $g_2^I(0)$.

 As discussed in the introduction, the hyperons ($\Lambda, \Sigma^0, \Sigma^-$) produced in the reactions given in 
 Eq.~(\ref{reaction1}) and (\ref{reaction2}) are the most appropriate candidates for doing the polarization 
 measurements as they are self analyzer of their polarization through the measurement of the asymmetry of the angular 
 distribution of the pions as their decay product. Out of the $\Lambda, \Sigma^0 \text{ and } \Sigma^-$ hyperons, 
 $\Lambda$ production cross sections are the largest. Therefore, we have studied the $Q^2$ dependence of the cross 
 section ($d \sigma/dQ^2$) and the polarization components $P_{L,P,T} (Q^2)$ for various antineutrino energies and 
 present the numerical results in the following.

 In Fig.~\ref{fig15}, we have presented the results for the $d \sigma/dQ^2$ for the process $\bar{\nu}_\mu + p 
 \longrightarrow \mu^+ + \Lambda$ at the different values of the antineutrino energies, viz. $E_{\bar{\nu}_\mu} = 0.7, 
 1$ and 3 GeV by varying $|g_2^{R} (0)|$ as well as $g_2^I (0)$ in the range $0-3$. We find that the $d \sigma/dQ^2$ is 
 not very sensitive to the presence of SCC.
 
 For the reaction ${\bar{\nu}_\mu + p \rightarrow \mu^+ + {\Lambda}}$, we have presented the results for $P_L (Q^2)$
 and $P_P (Q^2)$ as a function of $Q^2$ in Fig.~\ref{fig5}. The results are presented for the polarization components 
 using $g_2^{R} (0) =$ 0, $\pm$1 and $\pm$3 at the different values of $E_{\bar{\nu}_{\mu}} =$ 0.7, 1 and 3~GeV. We find
 that $P_L (Q^2)$ shows large variations as we change $|g_2^{R} (0)|$ from 0 to 3 at low antineutrino energies,
 $E_{\bar{\nu}_\mu}$ (say 0.7 GeV) in comparison to the higher energies (say 3 GeV). For example, in the peak region 
 of $Q^2$, the difference is 80$\%$ at $E_{\nu_{\mu}}$ = 0.7 GeV and it is 50$\%$ at $E_{\nu_{\mu}}$ = 3 GeV as 
 $|g_2^R (0)|$ is changed from 0 to 3. In the case of $P_P (Q^2)$ also, the $Q^2$ dependence is quite strong and 
 similar to $P_L (Q^2)$. 
 
 In Fig.~\ref{fig6}, the results are presented  for $P_L (Q^2)$, $P_P (Q^2)$ and $P_T (Q^2)$ as a function of $Q^2$
 using $g_2^{I} (0) =$ 0, 1 and 3 at the different values of $E_{\bar{\nu}_{\mu}} =$ 0.7, 1 and 3~GeV. We see that  
 while $P_L(Q^2)$ is less sensitive to $g_2^I (0)$ at low antineutrino energies, the sensitivity increases at higher 
 antineutrino energies. $P_P(Q^2)$ is sensitive to $g_2^{I}(0)$ in the antineutrino energy range $0.7-3$ GeV and the 
 difference increases with the increase in antineutrino energy as $g_2^{I} (0)$ increases from 0 to 3. Moreover, 
 $P_T(Q^2)$ is sensitive to $g_2^I(0)$ at all antineutrino energies. $P_T(Q^2)$ shows 25$\%$, 40$\%$ and 50$\%$ 
 variations at $Q^2 = $ 0.25, 0.4 and 1 GeV$^2$ at $E_{\bar{\nu}_{\mu}} =$ 0.7, 1 and 3 GeV, respectively, when 
 $g_2^{I} (0)$ is varied from 0 to 3.

\section{Summary and conclusions}\label{conclusions}
We have studied in this work the quasielastic scattering of neutrinos and antineutrinos from nucleons induced by the 
weak charged currents in the $\Delta S = 0$ and $\Delta S = 1$ sectors in the presence of the SCC with and without 
assuming the validity of TRI. The effect of the SCC has been studied on the total cross section $\sigma (E_{\nu_\mu 
(\bar{\nu}_\mu)})$ and the average polarization components of the leptons ($l$ = muon) and the baryons ($h = n, p, 
\Lambda, \Sigma^0, \Sigma^-$) in the longitudinal, perpendicular and transverse $\overline{P}_{L,P,T}^{l,h} 
(E_{\nu_\mu (\bar{\nu}_\mu)})$ components of these particles produced in the final state. In the case of the 
quasielastic production of $\Lambda$ hyperons by the antineutrinos which is the most suitable candidate for measuring 
the polarization observables of the hyperons, we have also studied the $Q^2$ dependence of the polarization components 
$P_{L,P,T}^{\Lambda} (Q^2)$ and the differential cross sections $d \sigma/dQ^2$.

The standard parameterizations of the electroweak form factors for the first class vector and axial vector currents in 
the $\Delta S = 0$ sector using the hypothesis of CVC and PCAC have been used for the numerical calculations. For the 
form factors corresponding to the SCC, the induced scalar form factor $f_3^{np} (Q^2)$ is assumed to be zero. The 
induced tensor form factor $g_2^{np} (Q^2)$ is parameterized in the dipole form with $g_2^{np} (0)$ taken to be either 
purely real or purely imaginary with $M_2 = M_A$. A purely real~(imaginary) value of $g_2^{np} (0)$ implies the 
presence of the SCC with~(without) TRI.

In the $\Delta S = 1$ sector, the electroweak form factors for the nucleon-hyperon transition corresponding to the 
first class vector and axial vector currents $f_1^{NY} (Q^2), f_2^{NY} (Q^2) \text{ and } g_1^{NY} (Q^2)$ for 
$Y=\Lambda, \Sigma^0, \Sigma^-$ have been calculated, respectively, in terms of the electroweak form factors of the 
nucleon and the parameters defined by the analysis of SHD using SU(3) symmetry. In the case of the induced pseudoscalar 
form factor $g_3^{NY} (Q^2)~( Y = \Lambda, \Sigma^0, \Sigma^-)$, a generalized GT relation given by 
Nambu~\cite{Nambu:1960xd} has been used to relate it to $g_1^{NY} (Q^2)$. In the case of the SCC, again we assume the 
induced scalar form factor $f_3^{NY} (Q^2) = 0~ (Y = \Lambda, \Sigma^0, \Sigma^-)$ and use the SU(3) relations to relate
the induced tensor form factor $g_2^{NY} (Q^2)$ to $g_2^{np} (Q^2)$ which is parameterized as described in the case of 
$\Delta S=0$ reactions.
\vspace{5mm}

We summarize our results in the following

\begin{itemize}
 \item [{\bf A.}] \underline{Total scattering cross section:} 
  \begin{enumerate}
  
 \item [(i)] The total cross section $\sigma (E_{\nu_\mu ({\bar{\nu}}_\mu)})$ due to the presence of the SCC in the 
 axial vector sector~(induced tensor term) i.e. $g_2^{NX} (Q^2) \ne 0 ~(X= p,n,\Lambda, \Sigma^0, \Sigma^-)$ increases 
 more with energy $E_{\nu_\mu (\bar{\nu}_\mu)}$ as compared to $g_2^{NX} (Q^2) = 0$. The additional increase is 
 significant only if $g_2^{NX} (0) \ge 1$. The presence of SCC with or without TRI would lead to a smaller value of 
 axial dipole mass $M_A$ by about $10\%$ in the case of $g_2^R (0) =2$ for the neutrino induced processes and $g_2^R 
 (0) = 1$ for the antineutrino induced processes. This will be in addition to the reduction due to the inclusion of 
 meson exchange current (MEC) and nucleon correlations~\cite{Amaro:2017eah, Megias:2016fjk, Butkevich:2017mnc, 
 Martini:2011wp, Martini:2013sha, Nieves:2011yp, Nieves:2013fr}.
 
 \item [(ii)] For a given value of $g_2^{np} (0)$ taken to be purely real in the case of SCC with TRI, the increase in 
 $\sigma (E_{\nu_\mu ({\bar{\nu}}_\mu)})$ is almost the same for the positive as well as the negative values of 
 $g_2^{np} (0)$ as the cross section depends mainly on $|g_2^{np} (Q^2)|^2$ and very mildly on the interference 
 between the first and second class currents involving $g_2^R (0)$. 
 
 \item [(iii)] In the case of $\Delta S = 1$ sector, for the case of SCC with TRI i.e. for the real values of 
 $g_2^{NY} (0)$ ($Y= \Lambda, \Sigma^0, \Sigma^-$), the additional increase in $\sigma (E_{{\bar{\nu}}_\mu})$ is larger 
 for $g_2^{NX} (0)<0$ than for $g_2^{NX} (0)>0$.
 
 \item [(iv)] For a given value of $g_2^{NX} (0)$ taken to be purely imaginary in the case of SCC without TRI, the 
 increase is the same and is independent of the sign of $g_2^{NX} (0)$ as its contribution is proportional to 
 $|g_2^{NX} (Q^2)|^2$.
 \end{enumerate}
 \vspace{5mm}
 
 \item [{\bf B.}] \underline{Average polarizations:}
 \begin{enumerate}
  \item [1.] Leptons

 \begin{enumerate}
  \item [(i)] The longitudinal component of the polarization $\overline{P}_L^{\mu} (E_{\nu_\mu})$ is almost always 
  close to -1 for $\mu^-$ and +1 for $\mu^+$ in reactions $\nu_\mu + n \longrightarrow \mu^- + p$ and $\bar{\nu}_\mu + 
  p \longrightarrow \mu^+ + n$.
  
  \item [(ii)] The perpendicular component of the polarization $\overline{P}_P^{\mu} (E_{\nu_\mu})$ is not very 
  sensitive to the value of $g_2^R (0)$ unless $g_2^R (0)$ (in the case of SCC with TRI) $\ge 1$. It shows no 
  sensitivity to $g_2^I (0)$ in the case of SCC without TRI.
  
  \item [(iii)] The transverse component of the polarization $\overline{P}_T^{\mu} (E_{\nu_\mu})$ is non-zero only in 
  the presence of SCC without TRI and is sensitive to the value of $g_2^I (0)$. It could be $1\% (3\%)$ corresponding 
  to $g_2^I (0) = 1(3)$ for the neutrino induced processes, and $3\% (9\%)$ for the antineutrino induced processes 
  around $E_{\nu_\mu} =$ 500 MeV and decreases with energy.
 \end{enumerate}

 \item [2.] Hadrons
 \begin{enumerate}
  \item [(i)] The longitudinal component of the polarization $\overline{P}_L^h (E_{\bar{\nu}_\mu});~ (h = n, p, 
  \Lambda, \Sigma^0, \Sigma^-)$ is negative for the proton in neutrino scattering and its absolute values decreases 
  with energy for the positive as well as the negative values of $g_2^R (0)$ in the case of SCC with TRI. In the case 
  of SCC without TRI, the positive as well as the negative values of $g_2^I (0)$ gives the same 
  results and it decreases for $g_2^I (0) \ge 0$. In the case of antineutrino scattering, the polarization of neutron 
  and hyperons $\overline{P}_L^{n,\Lambda, \Sigma^-} (E_{\bar{\nu}_\mu})$ is positive at lower energies and becomes 
  negative at higher energies and has similar behavior to the proton polarization $\overline{P}_L^{p} (E_{\nu_\mu})$ 
  with respect to $g_2^R (0)$ and $g_2^I (0)$.
  
  \item [(ii)] The perpendicular component of the polarization $\overline{P}_P^h (E_{\bar{\nu}_\mu})$ is always 
  negative for proton in the neutrino scattering and its absolute values decreases with $g_2^R (0) > 0$ and 
  $|g_2^I (0)| > 0$ and increases for $g_2^R (0) < 0$. In the case of antineutrino scattering, the perpendicular 
  polarization of neutrons and hyperons ($\Lambda, \Sigma^-$) have similar behavior to proton polarization with 
  respect to changes in $g_2^R (0)$ and $g_2^I (0)$.
  
  \item [(iii)] The transverse component of the polarization $\overline{P}_T^h (E_{{\nu}_\mu})$ of the proton in the 
  case of neutrino scattering is non-zero only in the presence of TRV and could be $15\% (45\%)$ for $g_2^I (0) = 1(3)$ 
  at $E_{{\nu}_\mu} = 1$ GeV. In the antineutrino scattering, the polarization of neutrons and hyperons have similar 
  behavior to the proton polarization, except that quantitatively the hyperon polarization ($\Lambda \text{ and } 
  \Sigma$) are smaller than proton polarization but the neutron polarization is comparable to the proton polarization. 
  For $g_2^I (0) < 0$, the transverse polarization $\overline{P}_T^{n,\Lambda, \Sigma^-}(E_{\bar{\nu}_\mu})$ changes 
  sign while the magnitude remains unchanged.
  
 \end{enumerate}
 \end{enumerate}

  \item [{\bf C.}] \underline{$Q^2$ dependence of $d \sigma/dQ^2$ and $P_{L,P,T}^{l,h} (Q^2)$:}
 \begin{enumerate}
  \item [(i)] Numerical results for the differential cross section and the polarization observables i.e. $d\sigma/dQ^2$ 
  and $P_{L,P,T}^{\Lambda} (Q^2)$ for $\bar{\nu}_\mu + p \longrightarrow \mu^+ + \Lambda$ are presented for 
  $E_{\bar{\nu}_\mu} = $ 0.7 GeV, 1 GeV and 3 GeV taking $M_A =$ 1.026 GeV and $M_2 = M_A$.
    
  \item [(ii)] The differential cross section $d \sigma/dQ^2$ for all the processes considered here is not very 
  sensitive to the presence of the SCC with or without TRI.
  
  \item [(iii)] All the polarization components are very sensitive to the numerical values of $g_2^R (0)$ and $g_2^I 
  (0)$ used for evaluating the polarization components $P_{L,P,T}^{\Lambda} (Q^2)$ and also to the sign of $g_2^R (0)$, 
  while in case of TRV, the positive and negative values of $g_2^I (0)$ give similar results.
  
 \end{enumerate}
  \item [{\bf D.}] These calculations are performed with the aim of estimating the effect of the SCC with and without 
  TRI in the quasielastic $\nu$ and $\bar{\nu}$ scattering from the nucleon target. However, presently almost all 
  the antineutrino experiments in the few GeV energy region are using nuclear targets. In the nuclear targets, these 
 results will be modified due to the nuclear medium effects. Therefore, in a realistic situation, these calculations 
 are required to be done in the nuclear targets relevant to the present experiments. Our future plan is to perform 
 such calculations with nuclear medium effects which will be reported elsewhere.
 \end{itemize}
\vspace{2mm}

We conclude that, in the era of precision experiments with neutrino and antineutrino beams, it is possible to get 
information on the SCC with and without TRI in the quasielastic reactions specially in the strangeness sector. 

\section*{Acknowledgment}   
M. S. A. and S. K. S. are thankful to Department of Science and Technology (DST), Government of India for providing 
financial assistance under Grant No. EMR/2016/002285.

\section*{Appendix-I}\label{appendix1}
The expressions $A^h(Q^2)$, $B^h(Q^2)$, $C^h(Q^2)$ and $N(Q^2)$ are expressed in terms of the Mandelstam variables and 
the form factors as:

\begin{eqnarray}\label{Ah}
 A^h(Q^2) &=& - 2\left[ f_1^2 (Q^2) \left(\pm \frac{1}{2} (M + M^\prime) \left(\Delta^2-t\right) \right) 
 \pm \frac{f_2^2 (Q^2)}{(M+M^\prime)^2} \left( \frac{1}{2} t (M + M^\prime) \left(\Delta^2-t\right) \right) \right. 
 \nonumber \\
 &\pm & g_1^2 (Q^2) \left( \frac{1}{2} \Delta \left((M + M^\prime)^2-t\right) \right) 
 \pm \frac{|g_2 (Q^2)|^2}{(M+M^\prime)^2} \left(\frac{1}{2} t \Delta \left((M + M^\prime)^2-t\right) \right) 
 \nonumber \\
 &\pm& \frac{f_1 (Q^2) f_2 (Q^2)}{(M+M^\prime)} \left( \frac{1}{2} \left(4 M M^\prime t+t^2- \Delta^2 \left(M + 
 M^\prime \right)^2 \right) \right) + f_1 (Q^2) g_1 (Q^2) \left( - M^\prime \left(M^2 + {M^\prime}^2-2 s-t\right) 
 \right) \nonumber \\
 &+& \frac{Re[f_1 (Q^2) g_2 (Q^2)]}{(M+M^\prime)} \left( \frac{1}{2} \left(\left( M^2 + {M^\prime}^2-2 s-t\right) 
 \left(-t -2 M^\prime \Delta + \Delta^2 \right) + m_\mu^2 \left(\Delta^2-t\right)\right) \right) \nonumber \\
 &+& \frac{f_1 (Q^2) g_3 (Q^2)}{(M+M^\prime)} \left( -m_\mu^2 \left(\Delta^2-t\right) \right) \nonumber \\
 &+& \frac{f_2 (Q^2) g_1 (Q^2)}{(M+M^\prime)} \left( \frac{1}{2} \left (\left(M^2 - {M^\prime}^2-t\right) \left(M^2 + 
 {M^\prime}^2-2 s-t\right) + m_\mu^2 \left((M + M^\prime)^2-t\right)\right) \right) \nonumber \\
 &+& \frac{Re[f_2 (Q^2) g_2 (Q^2)]}{(M+M^\prime)^2} \left( \frac{1}{2} \left((M + M^\prime) \left(\Delta^2-t\right) 
 \left(m_\mu^2 + M^2 + {M^\prime}^2-2 s-t\right) \right. \right. \nonumber \\
 &+&  \left. \Delta  \left(m_\mu^2 \left((M + M^\prime)^2-t\right)+\left(M^2 - {M^\prime}^2-t\right) 
 \left(M^2 + {M^\prime}^2-2 s-t\right)\right)\right) \Bigg) \nonumber \\ 
 &+& \frac{f_2 (Q^2) g_3 (Q^2)}{(M+M^\prime)^2} \left( -m_\mu^2 (M + M^\prime) \left(\Delta^2-t\right) \right) 
 \nonumber \\
 &\pm& \frac{Re[g_1 (Q^2) g_2 (Q^2)]}{(M+M^\prime)} \left( \frac{1}{2} \left((M + M^\prime)^2-t\right) 
 (- \Delta^2 -t) \right) \Bigg]
\end{eqnarray}

where $(+)-$ sign represents the (anti)neutrino induced scattering and the Mandelstam variables are defined as,

\begin{eqnarray}
 s &=& M^2 + 2 ME, \\
 t &=& M^2 + {M^\prime}^2 - 2 M E^\prime, 
\end{eqnarray}
with
\begin{equation}
 \Delta = M^\prime - M.
\end{equation}

\begin{eqnarray}\label{Bh}
 B^h(Q^2) &=& \frac{2}{M^\prime} \left[ f_1^2 (Q^2) \left(\pm \frac{1}{4} \left(t \left(\Delta^2-2 s\right)-t^2 - 2 
 M^\prime \Delta \left(M^2-s\right) +m_\mu^2 \left(M^2+2 M M^\prime - {M^\prime}^2+t\right) \right) \right) \right. 
 \nonumber \\
 &\pm& \frac{f_2^2 (Q^2)}{(M+M^\prime)^2} \left( \frac{1}{4} \left(t (M + M^\prime) \left(M^3 + M^2 M^\prime - M 
 \left({M^\prime}^2+2 s+t\right) + {M^\prime}^3 - M^\prime t \right) \right. \right. \nonumber \\
 &+& \left.  m_\mu^2 \left(M^4+t (M + M^\prime)^2 - {M^\prime}^4 \right) \right)\Bigg) \nonumber \\
 &\pm& g_1^2 (Q^2) \left( \frac{1}{4} ( \left(-2 M^\prime (M +  M^\prime) \left(M^2-s\right)+t \left((M + M^\prime)^2
 -2 s\right)-t^2 + m_\mu^2 \left(M^2-2 M M^\prime - {M^\prime}^2+t\right)\right) \right) \nonumber \\
 &\pm& \frac{|g_2 (Q^2)|^2}{(M+M^\prime)^2} \left( \frac{1}{4} \Delta  \left(2 M^\prime \left(-2 m_\mu^2 M^2 - M^4 + 
 M^2 \left({M^\prime}^2+s+t\right)+s \left(t- {M^\prime}^2\right) \right) \right. \right.  \nonumber \\
 &+& \left. \left. \Delta \left(- 2 M^\prime (M + M^\prime) \left(M^2-s\right)+t \left((M + M^\prime)^2-2 s\right)
 -t^2\right) + m_\mu^2 \left(M^2 - 2 M M^\prime - {M^\prime}^2+t\right) \right) \right) \nonumber \\
  &\pm& \frac{f_1(Q^2) f_2 (Q^2)}{(M+M^\prime)} \left( \frac{1}{2} \left(M^4 M^\prime + M^3 t- M^2 M^\prime 
 \left({M^\prime}^2+ s\right)-M t \left({M^\prime}^2+2 s+t\right)+ M^\prime \left({M^\prime}^2-t\right) (s+t) \right. 
 \right. \nonumber \\
&+& \left. \left. m_\mu^2 \left(M^3 + M^2 M^\prime + M \left({M^\prime}^2+t\right) - {M^\prime}^3 + M^\prime t\right) 
\right) \right) \nonumber \\
 &+& f_1 (Q^2) g_1 (Q^2) \left( \frac{1}{2} \left(t \left(M^2 + {M^\prime}^2-2 s\right) - 2 s \left(s-M^2\right)
 - t^2 - m_\mu^2 \left(M^2 + {M^\prime}^2-2 s-t\right) \right) \right) \nonumber
  \end{eqnarray}
 \begin{eqnarray}
 &+& \frac{Re[f_1 (Q^2) g_2 (Q^2)]}{(M+M^\prime)} \left( -\frac{1}{2} \left(M^4 M^\prime -2 M^3 s + M^2 \left(
 {M^\prime}^3 - M^\prime (s+t)-\Delta  (2 s+t)\right)+2 M s (s+t) - {M^\prime}^3 s \right. \right. \nonumber \\
 &-& \left. \left. {M^\prime}^2 \Delta  t + M^\prime s t + 2 \Delta  s^2+2 \Delta  s t+\Delta  t^2 + m_\mu^2 
 \left(M^3 + M^2 \Delta -M (3 s+t)+\Delta  \left({M^\prime}^2-2 s-t\right)\right)  + m_\mu^4 M \right) \right) 
 \nonumber \\ 
 &+& \frac{f_1(Q^2) g_3 (Q^2)}{(M+M^\prime)} \left( m_\mu^2 M \left(m_\mu^2 + M^2 - s - t\right) \right) \nonumber \\
 &+& \frac{f_2 (Q^2) g_1 (Q^2)}{(M+M^\prime)} \left( \frac{1}{2} \left( - M^\prime \left(M^2 - s\right) 
 \left(M^2 + {M^\prime}^2 - 2 s\right)+t \left(M^3+2 M^2 M^\prime + M {M^\prime}^2 + {M^\prime}^3 - 3 M^\prime s\right)
 \right. \right. \nonumber \\
 &-& \left. \left. t^2 (M + M^\prime) - m_\mu^2 \left(M^2 M^\prime + M \left({M^\prime}^2+s\right) + M^\prime 
 \left({M^\prime}^2 - 2 s - t\right)\right)  + m_\mu^4 M \right) \right) \nonumber \\
 &+& \frac{Re[f_2 (Q^2) g_2 (Q^2)]}{(M+M^\prime)^2} \left( \frac{1}{4} \left(2 \left(M^3 \left(- {M^\prime}^3 + 
 M^\prime (3 s+t) + \Delta  t\right) + M^2 \left({M^\prime}^4 - {M^\prime}^3 \Delta - {M^\prime}^2 (3 s+t) \right. 
 \right. \right. \right. \nonumber \\
 &+& \left. \left. \left. \left. M^\prime \Delta (3 s+2 t)+2 s t\right) + M M^\prime s \left( {M^\prime}^2 - 2 s-3 
 t\right) + s \left({M^\prime}^2-t\right) \left(2 (s+t)-{M^\prime}^2\right) + M \Delta  t \left({M^\prime}^2 
 -t\right) \right. \right. \right. \nonumber \\
 &+& \left. \left. \left. M^\prime \Delta  (s+t) \left( {M^\prime}^2 - 2 s-t\right) \right. \right. \right) \nonumber \\
 &+& \left. \left.  m_\mu^2 \left(-2 M^4 - 2 M^3 M^\prime + M^2 (-2 M^\prime  \Delta +2 s+t)-2 M \Delta  
 \left({M^\prime}^2 + s\right) + 2 M M^\prime (3 s+t) \right. \right. \right.  \nonumber \\
 &+& \left. \left.  2 M^\prime \Delta  \left(- {M^\prime}^2 + 2 s+t\right) - \left({M^\prime}^2-t\right) 
 (4 s+t)\right) - m_\mu^4 \left(3 M^2 - {M^\prime}^2 + t\right) \right) \Bigg)   \nonumber \\
 &+& \frac{f_2 (Q^2) g_3 (Q^2)}{(M+M^\prime)^2} \left( \frac{1}{2} m_\mu^2 \left(m_\mu^2 \left(M^2+2 M M^\prime - 
 {M^\prime}^2 + t\right) - 2 M^\prime \Delta \left(M^2-s\right)+t \left(\Delta^2 - 2 s\right) - t^2\right) 
 \right) \nonumber \\
 &\pm& \frac{Re[g_1(Q^2) g_2 (Q^2)]}{(M+M^\prime)} \left( \frac{1}{2} \left(M^\prime \left(-2 m_\mu^2 M^2 - M^4 + 
 M^2 \left({M^\prime}^2 + s + t\right) + s \left(t - {M^\prime}^2\right)\right) \right. \right. \nonumber \\
 &+& \left. \left. \Delta  \left(m_\mu^2 \left(M^2 - 2 M M^\prime - {M^\prime}^2 + t\right) - 2 M^\prime (M + 
 M^\prime) \left(M^2 - s\right) + t \left((M + M^\prime)^2 - 2 s\right) - t^2\right) \right) \right)  \Big] \\
 C^h(Q^2) &=& 2 \left[\pm \frac{Im[f_1(Q^2) g_2(Q^2)]}{(M+M^\prime)}\left(-t +2 M \Delta + \Delta^2  \right) \pm
 \frac{Im[f_2(Q^2) g_2(Q^2)]}{(M+M^\prime)}\left(-t +\frac{\Delta  \left(M^2-{M^\prime}^2+t\right)}{M+ M^\prime}+
 \Delta^2 \right) \right. \nonumber \\
 &+& \left. \frac{Im[g_1(Q^2) g_2(Q^2)]}{(M+M^\prime)}\left(M^2 + {M^\prime}^2-2 s-t +  m_\mu^2  \right) 
 + \frac{Im[g_3(Q^2) g_2(Q^2)]}{(M+M^\prime)^2}\left( 2 m_\mu^2 \Delta \right) \right] \\
N(Q^2) &=& f_1^2(Q^2)\left(\frac{1}{2} \left(2 \left(M^2-s\right) \left({M^\prime}^2-s\right)-t \left(\Delta^2-2 s
\right) +t^2 +  m_\mu^2 \left(\Delta^2-2 s-t\right) \right) \right) \nonumber \\
&+& \frac{f_2^2(Q^2)}{(M + M^\prime)^2} \left(\frac{1}{4} \left(-2 t \left(M^4-2 s \left(M^2 + {M^\prime}^2\right) + 
{M^\prime}^4+2 s^2\right)+2 t^2 \left((M + M^\prime)^2-2 s\right) \right. \right. \nonumber \\
&+& \left. \left. m_\mu^2 \left(2 \Delta (M + M^\prime) \left(M^2 + {M^\prime}^2-2 s\right)+t \left(
(M - 3 M^\prime) (M + M^\prime)+4 s)+t^2\right) \right. \right. \right. \nonumber \\
&+& \left. \left. m_\mu^4 (-((3 M-M^\prime) (M+ M^\prime)+t)) \right)\right) \nonumber \\
&+& g_1^2 (Q^2) \left(\frac{1}{2} \left(2 \left(M^2-s\right) \left({M^\prime}^2-s\right)-t \left((M + M^\prime)^2-2 s 
\right)+t^2 + m_\mu^2 \left((M + M^\prime)^2-2 s-t\right) \right) \right)\nonumber \\
&+& \frac{|g_2(Q^2)|^2}{(M + M^\prime)^2} \left( \frac{1}{4} \left(4 \left(\Delta^2-t\right) \left(\left(M^2-s\right) 
\left({M^\prime}^2-s\right)+s t\right)+ m_\mu^2 \left(4 \Delta \left(M^3+ M^2 M^\prime - M (3 s+t) + M^\prime s
\right) \right. \right. \right. \nonumber \\
&+& \left. \left. \left.2 \Delta ^2 \left((M + M^\prime)^2-2 s-t\right)-(4 s+t) \left(\Delta^2-t\right)\right) 
+ 2 \Delta ^2 \left(- 2 \left(M^2-s\right) \left({M^\prime}^2-s\right)-t \left((M + M^\prime)^2+2 s\right)
+t^2\right) \right. \right. \nonumber \\
&+& \left.  m_\mu^4 \left(\Delta^2+4 M \Delta -t\right) \right) \Big) \nonumber \\
&+& \frac{g_3^2(Q^2)}{(M+M^\prime)^2} \left(m_\mu^2 \left(m_\mu^2-t\right) \left(\Delta^2-t\right) \right) \nonumber \\
&+& \frac{f_1 (Q^2) f_2 (Q^2)}{(M+M^\prime)} \left(- \left(t (M + M^\prime) \left(\Delta^2-t\right) + m_\mu^2 
\left(- \Delta \left({M^\prime}^2-s \right)+M^\prime t\right) + m_\mu^4 M \right) \right) \nonumber \\
&\pm& f_1(Q^2) g_1 (Q^2) \left( - \left(t \left(M^2+{M^\prime}^2-2 s-t\right) + m_\mu^2 \left(M^2-{M^\prime}^2+t\right)
\right) \right) \nonumber 
\end{eqnarray}

\begin{eqnarray}
&\pm& \frac{Re[f_1 (Q^2) g_2(Q^2)]}{(M+M^\prime)} \left( - \Delta  \left(t \left(M^2+ {M^\prime}^2-2 s-t\right) + 
m_\mu^2 \left(M^2- {M^\prime}^2+t\right)\right) \right) \nonumber \\
&\pm& \frac{f_2 (Q^2) g_1 (Q^2)}{(M+M^\prime)} \left( -(M + M^\prime) \left(t \left(M^2 + {M^\prime}^2-2 s-t\right) + 
m_\mu^2 \left(M^2 - {M^\prime}^2+t\right)\right) \right) \nonumber \\
&\pm& \frac{Re[f_2 (Q^2) g_2(Q^2)]}{(M+M^\prime)^2} \left( \Delta  (-(M + M^\prime)) \left( t \left(M^2 + {M^\prime}^2 
-2 s-t\right) + m_\mu^2 \left(M^2 - {M^\prime}^2+t\right)\right) \right) \nonumber \\
&+& \frac{Re[g_1 (Q^2) g_2(Q^2)]}{(M+M^\prime)} \left( \left(\Delta  \left(-t (M + M^\prime)^2 +t^2\right)  + 
m_\mu^2 \left(M^3 + M^2 M^\prime + \Delta \left((M + M^\prime)^2-2 s-t\right) \right. \right. \right. \nonumber \\
&-& \left. 3 M s- M t+ M^\prime s\right) +m_\mu^4 M \left. \left. \right) \right) \nonumber \\
&+& \frac{g_1(Q^2) g_3 (Q^2)}{(M+M^\prime)} \left( -2  m_\mu^2 \left(m_\mu^2 M + M^3 - M^2 M^\prime - M (s+t)+
M^\prime s\right) \right) \nonumber \\
&+& \frac{Re[g_2 (Q^2) g_3 (Q^2)]}{(M+M^\prime)^2} \left( m_\mu^2 \left(- 2 \Delta \left(m_\mu^2 M + M^3 - 
M^2 M^\prime - M (s+t ) + M^\prime s \right) \right. \right. \nonumber \\
&-& \left. \left. \left(\Delta^2-t\right) \left(m_\mu^2+2 M^2-2 s-t\right)   \right) \right) 
\end{eqnarray}

\section*{Appendix-II}\label{appendix2}
The expressions $A^l(Q^2)$, $B^l(Q^2)$ and $C^l(Q^2)$ are expressed in terms of the Mandelstam variables and the form 
factors as:
\begin{eqnarray}\label{Al}
 A^l(Q^2)&=& 2 \left[f_1^2(Q^2) \left(\frac{1}{2} m_\mu \left(M^2+2 M M^\prime-s\right) \right) \right. \nonumber \\
 &+&  \frac{f_2^2(Q^2)}{(M + M^\prime)^2} \left( \frac{1}{4} m_\mu \left(-m_\mu^2 ((3 M-M^\prime) 
 (M+M^\prime)+t)-2  M^4+2 M^2 \left({M^\prime}^2+s\right)+4 M M^\prime t \right. \right.  \nonumber \\
 &-& \left. \left. 2 \left({M^\prime}^2-t\right) (s+t)\right) \right) \nonumber \\
 &+& g_1^2(Q^2) \left( \frac{1}{2} m_\mu \left(M^2-2 M M^\prime-s\right)\right) \nonumber \\
 &+& \frac{|g_2 (Q^2)|^2}{(M+M^\prime)^2} \left(\frac{1}{4} m_\mu \left(\left(\Delta^2-t\right) \left(m_\mu^2
 +4 M^2+2 {M^\prime}^2-2 s-2 t\right)+2 \Delta  \left(2 m_\mu^2 M - 4M^2 \Delta \right. \right. \right. 
 \nonumber \\
 &+& \left. \left.  3 M \left({M^\prime}^2-s- t\right)+M^\prime \left(-{M^\prime}^2+s+t\right)\right)+2 
 \Delta ^2 \left(M^2-2 M M^\prime-s\right)\right) \Bigg) \nonumber \\
 &+& \frac{g_3^2 (Q^2)}{(M+M^\prime)^2} \left(m_\mu^3 \left(\Delta^2-t\right) \right) \nonumber \\
 &+& \frac{f_1(Q^2) f_2(Q^2)}{(M+M^\prime)} \left(-\frac{1}{2} m_\mu \left(2 m_\mu^2 M - \Delta \left(2 M^2-
 {M^\prime}^2-s\right)-t (3 M+M^\prime)\right) \right) \nonumber \\
 &+& f_1(Q^2) g_1(Q^2) \left(-m_\mu \left(M^2-s\right) \right) + \frac{Re[f_1(Q^2) g_2(Q^2)]}{(M+M^\prime)} 
 \left(-m_\mu \Delta  \left(M^2-s\right) \right) \nonumber \\
 &+& \frac{f_2(Q^2) g_1(Q^2)}{(M+M^\prime)} \left(-m_\mu (M+M^\prime) \left(M^2-s\right) \right) + \frac{Re[f_2(Q^2)
 g_2 (Q^2)]}{(M+M^\prime)^2} \left( -m_\mu \Delta  (M+M^\prime) \left(M^2-s\right) \right) \nonumber \\
  &+& \frac{Re[g_1(Q^2) g_2(Q^2)]}{(M+M^\prime)} \left(\frac{1}{2} m_\mu  \left(2 m_\mu^2 M +4 M^3+2 M^2 (\Delta -
 2 M^\prime)+M M^\prime (3 M^\prime-4 \Delta ) \right. \right. \nonumber \\
 &-& \left. \left. 3 M (s+t)-{M^\prime}^3+M^\prime (s+ t)-2 \Delta  s\right) \right) \nonumber \\
  &+& \frac{g_1(Q^2) g_3(Q^2)}{(M+M^\prime)} \left(- \left(2 m_\mu^3 M - m_\mu \Delta \left({M^\prime}^2-s-t\right)
 \right) \right) \nonumber 
 \end{eqnarray}
 \begin{eqnarray}
 &+& \frac{Re[g_3(Q^2) g_2(Q^2)]}{(M+M^\prime)^2} \left(-m_\mu \left(m_\mu^2 \left(\Delta^2+2 M \Delta -
 t\right) - t({M^\prime}^2-s-t)\right) \right) \Big] \\
 B^l(Q^2)&=& \frac{2}{M^\prime} \left[f_1^2 (Q^2) \left( \frac{1}{4 m_\mu} M^\prime \left(m_\mu^2 \left(-\left(M^2+
 2 M M^\prime-{M^\prime}^2+t\right)\right)+2 \left(M^2-s\right) \left({M^\prime}^2-s\right)-t \Delta^2+2 s t+t^2\right)
 \right) \right. \nonumber \\
 &+& \frac{f_2^2 (Q^2)}{(M+M^\prime)^2} \left(\frac{1}{8 m_\mu} M^\prime \left(m_\mu^4 ((3 M-M^\prime) (M+
 M^\prime)+t) + m_\mu^2 \left(\Delta^2-t\right) \left(2 (M+M^\prime)^2+t\right) \right. \right. \nonumber \\
 &-& \left. \left. 2 t \left(M^4-2 s \left(M^2+{M^\prime}^2\right)+{M^\prime}^4+2 s^2\right)+2 t^2 \left((M+
 M^\prime)^2-2 s\right)\right) \right) \nonumber \\
 &+& g_1^2 (Q^2) \left( \frac{1}{4 m_\mu}M^\prime \left(m_\mu^2 \left(-\left(M^2-2 M M^\prime-{M^\prime}^2+t\right)
 \right) +2 \left(M^2-s\right) \left({M^\prime}^2-s\right)-t (M+M^\prime)^2+2 s t+t^2\right) \right) \nonumber \\
 &+& \frac{|g_2(Q^2)|^2}{(M+M^\prime)^2} \left(\frac{1}{8 m_\mu} M^\prime \left(m_\mu^4 \left(-\left(\Delta^2
 +4 M \Delta -t\right)\right)+m_\mu^2 \left(-2 \Delta ^2 \left(M^2-2 M M^\prime-{M^\prime}^2+t\right) \right. \right. 
 \right. \nonumber \\
 &-& \left. \left. \left. \left(4 M^2 -t\right) \left(\Delta^2-t\right)+4 M \Delta (2 M \Delta + t)\right)+2 \Delta^2
 \left(2 \left(M^2-s\right) \left({M^\prime}^2-s\right)-t (M+M^\prime)^2+2 s t+t^2\right) \right. \right. \nonumber \\
 &-& \left. 8 \Delta^2 \left(\left(M^2 -s\right) \left({M^\prime}^2-s\right)+s t\right)+4 \left(\Delta^2-
 t\right) \left(\left(M^2-s\right) \left({M^\prime}^2-s\right)+s t\right)\right) \Big) \nonumber \\
 &+& \frac{g_3^2 (Q^2)}{(M+M^\prime)^2} \left(-\frac{1}{2} m_\mu M^\prime \left(m_\mu^2-t\right) \left(\Delta^2
 -t\right) \right) \nonumber \\
 &+& \frac{f_1(Q^2) f_2(Q^2)}{(M+M^\prime)} \left(\frac{1}{2 m_\mu} M^\prime \left(m_\mu^2-t\right) \left(m_\mu^2 M+
 (M+M^\prime) \left((M-M^\prime)^2-t\right)\right) \right) \nonumber \\
 &+& f_1(Q^2) g_1 (Q^2) \left(\frac{1}{2 m_\mu} M^\prime \left(m_\mu^2-t\right) \left(M^2+{M^\prime}^2-2 s-t\right)
 \right) \nonumber \\
 &+& \frac{Re[f_1(Q^2) g_2 (Q^2)]}{(M+M^\prime)} \left(\frac{1}{2 m_\mu} M^\prime \Delta  \left(m_\mu^2-t\right) 
 \left(M^2+{M^\prime}^2-2 s-t\right) \right) \nonumber \\
 &+& \frac{f_2 (Q^2) g_1(Q^2)}{(M+M^\prime)} \left(\frac{1}{2 m_\mu} M^\prime \left(m_\mu^2-t\right) (M+M^\prime) 
 \left(M^2+{M^\prime}^2-2 s-t\right) \right) \nonumber \\
 &+& \frac{Re[f_2 (Q^2) g_2 (Q^2)]}{(M+M^\prime)^2} \left(\frac{1}{2 m_\mu} M^\prime \Delta  \left(m_\mu^2-t\right) 
 (M+M^\prime)\left(M^2+{M^\prime}^2-2 s-t\right) \right) \nonumber \\
 &+& \frac{Re[g_1(Q^2) g_2(Q^2)]}{(M+M^\prime)} \left( \frac{1}{2 m_\mu} M^\prime \left(-m_\mu^4 M+m_\mu^2 
 \left(-2 M^3+M^2 (2 M^\prime-\Delta )+M (2 M^\prime \Delta +t)+\Delta  \left({M^\prime}^2-t\right)\right) \right. 
 \right. \nonumber \\
 &+&  \left. \Delta  \left(2 \left(M^2-s\right) \left({M^\prime}^2-s\right)-t \left((M+M^\prime)^2-2 s \right)+t^2
 \right) -2 \Delta \left(\left(M^2-s\right) \left({M^\prime}^2-s\right)+s t\right)\right) \Bigg) \nonumber \\
 &+& \frac{g_1(Q^2) g_3(Q^2)}{(M+M^\prime)} \left(m_\mu M M^\prime \left(m_\mu^2-t\right) \right) \nonumber \\
 &+& \frac{Re[g_2 (Q^2) g_3 (Q^2)]}{(M+M^\prime)^2} \left(\frac{1}{2} m_\mu M^\prime \left(m_\mu^2-t\right) 
 \left(\Delta^2+2 M \Delta -t\right) \right) \Bigg] \\
 \label{Cl}
 C^l(Q^2) &=& 2 \left[\frac{Im[g_1(Q^2) g_2(Q^2)]}{(M+M^\prime)} \left(m_\mu (M+M^\prime) \right) + 
 \frac{Im[g_3(Q^2) g_2(Q^2)]}{(M+M^\prime)^2} (2 m_\mu t) \right]
\end{eqnarray}

\end{document}